\documentclass[English,superscriptaddress,prl,twocolumn]{revtex4-2}
\usepackage{graphicx}
\usepackage{epstopdf}
\usepackage[ansinew]{inputenc}
\usepackage{array}
\usepackage{color}
\usepackage{amsmath}
\usepackage{amsxtra}
\usepackage{amstext}
\usepackage{amssymb}
\usepackage{latexsym}
\usepackage{float}
\usepackage{soul}
\usepackage[colorlinks=true, linkcolor=black,urlcolor=black,citecolor=black]{hyperref}
\usepackage{lineno}
\begin{document}

%%%% 75 characters including spaces
\title{Non-identical moir\'e twins in bilayer graphene}

\author{E. Arrighi$^{\dagger}$}
\affiliation{Universit\'e Paris-Saclay, CNRS, Centre de Nanosciences et de Nanotechnologies (C2N), 91120 Palaiseau, France}
\author{V.-H. Nguyen$^{\dagger}$}
\affiliation{Institute of Condensed Matter and Nanosciences, Universit\'e catholique de Louvain (UCLouvain), 1348 Louvain-la-Neuve, Belgium}
\author{M. Di Luca}
\affiliation{Universit\'e Paris-Saclay, CNRS, Centre de Nanosciences et de Nanotechnologies (C2N), 91120 Palaiseau, France}
\author{G. Maffione}
\affiliation{Universit\'e Paris-Saclay, CNRS, Centre de Nanosciences et de Nanotechnologies (C2N), 91120 Palaiseau, France}
\author{Y. Hong}
\affiliation{Universit\'e Paris-Saclay, CNRS, Centre de Nanosciences et de Nanotechnologies (C2N), 91120 Palaiseau, France}
\author{L. Farrar}
\affiliation{Universit\'e Paris-Saclay, CNRS, Centre de Nanosciences et de Nanotechnologies (C2N), 91120 Palaiseau, France}
\author{K. Watanabe}
\affiliation{National Institute for Materials Science, 1-1 Namiki, Tsukuba, Japan}
\author{T. Taniguchi}
\affiliation{National Institute for Materials Science, 1-1 Namiki, Tsukuba, Japan}
\author{D. Mailly}
\affiliation{Universit\'e Paris-Saclay, CNRS, Centre de Nanosciences et de Nanotechnologies (C2N), 91120 Palaiseau, France}
\author{J.-C. Charlier}
\affiliation{Institute of Condensed Matter and Nanosciences, Universit\'e catholique de Louvain (UCLouvain), 1348 Louvain-la-Neuve, Belgium}
\author{R. Ribeiro-Palau*}
\affiliation{Universit\'e Paris-Saclay, CNRS, Centre de Nanosciences et de Nanotechnologies (C2N), 91120 Palaiseau, France}

\maketitle

\noindent{\bf \textcolor{black}{The superlattice obtained by aligning a monolayer graphene and boron nitride (BN) inherits from the hexagonal lattice a sixty degrees periodicity with the layer alignment. It implies that, in principle, the properties of the heterostructure must be identical for 0$^{\circ}$ and 60$^{\circ}$ of layer alignment. Here, we demonstrate, using dynamically rotatable van der Waals heterostructures, that the moir\'e superlattice formed in a bilayer graphene/BN has different electronic properties at 0$^{\circ}$ and 60$^{\circ}$ of alignment. Although the existence of these non-identical moir\'e twins is explained by different relaxation of the atomic structures for each alignment, the origin of the observed valley Hall effect remains to be explained. A simple Berry curvature argument do not hold to explain the hundred and twenty degrees periodicity of this observation. Our results highlight the complexity of the interplay between mechanical and electronic properties on moir\'e structure and the importance of taking into account atomic structure relaxation to understand its electronic properties.}}\\

%%%%%%%%%%%%%%%% INTRO
\noindent\textbf{INTRODUCTION}\\

When the crystallographic alignment of monolayer graphene and BN is almost perfect (close to zero degrees between layers), the electronic, mechanical and optical properties of graphene are strongly modified \cite{Hunt2013,Wang2015,Ponomarenko2013,Ribeiro-Palau2018}. This is caused by the combination of two effects: i) a long-wavelength geometric interference pattern, called  a moir\'e pattern, which effectively acts as a periodic superlattice, and ii) a local enlargement of the lattice constant of graphene to match the one of BN at the inner part of the moir\'e pattern, leading to a local commensurate state. Outside of the commensurate areas the accumulated stress, due to the stretching of the lattice constant, is released in the form of out-of-plane corrugations where the stacking order changes rapidly in space \cite{Woods2014}. These corrugations have the same periodicity as the moir\'e pattern. For monolayer graphene, both the long-wavelength pattern and the commensurate state are observed every time one of the layers is rotated by sixty degrees.

 The commensurate state creates an imbalance of the interaction that the carbon atoms have with the BN substrate breaking the sublattice symmetry \cite{Woods2014}. In monolayer graphene/BN the breaking of inversion symmetry has been proposed as the origin of the opening of an energy gap at the charge neutrality point (CNP) \cite{Jung2015}  and non-trivial quantum geometry characteristics of the electronic band structure \cite{Song2015}. However, very little is known about how the graphene/BN alignment  affects systems with more than one layer, such as bilayer graphene.  

Here, we demonstrate that the electronic properties of the commensurate state \textcolor{black}{in a bilayer graphene/BN heterostructure have a hundred and twenty degrees periodicity. We present experimental electron transport measurements in dynamically rotatable van der Waals heterostructures \cite{Ribeiro-Palau2018} made of Bernal stacked bilayer graphene and BN. Our measurements reveal distintic behaviors for 0$^{\circ}$ and 60$^{\circ}$ which we attribute to different electronic band structures generated by different atomic displacements inside the moir\'e superlattice. However, the observation of the valley Hall effect, only present for 0$^{\circ}$ of alignment remains to be explained given that the current theoretical model fail to explain this hundred and twenty degrees periodicity.}\\

\begin{figure*}[!ht]
\centering
\includegraphics[width=\textwidth]{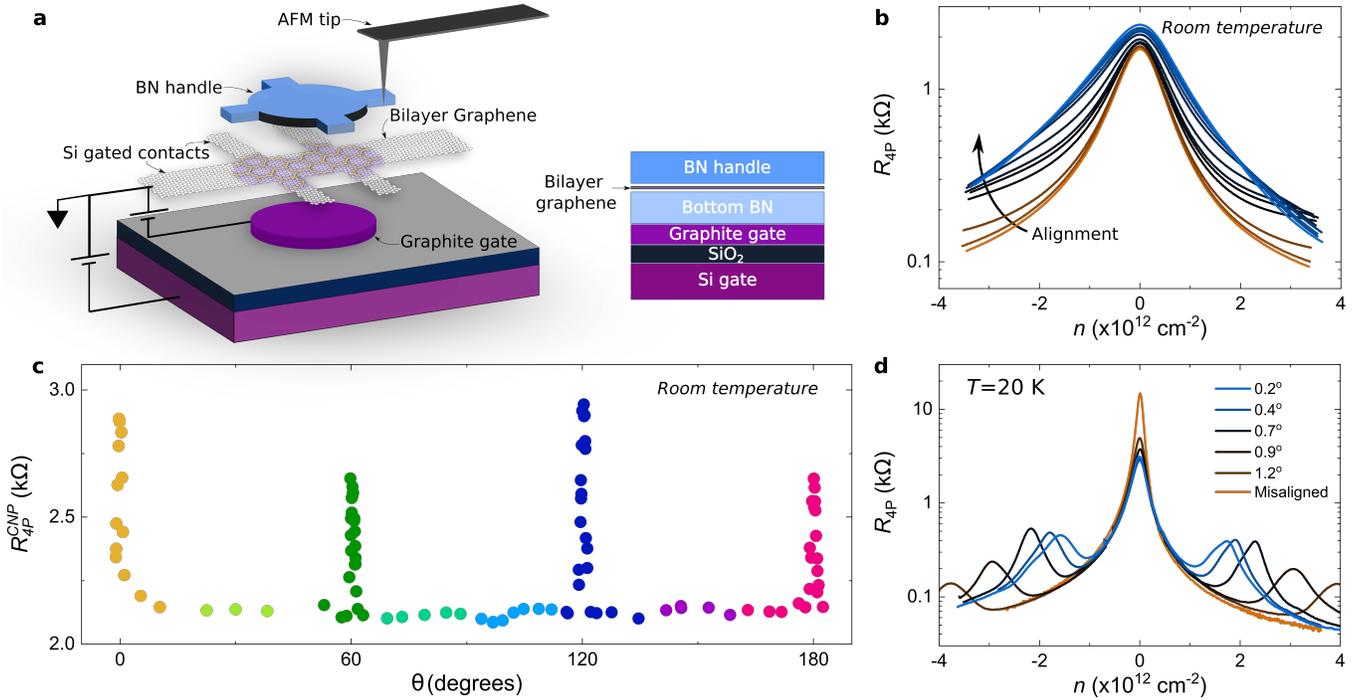}  
\caption{{\bf Dynamically rotatable heterostructure}. \textbf{a}, Schematic representation (left) of a dynamically rotatable van der Waals heterostructure, the BN between the graphite and graphene layers has been omitted for clarity. The central circular shape on the graphene represents the range of action of the graphite gate and the moir\'e superlattice. Cross section of the same heterostructure  is presented in the right pannel. \textbf{b}, Four probes resistance measurement as a function of carrier density for several angular alignment between the bilayer graphene and the BN handle, from completely misaligned (brown) to fully aligned (light black)  at room temperature, measurements of sample II.  \textbf{c}, Resistance of the CNP as a function of the angular alignment,  measured with the AFM, from -2$^{\circ}$ to 182$^{\circ}$, data taken in sample III.  \textbf{d}, Four probes resistance measurement as a function of carrier density for a selection of angular alignments of b at 20 K, sample II. The angular alignment is calculated from the position in energy of the satellite peaks.}
\label{Device}
\end{figure*}

\noindent\textbf{RESULTS}\\

\noindent\textbf{Angle calibration and room temperature experiments}\\

A schematics of our device and its cross section is shown in Fig. \ref{Device}a. The dynamically rotatable van der Waals heterostructures are realized as described in \cite{Ribeiro-Palau2018}, with the improvement of having a pre-shaped local graphite gate. The latter controls the carrier density only in the central area of our device, and has the same dimensions as the BN structure used to create the moir\'e pattern. It is important to mention that the bottom BN and graphene layers are intentionally  misaligned, to more than $10^{\circ}$, to avoid the formation of a double moir\'e \cite{Sun2021,Wang2019,Finney2019}. The carrier density of the external parts of graphene is tuned by the global Si gate, acting effectively as a tunable contact resistance \cite{Ribeiro-Palau2019}. The angular alignment of the bilayer graphene/BN heterostructure is controlled {\it in situ} by means of a pre-shaped BN handle deposited on top of graphene. This handle can be rotated by applying a lateral force with the tip of an atomic force microscope (AFM), Fig. \ref{Device}a. The AFM images of the three main positions described in this report: $0^{\circ}$, $30^{\circ}$ and $60^{\circ}$ can be see in Fig. S1 of the Supplementary Information. The alignment is fixed at room temperature inside the AFM using as a reference  electron transport measurements (see below), after which the sample is moved to a cryostat for low temperature experiments. The carrier mobility of \textcolor{black}{our samples} ranged from 150.000 to 200.000 cm$^{2}$V$^{-1}$s$^{-1}$ for intermediate densities $\pm0.65$x$10^{12}$ cm$^{-2}$ at $T<10$ K. The mean free path was calculated to be between 1.2 $\mu$m and 2 $\mu$m for the same carrier density and temperature range (see Fig. S8 of the SI for details). \textcolor{black}{These values of the mean free path are comparable with the nominal dimension of our samples $W=1.7~\mu$m and $L=2.3~\mu$m, reflecting a ballistic transport regime.} All measurements presented here were taken using lock-in amplifiers at $f\approx33.37$ Hz and applied currents of 10 nA. The non-local voltages were measured using a high input impedance voltage amplifiers to ensure the measurement had no leaking current effects. \textcolor{black}{We also ensure that this is not a heating effect by performing the same measurements at different currents (for more details see Fig. S10 and S11 of the SI).}

%%%%%%%%%%%%%%%%
\begin{figure*}[!ht]
%\centering
\includegraphics[width=\textwidth]{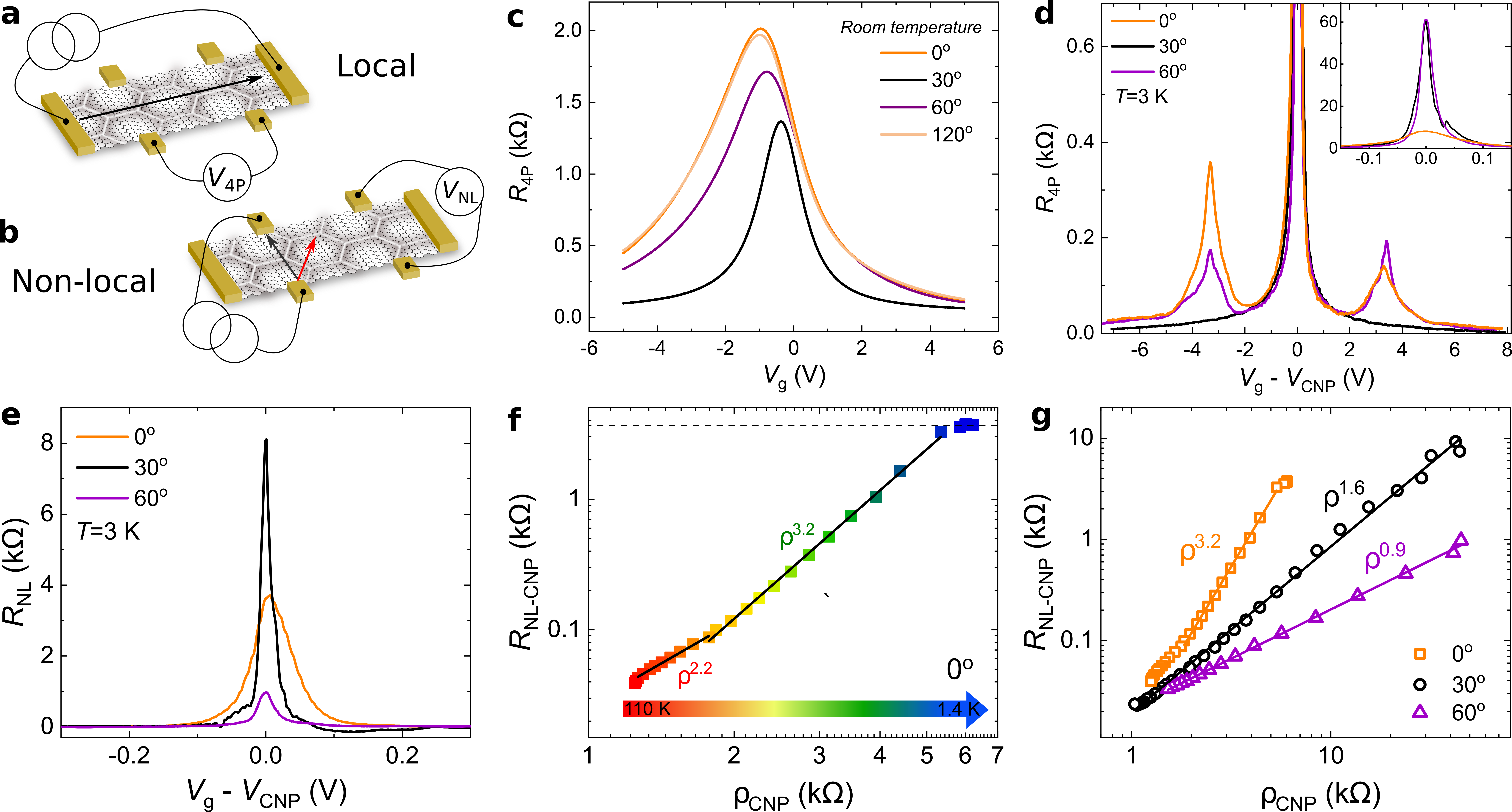} \caption{{\bf Local and non-local transport measurements.} \textbf{a-b} Artistic representation of the local \textcolor{black}{(a) and non-local (b) measurement configurations. \textbf{c-d}, Local measurements as a function of the applied gate voltage at room temperature (c) and $T=$3 K, insert: zoom around the CNP. \textbf{e}, Non-local resistance $R_{\mathrm{NL}}=V_{\mathrm{NL}}/I$ as a function of the gate voltage around the CNP at $T=3$ K. \textbf{f}, Non-local resistance as a function of the local resistivity $\rho=(W/L)R_{\mathrm{4P}}$, both  at the CNP, for 0$^{\circ}$. The color scale indicates different temperatures. Dashed line indicates the expected value of the non-local resistance for a valley angle of $\pi/2$, see text. Solid black lines are fit to the experimental data. \textbf{g}, Non-local resistance as a function of the local resistivity, both at the CNP, for different angular alignments. Solid lines are linear fits to the experimental data meant to extract the power law dependence. Both \textbf{f} and \textbf{g} are measurements between $T=1.4$ K and $T=110$ K.}}
\label{Local-NonLocal}
\end{figure*}

In contrast with monolayer graphene \cite{Ribeiro-Palau2018}, for bilayer aligned with BN the presence of satellite peaks in charge transport measurements - a clear signature of the moir\'e superlattice - becomes evident only at low temperatures. At room temperature, these satellite peaks are not visible. \textcolor{black}{This is explained by a smaller intensity of these satellite peaks in the bilayer case, which makes them indistinguishable from the CNP at room temperature due to thermal broadening, as  can be seen in the full temperature dependence curves of Figs. S5 and S6 in the SI. In the case of the bilayer,} the signature of crystallographic alignment is then given by a  broadening of the resistance peak around the charge neutrality point (CNP), Fig. \ref{Device}b. The combination of room and low temperature measurements, Fig. \ref{Device}b and \ref{Device}d respectively, allows us to have a calibration of the angular alignment at room temperature. \textcolor{black}{The broadening of the resistance peak and its corresponding increase in magnitude are observed every sixty degrees of alignment at room temperature, Fig. \ref{Device}c. However, the maximum of the resistance at the CNP for the aligned position is in fact periodic every hundred and twenty degrees of rotation, Fig. \ref{Device}c and \ref{Local-NonLocal}c, with a slow decrease as the moir\'e length is reduced. Other values, such as the position of the CNP in gate voltage are also affected and hold the same periodicity, see Fig. S3 and S4 of the SI. For reference, and to be consistent among all our samples, we establish that the aligned position with the highest resistance at the CNP will be named 0$^{\circ}$ of alignment. The features of alignment presented here, such as the observation of the valley Hall effect, are all consistently observed in the alignment with highest resistant at room temperature.}\\

\noindent\textbf{Local and Non-local charge transport response}\\

At low temperatures, for both 0$^{\circ}$ and 60$^{\circ}$  of alignment,  the local charge transport measurement shows the presence of very well pronounced satellite peaks at both sides of the CNP, Fig. \ref{Local-NonLocal}d. These satellite  peaks  are accompanied  by  a  sign  reversal  of  the Hall resistance $R_{\mathrm{xy}}$ when a low magnetic field is applied (see Fig. S7 of the SI). In the case of 30$^{\circ}$, as expected, these satellite peaks are not present since graphene and BN are completely misaligned and the moir\'e pattern is absent. From magneto-transport measurements we extract a moir\'e wavelength of $\lambda$=14.1$\pm$0.4 nm and $\lambda$=14.5$\pm$0.3 nm for $0^{\circ}$ and $60^{\circ}$, respectively (see Fig. S12 of the SI). The good coincidence between the positions of the satellite peaks and very close values of the extracted moir\'e wavelength allow us to say that a good alignment is reached in both cases. Notice that at low temperatures  \textcolor{black}{the difference between these two alignments can be reduced to a different magnitude of the local resistance at the satellite peak and CNP (see Fig. \ref{Local-NonLocal}d), which in an experiment using samples with fixed angles would be attributed to a sample-to-sample dependence.}

 \begin{figure*}[!ht]
\centering
\includegraphics[width=\textwidth]{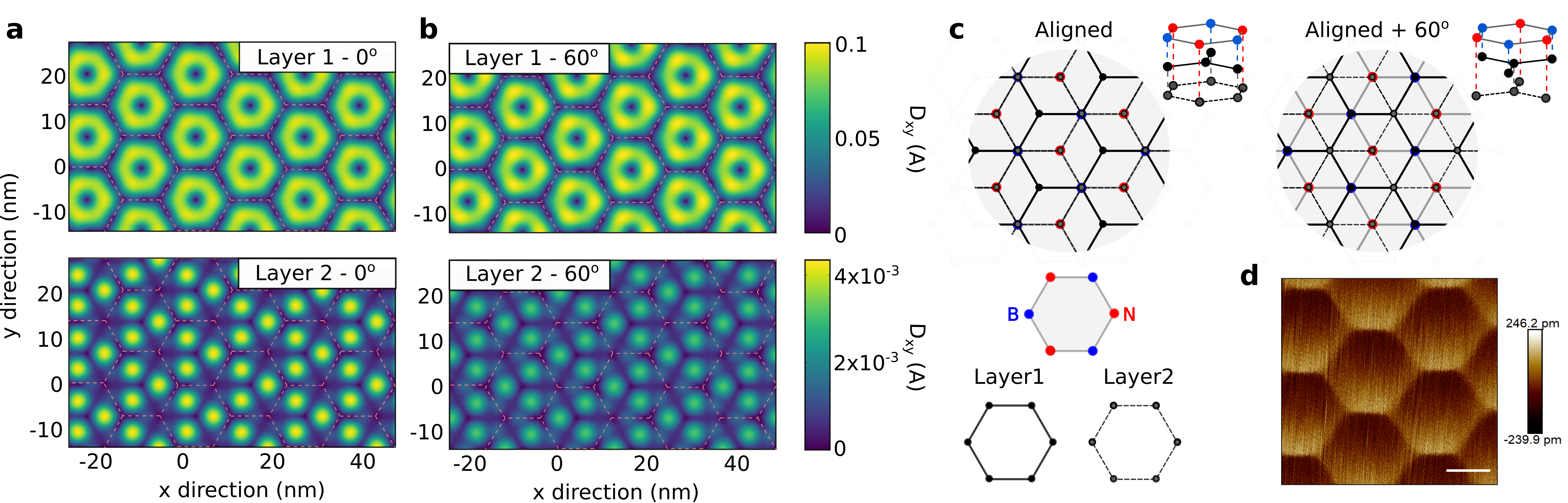} 
\caption{{\bf Atomic structure relaxation for bilayer graphene/BN.}  \textbf{a-b}, In-plane displacement, D$_{xy}$, of the carbon atoms for both graphene layers at $0^{\circ}$ and $60^{\circ}$ of alignment.  Pink dashed lines in a and b are guides for the eye to follow the moir\'e superlattice. \textbf{c}, Atomistic sketch of a BN/bilayer graphene for $0^{\circ}$ (left) and  $60^{\circ}$ (right) alignments. Description: black dots boron, red dots nitrogen, black dots carbon atoms of layer 1 and grey dots carbon atoms of layer two. gray lines represent bonds between boron and nitrogen, solid black line bonds between carbon atoms of layer 1 and black dashed line represent the bonds between carbon atoms of layer 2. \textbf{d}, AFM image in PeakForce mode, height sensor, of an aligned bilayer graphene BN. Scale bar 7 nm. The angular alignment (0$^{\circ}$ or 60$^{\circ}$) is unknown. }
\label{Simulations}
\end{figure*}

%%%%%%%%%%%%%%%%%%%%

\textcolor{black}{In order to explore more subtle modifications of the properties of this system, we changed the measurement configuration to a non-local one,  represented in Fig. \ref{Local-NonLocal}b}. The non-local electrical signal refers to the appearance of a voltage across contacts that are well outside the expected path of the current. This technique is largely used to detect spin/pseudospin signals \cite{Wang2015Jan,Gorbachev2014,Sui2015,Shimazaki2015,Brune2010}. Non-local signals, attributed to the existence of valley currents,  have been measured previously in aligned monolayer graphene/BN  \cite{Gorbachev2014,Komatsu2018,Li2020}, in bilayer graphene aligned with BN \cite{Endo2019} and in the presence of a strong displacement field \cite{Shimazaki2015,Sui2015} as well as in other 2D materials \cite{Wu2019}. In this report we focus only on non-local signals at the CNP, since the non-local signals around the satellite peaks are too weak to be studied systematically with our current experimental setup. As in previous reports \cite{Gorbachev2014,Sui2015}, the non-local resistance, $R_{\mathrm{NL}}$ - Fig. \ref{Local-NonLocal}e, decays rapidly with carrier density to values lower than our experimental measurement noise. Furthermore, we observe a very strong dependence of the non-local signal with the angular alignment. Indeed, the maximum value of the non-local resistance at the CNP decreases by a factor four between the measurements at 0$^{\circ}$ and the one at 60$^{\circ}$, Fig. \ref{Local-NonLocal}e, \textcolor{black}{a contrasting behavior with respect to the local signal where the 0$^{\circ}$ of alignment has a much smaller signal than the 60$^{\circ}$, see Fig.\ref{Local-NonLocal}d-insert. This suggest that the non-local signal is independent of the local one. In other words that this is not a simple ohmic response.}

Plotting the non-local resistance versus local resistivity, $\rho$, at the CNP for different temperatures, for 0$^{\circ}$ of alignment (Fig. \ref{Local-NonLocal}f), we observe three regimes: for $T>40$ K  an approximately quadratic dependence $R_{\mathrm{NL}}\propto\rho^{2.2}$; for 40 K $\geq T\geq12$ K a near-cubic relation, $R_{\mathrm{NL}}\propto\rho^{3.2}$, and a saturation regime for  $<12$ K. Let us start the discussion with what happens for $T\leq40$ K. In analogy with the spin Hall effect \cite{Abanin2009}, this cubic relation is expected in graphene when the valley Hall and inverse valley Hall effects are in operation. In particular,  the non-local resistance and local resistivity are related by \cite{Beconcini2016}:

\begin{equation}
    R_{\mathrm{NL}}=\frac{W}{2l_{\mathrm{v}}}(\sigma^{\mathrm{v}}_{\mathrm{xy}})^{2}\rho^{3}e^{-L/l_{\mathrm{v}}},
    \label{cubic}
\end{equation}

\noindent when $\sigma^{\mathrm{v}}_{\mathrm{xy}}\ll\sigma$. Here, $l_{\mathrm{v}}$ is the inter-valley scattering length, $\sigma^{\mathrm{v}}_{\mathrm{xy}}$ is the valley Hall conductivity, $\rho=1/\sigma$ is the local resistivity and $W$ and $L$ are the width and length of the sample, respectively. 

For $T\leq12$ K, Fig. \ref{Local-NonLocal}f, there is a clear change of behavior, characterized by a saturation of the non-local resistance. This is consistent with the regime where $\sigma^{\mathrm{v}}_{\mathrm{xy}}\gg\sigma$, and the non-local response becomes independent of the local resistivity \cite{Beconcini2016}:

\begin{equation}
    R_{\mathrm{NL}}=\frac{W}{2l_{\mathrm{v}}}\frac{1}{\sigma^{\mathrm{v}}_{\mathrm{xy}}}.
    \label{saturation}
\end{equation}

 Since the Fermi energy is at the CNP, and the temperature of the system is much lower than the energy gap, the valley Hall conductivity is maximal having a value of $\sigma^{\mathrm{v}}_{\mathrm{xy}}=4e^{2}/h$ ($e$ is the electron's charge and $h$ is Planck's constant). In this case all occupied states in the valence band contribute to the valley Hall effect \cite{Yamamoto2015,Beconcini2016,Gorbachev2014,Sui2015,Shimazaki2015,Li2020}. Using the saturation value of the non-local resistance, dashed horizontal line of Fig. \ref{Local-NonLocal}f, and maximum valley Hall conductivity we obtain an inter-valley scattering length $l_{\mathrm{v}}=1.5~\mu$m, in good agreement with previous reports \cite{Gorbachev2014,Shimazaki2015,Sui2015}. 
 
 This behavior, where the non-local signal is independent of the local response, is characteristic of a system where the valley conductivity is larger than the local conductivity, which implies a fully developed valley Hall effect, where the Hall angle becomes $\pi/2$. \textcolor{black}{For $T>40$ K we do not have a clear picture of why we observe a nearly quadratic behavior, we attribute this to a mixture of regimes where both the valley Hall effect and ohmic response compete. This regime needs more experimental and theoretical investigation.}

\begin{figure*}[!ht]
\centering
\includegraphics[width=\textwidth]{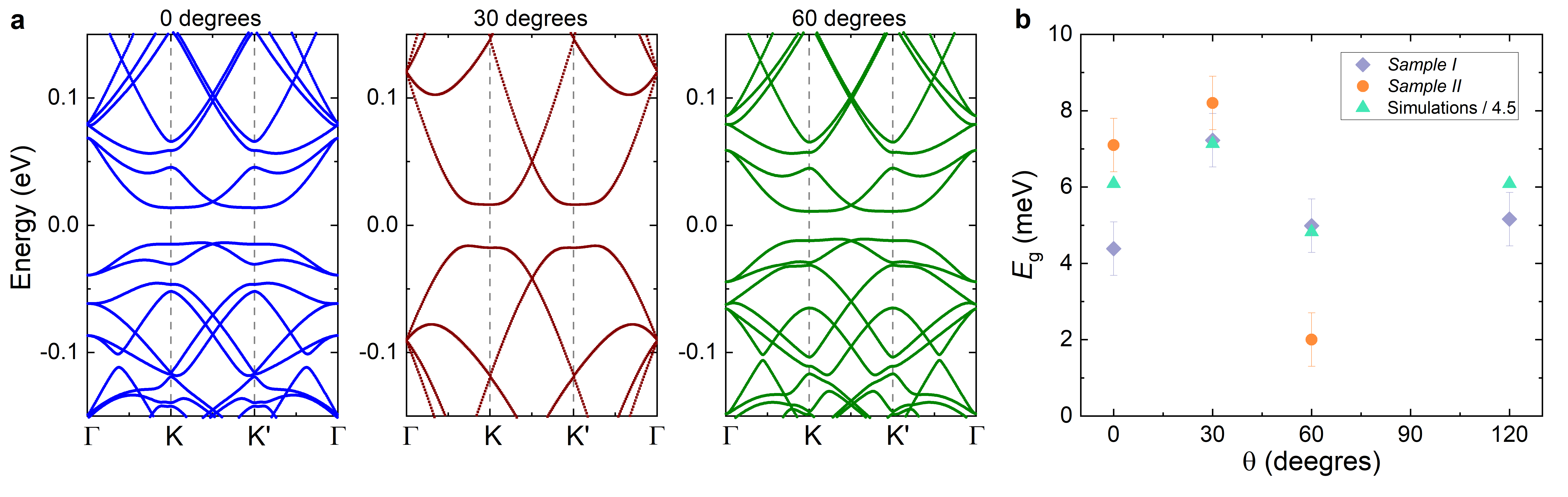} 
\caption{{\bf Electronic band structure for different crystallographic alignments.}  \textbf{a}, electronic band structure for a relaxed bilayer graphene aligned on BN with 0.1 V/nm of displacement field. \textbf{b}, Local energy gaps obtained by thermal activation at different crystallographic alignments (for sample I and II) and theoretical values obtained from (a) divided by a factor of 4.5. Error bars represent the standard deviation of our measurements. }
\label{SimulationsBS}
\end{figure*}
 
If we now change the crystallographic alignment of the layer, by \textit{in situ} rotation of the BN handle with the AFM tip at room temperature, we can see that the 60$^{\circ}$ case is very different, Fig. \ref{Local-NonLocal}g. In this case the relation between the local and non-local resistance is close to linear, $R_{\mathrm{NL}}\propto\rho^{0.9}$. This behavior is consistent with an ohmic contribution $R_{\mathrm{NL}}=4\rho e^{-\pi L/W}/\pi$, and can be adjusted by using only the geometry of our sample, as is expected given its van der Pauw geometry. \textcolor{black}{In contrast with the 0$^{\circ}$ of alignment this} linear behavior can be observed over the full temperature range. This striking difference of the development of the valley Hall effect at 0$^{\circ}$ but absent at  60$^{\circ}$ reveals that the consequences of the moir\'e patterns for the two alignments are non-identical. Changing the alignment further to $120^{\circ}$ restores the signatures of the valley Hall effect (for more details see Fig. S17 of the SI).

Interestingly, neither the valley Hall effect, observed at 0$^{\circ}$, or the ohmic behavior, observed at 60$^{\circ}$, are reproduced in the fully misaligned case, 30$^{\circ}$, Fig. \ref{Local-NonLocal}g. In this case, where no signature of alignment is observed in the local charge transport or magneto-transport measurements, we observe a $R_{\mathrm{NL}}\propto\rho^{1.6}$ relation. \textcolor{black}{This behavior, unrelated to the valley Hall effect, could be explained either by the existence of localized  and non-topological edge states resulting from edge disorder \cite{Aharon-Steinberg2021}. Or by the presence of electronic jets separated of $\approx60$ degrees between them, and consequence of the trigonal warping of the bilayer graphene electronic band structure \cite{Gold2021Jul}. The existence of these localized states  contrasts with the ohmic response observed for the 60$^{\circ}$ of alignment. However, in both cases we can hypothesize that the periodic moir\'e potential may prevent the formation of the localized states (in the same way as artificial disorder generated by a scanning gate does \cite{Aharon-Steinberg2021}) and that it will modify the trigonal warping of the electronic band structure. Further combinations of scanning gates and electron transport will be necessary to clarify our observation.}\\

%%%%%%%%%%%%%%%

\noindent \textbf{Atomic structure relaxation inside the moir\'e cell}\\

\textcolor{black}{To understand why these two angular alignments give rise to different behaviors} we investigated the  in-plane atomic structural relaxation of each layer, Fig. \ref{Simulations}a-b. For this we use atomistic numerical simulations to obtain the atomic structural relaxation of each layer and its consequence in the electronic band structure of both alignments. Classical molecular dynamics is used to relax the system {\it i.e.}, minimize forces and energy. In particular, intra-layer forces are computed using the optimized Tersoff and Brenner potentials \cite{Lindsay2010}, while inter-layer interactions are modeled using the Kolmogorov-Crespi potentials \cite{Leven2016,Kolmogorov2005}. A lattice mismatch ($\sim$1.8\%) between BN and graphene is taken into account. The atomic structure is optimized until all force components are smaller than 1 meV/atom. Similarly as the features discussed in \cite{Rickhaus2019Dec}, the presence of misaligned hBN substrate induces a small crystal field ($\approx$15 meV/nm) in bilayer graphene, leading to a small correction on the simulated bandgap. This correction was added in calculations of Fig.\ref{SimulationsBS}a. As illustrated in Fig. \ref{Simulations}a and b, the in-plane atomic displacement, $D_{\mathrm{xy}}$, clearly shows that, for the layer that is closer to the BN (layer 1), there is an almost circular symmetry around the centre of each moir\'e superlattice (marked by the pink dashed lines). On the other hand, the second layer shows a breaking of this symmetry into a $2\pi/3$ rotational symmetry. The in-plane atomic displacement of the second layer is also at least one order of magnitude smaller than for the first layer.  Additionally, we can see that the in-plane atomic displacement on the second layer is larger in the case of 0$^{\circ}$. The differences in the stretching of each layer results in the spatial variation of stacking structure of the bilayer graphene (initially, perfect AB stacking) as illustrated in Fig. S23 of the SI.

\textcolor{black}{The difference of the in-plane  atomic structure relaxations for 0$^{\circ}$ and 60$^{\circ}$ can be traced back to the  Bernal stacked configuration, see Fig. \ref{Simulations}c. We assume, that at the inner part of the moir\'e cell the atoms are arrange in a BA stacking, between layer 1 of graphene and the BN layer, here the carbon atoms of layer 1 are preferentially sitting on boron atoms since this is the most  energetically favorable configuration \cite{Jung2015}. Then the carbon atoms of layer 2 will be sitting on nitrogen atoms. We can see in Fig. \ref{Simulations}c that the two stacking configurations, 0$^{\circ}$ and 60$^{\circ}$, turn out to be nonequivalent given that the chemical bonds between them are not arranged in the same way, creating an inhomogeneous stretch of the second layer. This inhomogeneous atomic configuration is at the heart of our non-identical moir\'e twins. For examples of all the different stacking configurations see Fig. S21 of the SI.}

As we mentioned before, the stress generated by the commensurate state is released in the form of corrugations, as in the case of monolayer graphene \cite{Woods2014}. These corrugations are transmitted to the second layer and can be observed in the height sensor of our AFM measurements (PeakForce mode) of a different bilayer graphene/BN aligned sample, Fig. \ref{Simulations}d (for more details see Fig, S22 of the SI). In this AFM image, a moir\'e pattern of $\lambda\sim14$ nm is clearly observed. Note that the asymmetry on the AFM image and large value on the height of the deformation is an artefact given by the width of the AFM tip being comparable to the size of the features we want to measure ($\sim5$ nm). This confirms the  existence of a commensurate state and the transmission of the corrugations to the second layer in aligned bilayer graphene/BN heterostructures, supporting our numerical simulations.

\textcolor{black}{The different atomic structure relaxation of the layers results in different electronic band structures for 0$^{\circ}$ and 60$^{\circ}$ of alignment (see Fig. S24 of the SI). However, a direct comparison of these with our experimental results is more complicated than it seems since many parameters need to be taken into account, for example the intrinsic displacement field of our samples. Local charge transport measurements show the presence of an energy gap of $E_{\mathrm{g}}^{30^{\circ}}\approx7.5\pm1.5$ meV at 30$^{\circ}$ of alignment (for sample I), compared to literature \cite{Icking2022} this represents an unintentional displacement field of $\sim0.1$ V/nm. This is not surprising since our devices do not have a top gate to screen external doping deposited on top of the device. Taking into account this unintentional doping and the atomic structure relaxation we calculated the electronic band structures for 0$^{\circ}$ and 60$^{\circ}$, Fig. \ref{SimulationsBS}a. These electronic band structures share with our experimental results a small variation of the energy gap with alignment, Fig. \ref{SimulationsBS}b, even when the magnitude of the energy gap of the simulation is 4.5 times larger than what we measured in charge transport. Clear differences on the band structure can also be seen in the measurements of non-local resistance as a function of gate voltage and magnetic field (equivalent to magnetic focusing measurements) presented in Fig. S20 of the SI, the explanation of these is out of the scope of this manuscript.}

\textcolor{black}{A feature that we do not recover in our experimental measurements is the presence of an energy gap  in the valence band for 0$^{\circ}$ of alignment. Our hypothesis is that this energy gap is too small to be observed in our sample. Even when the quality of our samples is remarkable, compared to previous experiments, the calculated energy gap  is about four times smaller than the energy gap at the CNP, which will place it out of reach in our temperature range.}\\

\noindent \textbf{DISCUSSION}\\

\textcolor{black}{Let's start by discussing the basic charge transport properties of the system. The first sign we present here of the non-identical moir\'e twins is the difference in the resistance of the CNP at room temperature for 0$^{\circ}$ and 60$^{\circ}$ of alignment. Putting these results in the context of the Drude model, and taking into account that we are always working with the same sample, we can attribute this to a different effective mass for each alignment, coming from distinct electronic band structures, as reflected by our numerical simulations. We will therefore expect to have a heavier mass in the case of 0$^{\circ}$, as also suggested by our numerical simulations, see Fig. S26 of the SI. At low temperature the values of the resistance are inverted, now 0$^{\circ}$ of alignment has a resistance that is about six times lower than the 60$^{\circ}$ alignment. This can be explain by the presence of the valley Hall effect which will  reduce the scattering creating a much better conduction in the 0$^{\circ}$ case, at low temperatures.}

\textcolor{black}{Now we discuss the observation of the valley Hall effect. The most widely used explanation for the existence of the valley Hall effect in aligned graphene/BN is the presence of a Berry curvature \cite{Gorbachev2014,Shintaku2023,Li2020,Endo2019}, detailed in the SI. The Berry curvature has a dependence with the energy gap: it reaches its maximum value for small energy gaps and then it decays rapidly as the energy gap increases \cite{Yin2022}. Although this theory explains well the observation of the valley Hall effect in bilayer graphene in the presence of a displacement field \cite{Yin2022, Sui2015, Shimazaki2015}, it is in contradiction with our experimental results, where the energy gap amplitude has no incidence in the observation of the valley Hall effect. In Fig. \ref{SimulationsBS} we present two samples which show the valley Hall effect at 0$^{\circ}$ of alignment. For sample I, the energy gaps do not change significantly among the different alignments, Fig \ref{SimulationsBS}c and for sample II the smallest energy gap is observed for 60$^{\circ}$ of alignment. We have also perform numerical simulations of the Berry curvature for the obtained band structures and there are not remarkable difference that could explain our experimental results, see Fig. S25 of the SI.}

\textcolor{black}{An alternative explanation to our results could be found in the symmetry of the atomic structure relaxation patterns, Fig. \ref{Simulations}a and b. The atomic structure relaxation of the second layer creates a $2\pi/3$ pattern and therefore an anisotropic strain. The particular way in which this strain is applied has been predicted to create a strong gauge field that effectively acts as a uniform magnetic field \cite{Guinea2010Jan}.  This gauge field vector potential has opposite signs for each valley, making possible to have a valley separation and therefore a fully developed valley Hall effect. However, using transport measurements we do not have access to the values of pseudo-magnetic field. We observe a shift in gate voltage of the CNP when the devices are aligned, see Figs. S4 and S5 of the SI for more data. This shift is generated by the change in the work function of graphene induced by strain \cite{Wang2021,Choi2010}. Unfortunately, these measurements can only be taken as a signal of a larger strain but cannot be used to calculate the pseudo-magnetic field of the system given that they represent an average over the whole device. To prove this theory, local measurements such as scanning tunneling microscopy, will be required.}

\textcolor{black}{Spatially varying regions of broken sublattice symmetry: Recent theoretical calculations propose that the valley Hall effect observed in monolayer graphene aligned with BN \cite{Gorbachev2014,Komatsu2018,Li2020} originates from the spatial variation of the broken sublattice symmetry\cite{Aktor2021}. If this effect is at the origin of the valley Hall effect in monolayer graphene the picture becomes more complicated when dealing with bilayer graphene. Following the results of our numerical simulations we can  say that the spatial variations of broken sublattice symmetry will be different between the two layers, and it will always exist for the first layer. It is then not evident why the valley effect is observed for only one of the two layer alignments, and clearly, further numerical investigations would be needed to clarify the situation.}

\textcolor{black}{To conclude, our experimental results show the existence of non-identical moir\'es in bilayer graphene aligned with BN. We attribute this difference to the atomic structure relaxation of the commensurate state, which modifies the band structure of bilayer graphene in different ways for 0$^{\circ}$ and 60$^{\circ}$ of alignment. The observation of the valley Hall effect with a hundred and twenty degrees periodicity cannot be explained by current theoretical model. We hope that our experimental results inspire further theoretical and experimental developments to address the existence of the valley Hall effect in this system.}

\bibliography{references}

\begin{thebibliography}{43}%
\makeatletter
\providecommand \@ifxundefined [1]{%
 \@ifx{#1\undefined}
}%
\providecommand \@ifnum [1]{%
 \ifnum #1\expandafter \@firstoftwo
 \else \expandafter \@secondoftwo
 \fi
}%
\providecommand \@ifx [1]{%
 \ifx #1\expandafter \@firstoftwo
 \else \expandafter \@secondoftwo
 \fi
}%
\providecommand \natexlab [1]{#1}%
\providecommand \enquote  [1]{``#1''}%
\providecommand \bibnamefont  [1]{#1}%
\providecommand \bibfnamefont [1]{#1}%
\providecommand \citenamefont [1]{#1}%
\providecommand \href@noop [0]{\@secondoftwo}%
\providecommand \href [0]{\begingroup \@sanitize@url \@href}%
\providecommand \@href[1]{\@@startlink{#1}\@@href}%
\providecommand \@@href[1]{\endgroup#1\@@endlink}%
\providecommand \@sanitize@url [0]{\catcode `\\12\catcode `\$12\catcode
  `\&12\catcode `\#12\catcode `\^12\catcode `\_12\catcode `\%12\relax}%
\providecommand \@@startlink[1]{}%
\providecommand \@@endlink[0]{}%
\providecommand \url  [0]{\begingroup\@sanitize@url \@url }%
\providecommand \@url [1]{\endgroup\@href {#1}{\urlprefix }}%
\providecommand \urlprefix  [0]{URL }%
\providecommand \Eprint [0]{\href }%
\providecommand \doibase [0]{https://doi.org/}%
\providecommand \selectlanguage [0]{\@gobble}%
\providecommand \bibinfo  [0]{\@secondoftwo}%
\providecommand \bibfield  [0]{\@secondoftwo}%
\providecommand \translation [1]{[#1]}%
\providecommand \BibitemOpen [0]{}%
\providecommand \bibitemStop [0]{}%
\providecommand \bibitemNoStop [0]{.\EOS\space}%
\providecommand \EOS [0]{\spacefactor3000\relax}%
\providecommand \BibitemShut  [1]{\csname bibitem#1\endcsname}%
\let\auto@bib@innerbib\@empty
%</preamble>
\bibitem [{\citenamefont {Hunt}\ \emph {et~al.}(2013)\citenamefont {Hunt},
  \citenamefont {Sanchez-Yamagishi}, \citenamefont {Young}, \citenamefont
  {Yankowitz}, \citenamefont {LeRoy}, \citenamefont {Watanabe}, \citenamefont
  {Taniguchi}, \citenamefont {Moon}, \citenamefont {Koshino}, \citenamefont
  {Jarillo-Herrero},\ and\ \citenamefont {Ashoori}}]{Hunt2013}%
  \BibitemOpen
  \bibfield  {author} {\bibinfo {author} {\bibfnamefont {B.}~\bibnamefont
  {Hunt}}, \bibinfo {author} {\bibfnamefont {J.~D.}\ \bibnamefont
  {Sanchez-Yamagishi}}, \bibinfo {author} {\bibfnamefont {A.~F.}\ \bibnamefont
  {Young}}, \bibinfo {author} {\bibfnamefont {M.}~\bibnamefont {Yankowitz}},
  \bibinfo {author} {\bibfnamefont {B.~J.}\ \bibnamefont {LeRoy}}, \bibinfo
  {author} {\bibfnamefont {K.}~\bibnamefont {Watanabe}}, \bibinfo {author}
  {\bibfnamefont {T.}~\bibnamefont {Taniguchi}}, \bibinfo {author}
  {\bibfnamefont {P.}~\bibnamefont {Moon}}, \bibinfo {author} {\bibfnamefont
  {M.}~\bibnamefont {Koshino}}, \bibinfo {author} {\bibfnamefont
  {P.}~\bibnamefont {Jarillo-Herrero}},\ and\ \bibinfo {author} {\bibfnamefont
  {R.~C.}\ \bibnamefont {Ashoori}},\ }\bibfield  {title} {\bibinfo {title}
  {{Massive Dirac Fermions and Hofstadter Butterfly in a van der Waals
  Heterostructure}},\ }\href {https://doi.org/10.1126/science.1237240}
  {\bibfield  {journal} {\bibinfo  {journal} {Science}\ }\textbf {\bibinfo
  {volume} {340}},\ \bibinfo {pages} {1427} (\bibinfo {year}
  {2013})}\BibitemShut {NoStop}%
\bibitem [{\citenamefont {Wang}\ \emph
  {et~al.}(2015{\natexlab{a}})\citenamefont {Wang}, \citenamefont {Gao},
  \citenamefont {Wen}, \citenamefont {Han}, \citenamefont {Taniguchi},
  \citenamefont {Watanabe}, \citenamefont {Koshino}, \citenamefont {Hone},\
  and\ \citenamefont {Dean}}]{Wang2015}%
  \BibitemOpen
  \bibfield  {author} {\bibinfo {author} {\bibfnamefont {L.}~\bibnamefont
  {Wang}}, \bibinfo {author} {\bibfnamefont {Y.}~\bibnamefont {Gao}}, \bibinfo
  {author} {\bibfnamefont {B.}~\bibnamefont {Wen}}, \bibinfo {author}
  {\bibfnamefont {Z.}~\bibnamefont {Han}}, \bibinfo {author} {\bibfnamefont
  {T.}~\bibnamefont {Taniguchi}}, \bibinfo {author} {\bibfnamefont
  {K.}~\bibnamefont {Watanabe}}, \bibinfo {author} {\bibfnamefont
  {M.}~\bibnamefont {Koshino}}, \bibinfo {author} {\bibfnamefont
  {J.}~\bibnamefont {Hone}},\ and\ \bibinfo {author} {\bibfnamefont {C.~R.}\
  \bibnamefont {Dean}},\ }\bibfield  {title} {\bibinfo {title} {{Evidence for a
  fractional fractal quantum Hall effect in graphene superlattices}},\ }\href
  {https://doi.org/10.1126/science.aad2102} {\bibfield  {journal} {\bibinfo
  {journal} {Science}\ }\textbf {\bibinfo {volume} {350}},\ \bibinfo {pages}
  {1231} (\bibinfo {year} {2015}{\natexlab{a}})}\BibitemShut {NoStop}%
\bibitem [{\citenamefont {Ponomarenko}\ \emph {et~al.}(2013)\citenamefont
  {Ponomarenko}, \citenamefont {Gorbachev}, \citenamefont {Yu}, \citenamefont
  {Elias}, \citenamefont {Jalil}, \citenamefont {Patel}, \citenamefont
  {Mishchenko}, \citenamefont {Mayorov}, \citenamefont {Woods}, \citenamefont
  {Wallbank}, \citenamefont {Mucha-Kruczynski}, \citenamefont {Piot},
  \citenamefont {Potemski}, \citenamefont {Grigorieva}, \citenamefont
  {Novoselov}, \citenamefont {Guinea}, \citenamefont {Fal{'}ko},\ and\
  \citenamefont {Geim}}]{Ponomarenko2013}%
  \BibitemOpen
  \bibfield  {author} {\bibinfo {author} {\bibfnamefont {L.~A.}\ \bibnamefont
  {Ponomarenko}}, \bibinfo {author} {\bibfnamefont {R.~V.}\ \bibnamefont
  {Gorbachev}}, \bibinfo {author} {\bibfnamefont {G.~L.}\ \bibnamefont {Yu}},
  \bibinfo {author} {\bibfnamefont {D.~C.}\ \bibnamefont {Elias}}, \bibinfo
  {author} {\bibfnamefont {R.}~\bibnamefont {Jalil}}, \bibinfo {author}
  {\bibfnamefont {A.~A.}\ \bibnamefont {Patel}}, \bibinfo {author}
  {\bibfnamefont {A.}~\bibnamefont {Mishchenko}}, \bibinfo {author}
  {\bibfnamefont {A.~S.}\ \bibnamefont {Mayorov}}, \bibinfo {author}
  {\bibfnamefont {C.~R.}\ \bibnamefont {Woods}}, \bibinfo {author}
  {\bibfnamefont {J.~R.}\ \bibnamefont {Wallbank}}, \bibinfo {author}
  {\bibfnamefont {M.}~\bibnamefont {Mucha-Kruczynski}}, \bibinfo {author}
  {\bibfnamefont {B.~A.}\ \bibnamefont {Piot}}, \bibinfo {author}
  {\bibfnamefont {M.}~\bibnamefont {Potemski}}, \bibinfo {author}
  {\bibfnamefont {I.~V.}\ \bibnamefont {Grigorieva}}, \bibinfo {author}
  {\bibfnamefont {K.~S.}\ \bibnamefont {Novoselov}}, \bibinfo {author}
  {\bibfnamefont {F.}~\bibnamefont {Guinea}}, \bibinfo {author} {\bibfnamefont
  {V.~I.}\ \bibnamefont {Fal{'}ko}},\ and\ \bibinfo {author} {\bibfnamefont
  {A.~K.}\ \bibnamefont {Geim}},\ }\bibfield  {title} {\bibinfo {title}
  {{Cloning of Dirac fermions in graphene superlattices}},\ }\href
  {https://doi.org/10.1038/nature12187} {\bibfield  {journal} {\bibinfo
  {journal} {Nature}\ }\textbf {\bibinfo {volume} {497}},\ \bibinfo {pages}
  {594} (\bibinfo {year} {2013})}\BibitemShut {NoStop}%
\bibitem [{\citenamefont {Ribeiro-Palau}\ \emph {et~al.}(2018)\citenamefont
  {Ribeiro-Palau}, \citenamefont {Zhang}, \citenamefont {Watanabe},
  \citenamefont {Taniguchi}, \citenamefont {Hone},\ and\ \citenamefont
  {Dean}}]{Ribeiro-Palau2018}%
  \BibitemOpen
  \bibfield  {author} {\bibinfo {author} {\bibfnamefont {R.}~\bibnamefont
  {Ribeiro-Palau}}, \bibinfo {author} {\bibfnamefont {C.}~\bibnamefont
  {Zhang}}, \bibinfo {author} {\bibfnamefont {K.}~\bibnamefont {Watanabe}},
  \bibinfo {author} {\bibfnamefont {T.}~\bibnamefont {Taniguchi}}, \bibinfo
  {author} {\bibfnamefont {J.}~\bibnamefont {Hone}},\ and\ \bibinfo {author}
  {\bibfnamefont {C.~R.}\ \bibnamefont {Dean}},\ }\bibfield  {title} {\bibinfo
  {title} {{Twistable electronics with dynamically rotatable
  heterostructures}},\ }\href
  {https://www.science.org/doi/10.1126/science.aat6981} {\bibfield  {journal}
  {\bibinfo  {journal} {Science}\ } (\bibinfo {year} {2018})}\BibitemShut
  {NoStop}%
\bibitem [{\citenamefont {Woods}\ \emph {et~al.}(2014)\citenamefont {Woods},
  \citenamefont {Britnell}, \citenamefont {Eckmann}, \citenamefont {Ma},
  \citenamefont {Lu}, \citenamefont {Guo}, \citenamefont {Lin}, \citenamefont
  {Yu}, \citenamefont {Cao}, \citenamefont {Gorbachev}, \citenamefont
  {Kretinin}, \citenamefont {Park}, \citenamefont {Ponomarenko}, \citenamefont
  {Katsnelson}, \citenamefont {Gornostyrev}, \citenamefont {Watanabe},
  \citenamefont {Taniguchi}, \citenamefont {Casiraghi}, \citenamefont {Gao},
  \citenamefont {Geim},\ and\ \citenamefont {Novoselov}}]{Woods2014}%
  \BibitemOpen
  \bibfield  {author} {\bibinfo {author} {\bibfnamefont {C.~R.}\ \bibnamefont
  {Woods}}, \bibinfo {author} {\bibfnamefont {L.}~\bibnamefont {Britnell}},
  \bibinfo {author} {\bibfnamefont {A.}~\bibnamefont {Eckmann}}, \bibinfo
  {author} {\bibfnamefont {R.~S.}\ \bibnamefont {Ma}}, \bibinfo {author}
  {\bibfnamefont {J.~C.}\ \bibnamefont {Lu}}, \bibinfo {author} {\bibfnamefont
  {H.~M.}\ \bibnamefont {Guo}}, \bibinfo {author} {\bibfnamefont
  {X.}~\bibnamefont {Lin}}, \bibinfo {author} {\bibfnamefont {G.~L.}\
  \bibnamefont {Yu}}, \bibinfo {author} {\bibfnamefont {Y.}~\bibnamefont
  {Cao}}, \bibinfo {author} {\bibfnamefont {R.~V.}\ \bibnamefont {Gorbachev}},
  \bibinfo {author} {\bibfnamefont {A.~V.}\ \bibnamefont {Kretinin}}, \bibinfo
  {author} {\bibfnamefont {J.}~\bibnamefont {Park}}, \bibinfo {author}
  {\bibfnamefont {L.~A.}\ \bibnamefont {Ponomarenko}}, \bibinfo {author}
  {\bibfnamefont {M.~I.}\ \bibnamefont {Katsnelson}}, \bibinfo {author}
  {\bibfnamefont {{\relax Yu}.~N.}\ \bibnamefont {Gornostyrev}}, \bibinfo
  {author} {\bibfnamefont {K.}~\bibnamefont {Watanabe}}, \bibinfo {author}
  {\bibfnamefont {T.}~\bibnamefont {Taniguchi}}, \bibinfo {author}
  {\bibfnamefont {C.}~\bibnamefont {Casiraghi}}, \bibinfo {author}
  {\bibfnamefont {H.-J.}\ \bibnamefont {Gao}}, \bibinfo {author} {\bibfnamefont
  {A.~K.}\ \bibnamefont {Geim}},\ and\ \bibinfo {author} {\bibfnamefont
  {K.~S.}\ \bibnamefont {Novoselov}},\ }\bibfield  {title} {\bibinfo {title}
  {{Commensurate{\textendash}incommensurate transition in graphene on hexagonal
  boron nitride}},\ }\href {https://doi.org/10.1038/nphys2954} {\bibfield
  {journal} {\bibinfo  {journal} {Nat. Phys.}\ }\textbf {\bibinfo {volume}
  {10}},\ \bibinfo {pages} {451} (\bibinfo {year} {2014})}\BibitemShut
  {NoStop}%
\bibitem [{\citenamefont {Jung}\ \emph {et~al.}(2015)\citenamefont {Jung},
  \citenamefont {DaSilva}, \citenamefont {MacDonald},\ and\ \citenamefont
  {Adam}}]{Jung2015}%
  \BibitemOpen
  \bibfield  {author} {\bibinfo {author} {\bibfnamefont {J.}~\bibnamefont
  {Jung}}, \bibinfo {author} {\bibfnamefont {A.~M.}\ \bibnamefont {DaSilva}},
  \bibinfo {author} {\bibfnamefont {A.~H.}\ \bibnamefont {MacDonald}},\ and\
  \bibinfo {author} {\bibfnamefont {S.}~\bibnamefont {Adam}},\ }\bibfield
  {title} {\bibinfo {title} {{Origin of band gaps in graphene on hexagonal
  boron nitride}},\ }\href {https://doi.org/10.1038/ncomms7308} {\bibfield
  {journal} {\bibinfo  {journal} {Nat. Commun.}\ }\textbf {\bibinfo {volume}
  {6}},\ \bibinfo {pages} {1} (\bibinfo {year} {2015})}\BibitemShut {NoStop}%
\bibitem [{\citenamefont {Song Justin~C.}\ \emph {et~al.}(2015)\citenamefont
  {Song Justin~C.}, \citenamefont {Polnop},\ and\ \citenamefont
  {Levitov~Leonid}}]{Song2015}%
  \BibitemOpen
  \bibfield  {author} {\bibinfo {author} {\bibfnamefont {W.}~\bibnamefont {Song
  Justin~C.}}, \bibinfo {author} {\bibfnamefont {S.}~\bibnamefont {Polnop}},\
  and\ \bibinfo {author} {\bibfnamefont {S.}~\bibnamefont {Levitov~Leonid}},\
  }\bibfield  {title} {\bibinfo {title} {{Topological Bloch bands in graphene
  superlattices}},\ }\href {https://doi.org/10.1073/pnas.1424760112} {\bibfield
   {journal} {\bibinfo  {journal} {Proc. Natl. Acad. Sci. U.S.A.}\ }\textbf
  {\bibinfo {volume} {112}},\ \bibinfo {pages} {10879} (\bibinfo {year}
  {2015})}\BibitemShut {NoStop}%
\bibitem [{\citenamefont {Sun}\ \emph {et~al.}(2021)\citenamefont {Sun},
  \citenamefont {Zhang}, \citenamefont {Liu}, \citenamefont {Zhu},
  \citenamefont {Huang}, \citenamefont {Yuan}, \citenamefont {Wang},
  \citenamefont {Watanabe}, \citenamefont {Taniguchi}, \citenamefont {Li},
  \citenamefont {Zhu}, \citenamefont {Mao}, \citenamefont {Yang}, \citenamefont
  {Kang}, \citenamefont {Liu}, \citenamefont {Ye}, \citenamefont {Han},\ and\
  \citenamefont {Zhang}}]{Sun2021}%
  \BibitemOpen
  \bibfield  {author} {\bibinfo {author} {\bibfnamefont {X.}~\bibnamefont
  {Sun}}, \bibinfo {author} {\bibfnamefont {S.}~\bibnamefont {Zhang}}, \bibinfo
  {author} {\bibfnamefont {Z.}~\bibnamefont {Liu}}, \bibinfo {author}
  {\bibfnamefont {H.}~\bibnamefont {Zhu}}, \bibinfo {author} {\bibfnamefont
  {J.}~\bibnamefont {Huang}}, \bibinfo {author} {\bibfnamefont
  {K.}~\bibnamefont {Yuan}}, \bibinfo {author} {\bibfnamefont {Z.}~\bibnamefont
  {Wang}}, \bibinfo {author} {\bibfnamefont {K.}~\bibnamefont {Watanabe}},
  \bibinfo {author} {\bibfnamefont {T.}~\bibnamefont {Taniguchi}}, \bibinfo
  {author} {\bibfnamefont {X.}~\bibnamefont {Li}}, \bibinfo {author}
  {\bibfnamefont {M.}~\bibnamefont {Zhu}}, \bibinfo {author} {\bibfnamefont
  {J.}~\bibnamefont {Mao}}, \bibinfo {author} {\bibfnamefont {T.}~\bibnamefont
  {Yang}}, \bibinfo {author} {\bibfnamefont {J.}~\bibnamefont {Kang}}, \bibinfo
  {author} {\bibfnamefont {J.}~\bibnamefont {Liu}}, \bibinfo {author}
  {\bibfnamefont {Y.}~\bibnamefont {Ye}}, \bibinfo {author} {\bibfnamefont
  {Z.~V.}\ \bibnamefont {Han}},\ and\ \bibinfo {author} {\bibfnamefont
  {Z.}~\bibnamefont {Zhang}},\ }\bibfield  {title} {\bibinfo {title}
  {{Correlated states in doubly-aligned hBN/graphene/hBN heterostructures}},\
  }\href {https://doi.org/10.1038/s41467-021-27514-y} {\bibfield  {journal}
  {\bibinfo  {journal} {Nat. Commun.}\ }\textbf {\bibinfo {volume} {12}},\
  \bibinfo {pages} {1} (\bibinfo {year} {2021})}\BibitemShut {NoStop}%
\bibitem [{\citenamefont {Wang}\ \emph {et~al.}(2019)\citenamefont {Wang},
  \citenamefont {Zihlmann}, \citenamefont {Liu}, \citenamefont {Makk},
  \citenamefont {Watanabe}, \citenamefont {Taniguchi}, \citenamefont
  {Baumgartner},\ and\ \citenamefont
  {Sch{\ifmmode\ddot{o}\else\"{o}\fi}nenberger}}]{Wang2019}%
  \BibitemOpen
  \bibfield  {author} {\bibinfo {author} {\bibfnamefont {L.}~\bibnamefont
  {Wang}}, \bibinfo {author} {\bibfnamefont {S.}~\bibnamefont {Zihlmann}},
  \bibinfo {author} {\bibfnamefont {M.-H.}\ \bibnamefont {Liu}}, \bibinfo
  {author} {\bibfnamefont {P.}~\bibnamefont {Makk}}, \bibinfo {author}
  {\bibfnamefont {K.}~\bibnamefont {Watanabe}}, \bibinfo {author}
  {\bibfnamefont {T.}~\bibnamefont {Taniguchi}}, \bibinfo {author}
  {\bibfnamefont {A.}~\bibnamefont {Baumgartner}},\ and\ \bibinfo {author}
  {\bibfnamefont {C.}~\bibnamefont
  {Sch{\ifmmode\ddot{o}\else\"{o}\fi}nenberger}},\ }\bibfield  {title}
  {\bibinfo {title} {{New Generation of Moir{\ifmmode\acute{e}\else\'{e}\fi}
  Superlattices in Doubly Aligned hBN/Graphene/hBN Heterostructures}},\ }\href
  {https://doi.org/10.1021/acs.nanolett.8b05061} {\bibfield  {journal}
  {\bibinfo  {journal} {Nano Lett.}\ }\textbf {\bibinfo {volume} {19}},\
  \bibinfo {pages} {2371} (\bibinfo {year} {2019})}\BibitemShut {NoStop}%
\bibitem [{\citenamefont {Finney}\ \emph {et~al.}(2019)\citenamefont {Finney},
  \citenamefont {Yankowitz}, \citenamefont {Muraleetharan}, \citenamefont
  {Watanabe}, \citenamefont {Taniguchi}, \citenamefont {Dean},\ and\
  \citenamefont {Hone}}]{Finney2019}%
  \BibitemOpen
  \bibfield  {author} {\bibinfo {author} {\bibfnamefont {N.~R.}\ \bibnamefont
  {Finney}}, \bibinfo {author} {\bibfnamefont {M.}~\bibnamefont {Yankowitz}},
  \bibinfo {author} {\bibfnamefont {L.}~\bibnamefont {Muraleetharan}}, \bibinfo
  {author} {\bibfnamefont {K.}~\bibnamefont {Watanabe}}, \bibinfo {author}
  {\bibfnamefont {T.}~\bibnamefont {Taniguchi}}, \bibinfo {author}
  {\bibfnamefont {C.~R.}\ \bibnamefont {Dean}},\ and\ \bibinfo {author}
  {\bibfnamefont {J.}~\bibnamefont {Hone}},\ }\bibfield  {title} {\bibinfo
  {title} {{Tunable crystal symmetry in graphene{\textendash}boron nitride
  heterostructures with coexisting moir{\ifmmode\acute{e}\else\'{e}\fi}
  superlattices}},\ }\href {https://doi.org/10.1038/s41565-019-0547-2}
  {\bibfield  {journal} {\bibinfo  {journal} {Nat. Nanotechnol.}\ }\textbf
  {\bibinfo {volume} {14}},\ \bibinfo {pages} {1029} (\bibinfo {year}
  {2019})}\BibitemShut {NoStop}%
\bibitem [{\citenamefont {Ribeiro-Palau}\ \emph {et~al.}(2019)\citenamefont
  {Ribeiro-Palau}, \citenamefont {Chen}, \citenamefont {Zeng}, \citenamefont
  {Watanabe}, \citenamefont {Taniguchi}, \citenamefont {Hone},\ and\
  \citenamefont {Dean}}]{Ribeiro-Palau2019}%
  \BibitemOpen
  \bibfield  {author} {\bibinfo {author} {\bibfnamefont {R.}~\bibnamefont
  {Ribeiro-Palau}}, \bibinfo {author} {\bibfnamefont {S.}~\bibnamefont {Chen}},
  \bibinfo {author} {\bibfnamefont {Y.}~\bibnamefont {Zeng}}, \bibinfo {author}
  {\bibfnamefont {K.}~\bibnamefont {Watanabe}}, \bibinfo {author}
  {\bibfnamefont {T.}~\bibnamefont {Taniguchi}}, \bibinfo {author}
  {\bibfnamefont {J.}~\bibnamefont {Hone}},\ and\ \bibinfo {author}
  {\bibfnamefont {C.~R.}\ \bibnamefont {Dean}},\ }\bibfield  {title} {\bibinfo
  {title} {{High-Quality Electrostatically Defined Hall Bars in Monolayer
  Graphene}},\ }\href {https://doi.org/10.1021/acs.nanolett.9b00351} {\bibfield
   {journal} {\bibinfo  {journal} {Nano Lett.}\ }\textbf {\bibinfo {volume}
  {19}},\ \bibinfo {pages} {2583} (\bibinfo {year} {2019})}\BibitemShut
  {NoStop}%
\bibitem [{\citenamefont {Wang}\ \emph
  {et~al.}(2015{\natexlab{b}})\citenamefont {Wang}, \citenamefont {Tang},
  \citenamefont {Sachs}, \citenamefont {Barlas},\ and\ \citenamefont
  {Shi}}]{Wang2015Jan}%
  \BibitemOpen
  \bibfield  {author} {\bibinfo {author} {\bibfnamefont {Z.}~\bibnamefont
  {Wang}}, \bibinfo {author} {\bibfnamefont {C.}~\bibnamefont {Tang}}, \bibinfo
  {author} {\bibfnamefont {R.}~\bibnamefont {Sachs}}, \bibinfo {author}
  {\bibfnamefont {Y.}~\bibnamefont {Barlas}},\ and\ \bibinfo {author}
  {\bibfnamefont {J.}~\bibnamefont {Shi}},\ }\bibfield  {title} {\bibinfo
  {title} {{Proximity-Induced Ferromagnetism in Graphene Revealed by the
  Anomalous Hall Effect}},\ }\href
  {https://doi.org/10.1103/PhysRevLett.114.016603} {\bibfield  {journal}
  {\bibinfo  {journal} {Phys. Rev. Lett.}\ }\textbf {\bibinfo {volume} {114}},\
  \bibinfo {pages} {016603} (\bibinfo {year} {2015}{\natexlab{b}})}\BibitemShut
  {NoStop}%
\bibitem [{\citenamefont {Gorbachev}\ \emph {et~al.}(2014)\citenamefont
  {Gorbachev}, \citenamefont {Song}, \citenamefont {Yu}, \citenamefont
  {Kretinin}, \citenamefont {Withers}, \citenamefont {Cao}, \citenamefont
  {Mishchenko}, \citenamefont {Grigorieva}, \citenamefont {Novoselov},
  \citenamefont {Levitov},\ and\ \citenamefont {Geim}}]{Gorbachev2014}%
  \BibitemOpen
  \bibfield  {author} {\bibinfo {author} {\bibfnamefont {R.~V.}\ \bibnamefont
  {Gorbachev}}, \bibinfo {author} {\bibfnamefont {J.~C.~W.}\ \bibnamefont
  {Song}}, \bibinfo {author} {\bibfnamefont {G.~L.}\ \bibnamefont {Yu}},
  \bibinfo {author} {\bibfnamefont {A.~V.}\ \bibnamefont {Kretinin}}, \bibinfo
  {author} {\bibfnamefont {F.}~\bibnamefont {Withers}}, \bibinfo {author}
  {\bibfnamefont {Y.}~\bibnamefont {Cao}}, \bibinfo {author} {\bibfnamefont
  {A.}~\bibnamefont {Mishchenko}}, \bibinfo {author} {\bibfnamefont {I.~V.}\
  \bibnamefont {Grigorieva}}, \bibinfo {author} {\bibfnamefont {K.~S.}\
  \bibnamefont {Novoselov}}, \bibinfo {author} {\bibfnamefont {L.~S.}\
  \bibnamefont {Levitov}},\ and\ \bibinfo {author} {\bibfnamefont {A.~K.}\
  \bibnamefont {Geim}},\ }\bibfield  {title} {\bibinfo {title} {{Detecting
  topological currents in graphene superlattices}},\ }\href
  {https://www.science.org/doi/abs/10.1126/science.1254966} {\bibfield
  {journal} {\bibinfo  {journal} {Science}\ } (\bibinfo {year}
  {2014})}\BibitemShut {NoStop}%
\bibitem [{\citenamefont {Sui}\ \emph {et~al.}(2015)\citenamefont {Sui},
  \citenamefont {Chen}, \citenamefont {Ma}, \citenamefont {Shan}, \citenamefont
  {Tian}, \citenamefont {Watanabe}, \citenamefont {Taniguchi}, \citenamefont
  {Jin}, \citenamefont {Yao}, \citenamefont {Xiao},\ and\ \citenamefont
  {Zhang}}]{Sui2015}%
  \BibitemOpen
  \bibfield  {author} {\bibinfo {author} {\bibfnamefont {M.}~\bibnamefont
  {Sui}}, \bibinfo {author} {\bibfnamefont {G.}~\bibnamefont {Chen}}, \bibinfo
  {author} {\bibfnamefont {L.}~\bibnamefont {Ma}}, \bibinfo {author}
  {\bibfnamefont {W.-Y.}\ \bibnamefont {Shan}}, \bibinfo {author}
  {\bibfnamefont {D.}~\bibnamefont {Tian}}, \bibinfo {author} {\bibfnamefont
  {K.}~\bibnamefont {Watanabe}}, \bibinfo {author} {\bibfnamefont
  {T.}~\bibnamefont {Taniguchi}}, \bibinfo {author} {\bibfnamefont
  {X.}~\bibnamefont {Jin}}, \bibinfo {author} {\bibfnamefont {W.}~\bibnamefont
  {Yao}}, \bibinfo {author} {\bibfnamefont {D.}~\bibnamefont {Xiao}},\ and\
  \bibinfo {author} {\bibfnamefont {Y.}~\bibnamefont {Zhang}},\ }\bibfield
  {title} {\bibinfo {title} {{Gate-tunable topological valley transport in
  bilayer graphene - Nature Physics}},\ }\href
  {https://doi.org/10.1038/nphys3485} {\bibfield  {journal} {\bibinfo
  {journal} {Nat. Phys.}\ }\textbf {\bibinfo {volume} {11}},\ \bibinfo {pages}
  {1027} (\bibinfo {year} {2015})}\BibitemShut {NoStop}%
\bibitem [{\citenamefont {Shimazaki}\ \emph {et~al.}(2015)\citenamefont
  {Shimazaki}, \citenamefont {Yamamoto}, \citenamefont {Borzenets},
  \citenamefont {Watanabe}, \citenamefont {Taniguchi},\ and\ \citenamefont
  {Tarucha}}]{Shimazaki2015}%
  \BibitemOpen
  \bibfield  {author} {\bibinfo {author} {\bibfnamefont {Y.}~\bibnamefont
  {Shimazaki}}, \bibinfo {author} {\bibfnamefont {M.}~\bibnamefont {Yamamoto}},
  \bibinfo {author} {\bibfnamefont {I.~V.}\ \bibnamefont {Borzenets}}, \bibinfo
  {author} {\bibfnamefont {K.}~\bibnamefont {Watanabe}}, \bibinfo {author}
  {\bibfnamefont {T.}~\bibnamefont {Taniguchi}},\ and\ \bibinfo {author}
  {\bibfnamefont {S.}~\bibnamefont {Tarucha}},\ }\bibfield  {title} {\bibinfo
  {title} {{Generation and detection of pure valley current by electrically
  induced Berry curvature in bilayer graphene - Nature Physics}},\ }\href
  {https://doi.org/10.1038/nphys3551} {\bibfield  {journal} {\bibinfo
  {journal} {Nat. Phys.}\ }\textbf {\bibinfo {volume} {11}},\ \bibinfo {pages}
  {1032} (\bibinfo {year} {2015})}\BibitemShut {NoStop}%
\bibitem [{\citenamefont {Br{\ifmmode\ddot{u}\else\"{u}\fi}ne}\ \emph
  {et~al.}(2010)\citenamefont {Br{\ifmmode\ddot{u}\else\"{u}\fi}ne},
  \citenamefont {Roth}, \citenamefont {Novik}, \citenamefont
  {K{\ifmmode\ddot{o}\else\"{o}\fi}nig}, \citenamefont {Buhmann}, \citenamefont
  {Hankiewicz}, \citenamefont {Hanke}, \citenamefont {Sinova},\ and\
  \citenamefont {Molenkamp}}]{Brune2010}%
  \BibitemOpen
  \bibfield  {author} {\bibinfo {author} {\bibfnamefont {C.}~\bibnamefont
  {Br{\ifmmode\ddot{u}\else\"{u}\fi}ne}}, \bibinfo {author} {\bibfnamefont
  {A.}~\bibnamefont {Roth}}, \bibinfo {author} {\bibfnamefont {E.~G.}\
  \bibnamefont {Novik}}, \bibinfo {author} {\bibfnamefont {M.}~\bibnamefont
  {K{\ifmmode\ddot{o}\else\"{o}\fi}nig}}, \bibinfo {author} {\bibfnamefont
  {H.}~\bibnamefont {Buhmann}}, \bibinfo {author} {\bibfnamefont {E.~M.}\
  \bibnamefont {Hankiewicz}}, \bibinfo {author} {\bibfnamefont
  {W.}~\bibnamefont {Hanke}}, \bibinfo {author} {\bibfnamefont
  {J.}~\bibnamefont {Sinova}},\ and\ \bibinfo {author} {\bibfnamefont {L.~W.}\
  \bibnamefont {Molenkamp}},\ }\bibfield  {title} {\bibinfo {title} {{Evidence
  for the ballistic intrinsic spin Hall effect in HgTe nanostructures}},\
  }\href {https://doi.org/10.1038/nphys1655} {\bibfield  {journal} {\bibinfo
  {journal} {Nat. Phys.}\ }\textbf {\bibinfo {volume} {6}},\ \bibinfo {pages}
  {448} (\bibinfo {year} {2010})}\BibitemShut {NoStop}%
\bibitem [{\citenamefont {Komatsu}\ \emph {et~al.}(2018)\citenamefont
  {Komatsu}, \citenamefont {Morita}, \citenamefont {Watanabe}, \citenamefont
  {Tsuya}, \citenamefont {Watanabe}, \citenamefont {Taniguchi},\ and\
  \citenamefont {Moriyama}}]{Komatsu2018}%
  \BibitemOpen
  \bibfield  {author} {\bibinfo {author} {\bibfnamefont {K.}~\bibnamefont
  {Komatsu}}, \bibinfo {author} {\bibfnamefont {Y.}~\bibnamefont {Morita}},
  \bibinfo {author} {\bibfnamefont {E.}~\bibnamefont {Watanabe}}, \bibinfo
  {author} {\bibfnamefont {D.}~\bibnamefont {Tsuya}}, \bibinfo {author}
  {\bibfnamefont {K.}~\bibnamefont {Watanabe}}, \bibinfo {author}
  {\bibfnamefont {T.}~\bibnamefont {Taniguchi}},\ and\ \bibinfo {author}
  {\bibfnamefont {S.}~\bibnamefont {Moriyama}},\ }\bibfield  {title} {\bibinfo
  {title} {{Observation of the quantum valley Hall state in ballistic graphene
  superlattices}},\ }\href {https://www.science.org/doi/10.1126/sciadv.aaq0194}
  {\bibfield  {journal} {\bibinfo  {journal} {Sci. Adv.}\ } (\bibinfo {year}
  {2018})}\BibitemShut {NoStop}%
\bibitem [{\citenamefont {Li}\ \emph {et~al.}(2020)\citenamefont {Li},
  \citenamefont {Amado}, \citenamefont {Hyart}, \citenamefont {Mazur},\ and\
  \citenamefont {Robinson}}]{Li2020}%
  \BibitemOpen
  \bibfield  {author} {\bibinfo {author} {\bibfnamefont {Y.}~\bibnamefont
  {Li}}, \bibinfo {author} {\bibfnamefont {M.}~\bibnamefont {Amado}}, \bibinfo
  {author} {\bibfnamefont {T.}~\bibnamefont {Hyart}}, \bibinfo {author}
  {\bibfnamefont {{\relax Grzegorz}.~P.}\ \bibnamefont {Mazur}},\ and\ \bibinfo
  {author} {\bibfnamefont {J.~W.~A.}\ \bibnamefont {Robinson}},\ }\bibfield
  {title} {\bibinfo {title} {{Topological valley currents via ballistic edge
  modes in graphene superlattices near the primary Dirac point - Communications
  Physics}},\ }\href {https://doi.org/10.1038/s42005-020-00495-y} {\bibfield
  {journal} {\bibinfo  {journal} {Commun. Phys.}\ }\textbf {\bibinfo {volume}
  {3}},\ \bibinfo {pages} {1} (\bibinfo {year} {2020})}\BibitemShut {NoStop}%
\bibitem [{\citenamefont {Endo}\ \emph {et~al.}(2019)\citenamefont {Endo},
  \citenamefont {Komatsu}, \citenamefont {Iwasaki}, \citenamefont {Watanabe},
  \citenamefont {Tsuya}, \citenamefont {Watanabe}, \citenamefont {Taniguchi},
  \citenamefont {Noguchi}, \citenamefont {Wakayama}, \citenamefont {Morita},\
  and\ \citenamefont {Moriyama}}]{Endo2019}%
  \BibitemOpen
  \bibfield  {author} {\bibinfo {author} {\bibfnamefont {K.}~\bibnamefont
  {Endo}}, \bibinfo {author} {\bibfnamefont {K.}~\bibnamefont {Komatsu}},
  \bibinfo {author} {\bibfnamefont {T.}~\bibnamefont {Iwasaki}}, \bibinfo
  {author} {\bibfnamefont {E.}~\bibnamefont {Watanabe}}, \bibinfo {author}
  {\bibfnamefont {D.}~\bibnamefont {Tsuya}}, \bibinfo {author} {\bibfnamefont
  {K.}~\bibnamefont {Watanabe}}, \bibinfo {author} {\bibfnamefont
  {T.}~\bibnamefont {Taniguchi}}, \bibinfo {author} {\bibfnamefont
  {Y.}~\bibnamefont {Noguchi}}, \bibinfo {author} {\bibfnamefont
  {Y.}~\bibnamefont {Wakayama}}, \bibinfo {author} {\bibfnamefont
  {Y.}~\bibnamefont {Morita}},\ and\ \bibinfo {author} {\bibfnamefont
  {S.}~\bibnamefont {Moriyama}},\ }\bibfield  {title} {\bibinfo {title}
  {{Topological valley currents in bilayer graphene/hexagonal boron nitride
  superlattices}},\ }\href {https://doi.org/10.1063/1.5094456} {\bibfield
  {journal} {\bibinfo  {journal} {Appl. Phys. Lett.}\ }\textbf {\bibinfo
  {volume} {114}},\ \bibinfo {pages} {243105} (\bibinfo {year}
  {2019})}\BibitemShut {NoStop}%
\bibitem [{\citenamefont {Wu}\ \emph {et~al.}(2019)\citenamefont {Wu},
  \citenamefont {Zhou}, \citenamefont {Cai}, \citenamefont {Cheung},
  \citenamefont {Liu}, \citenamefont {Huang}, \citenamefont {Lin},
  \citenamefont {Han}, \citenamefont {An}, \citenamefont {Wang}, \citenamefont
  {Xu}, \citenamefont {Long}, \citenamefont {Cheng}, \citenamefont {Law},
  \citenamefont {Zhang},\ and\ \citenamefont {Wang}}]{Wu2019}%
  \BibitemOpen
  \bibfield  {author} {\bibinfo {author} {\bibfnamefont {Z.}~\bibnamefont
  {Wu}}, \bibinfo {author} {\bibfnamefont {B.~T.}\ \bibnamefont {Zhou}},
  \bibinfo {author} {\bibfnamefont {X.}~\bibnamefont {Cai}}, \bibinfo {author}
  {\bibfnamefont {P.}~\bibnamefont {Cheung}}, \bibinfo {author} {\bibfnamefont
  {G.-B.}\ \bibnamefont {Liu}}, \bibinfo {author} {\bibfnamefont
  {M.}~\bibnamefont {Huang}}, \bibinfo {author} {\bibfnamefont
  {J.}~\bibnamefont {Lin}}, \bibinfo {author} {\bibfnamefont {T.}~\bibnamefont
  {Han}}, \bibinfo {author} {\bibfnamefont {L.}~\bibnamefont {An}}, \bibinfo
  {author} {\bibfnamefont {Y.}~\bibnamefont {Wang}}, \bibinfo {author}
  {\bibfnamefont {S.}~\bibnamefont {Xu}}, \bibinfo {author} {\bibfnamefont
  {G.}~\bibnamefont {Long}}, \bibinfo {author} {\bibfnamefont {C.}~\bibnamefont
  {Cheng}}, \bibinfo {author} {\bibfnamefont {K.~T.}\ \bibnamefont {Law}},
  \bibinfo {author} {\bibfnamefont {F.}~\bibnamefont {Zhang}},\ and\ \bibinfo
  {author} {\bibfnamefont {N.}~\bibnamefont {Wang}},\ }\bibfield  {title}
  {\bibinfo {title} {{Intrinsic valley Hall transport in atomically thin
  MoS2}},\ }\href {https://doi.org/10.1038/s41467-019-08629-9} {\bibfield
  {journal} {\bibinfo  {journal} {Nat. Commun.}\ }\textbf {\bibinfo {volume}
  {10}},\ \bibinfo {pages} {1} (\bibinfo {year} {2019})}\BibitemShut {NoStop}%
\bibitem [{\citenamefont {Abanin}\ \emph {et~al.}(2009)\citenamefont {Abanin},
  \citenamefont {Shytov}, \citenamefont {Levitov},\ and\ \citenamefont
  {Halperin}}]{Abanin2009}%
  \BibitemOpen
  \bibfield  {author} {\bibinfo {author} {\bibfnamefont {D.~A.}\ \bibnamefont
  {Abanin}}, \bibinfo {author} {\bibfnamefont {A.~V.}\ \bibnamefont {Shytov}},
  \bibinfo {author} {\bibfnamefont {L.~S.}\ \bibnamefont {Levitov}},\ and\
  \bibinfo {author} {\bibfnamefont {B.~I.}\ \bibnamefont {Halperin}},\
  }\bibfield  {title} {\bibinfo {title} {{Nonlocal charge transport mediated by
  spin diffusion in the spin Hall effect regime}},\ }\href
  {https://doi.org/10.1103/PhysRevB.79.035304} {\bibfield  {journal} {\bibinfo
  {journal} {Phys. Rev. B}\ }\textbf {\bibinfo {volume} {79}},\ \bibinfo
  {pages} {035304} (\bibinfo {year} {2009})}\BibitemShut {NoStop}%
\bibitem [{\citenamefont {Beconcini}\ \emph {et~al.}(2016)\citenamefont
  {Beconcini}, \citenamefont {Taddei},\ and\ \citenamefont
  {Polini}}]{Beconcini2016}%
  \BibitemOpen
  \bibfield  {author} {\bibinfo {author} {\bibfnamefont {M.}~\bibnamefont
  {Beconcini}}, \bibinfo {author} {\bibfnamefont {F.}~\bibnamefont {Taddei}},\
  and\ \bibinfo {author} {\bibfnamefont {M.}~\bibnamefont {Polini}},\
  }\bibfield  {title} {\bibinfo {title} {{Nonlocal topological valley transport
  at large valley Hall angles}},\ }\href
  {https://doi.org/10.1103/PhysRevB.94.121408} {\bibfield  {journal} {\bibinfo
  {journal} {Phys. Rev. B}\ }\textbf {\bibinfo {volume} {94}},\ \bibinfo
  {pages} {121408} (\bibinfo {year} {2016})}\BibitemShut {NoStop}%
\bibitem [{\citenamefont {Yamamoto}\ \emph {et~al.}(2015)\citenamefont
  {Yamamoto}, \citenamefont {Shimazaki}, \citenamefont {Borzenets},\ and\
  \citenamefont {Tarucha}}]{Yamamoto2015}%
  \BibitemOpen
  \bibfield  {author} {\bibinfo {author} {\bibfnamefont {M.}~\bibnamefont
  {Yamamoto}}, \bibinfo {author} {\bibfnamefont {Y.}~\bibnamefont {Shimazaki}},
  \bibinfo {author} {\bibfnamefont {I.~V.}\ \bibnamefont {Borzenets}},\ and\
  \bibinfo {author} {\bibfnamefont {S.}~\bibnamefont {Tarucha}},\ }\bibfield
  {title} {\bibinfo {title} {{Valley Hall Effect in Two-Dimensional Hexagonal
  Lattices}},\ }\href {https://doi.org/10.7566/JPSJ.84.121006} {\bibfield
  {journal} {\bibinfo  {journal} {J. Phys. Soc. Jpn.}\ }\textbf {\bibinfo
  {volume} {84}},\ \bibinfo {pages} {121006} (\bibinfo {year}
  {2015})}\BibitemShut {NoStop}%
\bibitem [{\citenamefont {Aharon-Steinberg}\ \emph {et~al.}(2021)\citenamefont
  {Aharon-Steinberg}, \citenamefont {Marguerite}, \citenamefont {Perello},
  \citenamefont {Bagani}, \citenamefont {Holder}, \citenamefont {Myasoedov},
  \citenamefont {Levitov}, \citenamefont {Geim},\ and\ \citenamefont
  {Zeldov}}]{Aharon-Steinberg2021}%
  \BibitemOpen
  \bibfield  {author} {\bibinfo {author} {\bibfnamefont {A.}~\bibnamefont
  {Aharon-Steinberg}}, \bibinfo {author} {\bibfnamefont {A.}~\bibnamefont
  {Marguerite}}, \bibinfo {author} {\bibfnamefont {D.~J.}\ \bibnamefont
  {Perello}}, \bibinfo {author} {\bibfnamefont {K.}~\bibnamefont {Bagani}},
  \bibinfo {author} {\bibfnamefont {T.}~\bibnamefont {Holder}}, \bibinfo
  {author} {\bibfnamefont {Y.}~\bibnamefont {Myasoedov}}, \bibinfo {author}
  {\bibfnamefont {L.~S.}\ \bibnamefont {Levitov}}, \bibinfo {author}
  {\bibfnamefont {A.~K.}\ \bibnamefont {Geim}},\ and\ \bibinfo {author}
  {\bibfnamefont {E.}~\bibnamefont {Zeldov}},\ }\bibfield  {title} {\bibinfo
  {title} {{Long-range nontopological edge currents in charge-neutral graphene
  - Nature}},\ }\href {https://doi.org/10.1038/s41586-021-03501-7} {\bibfield
  {journal} {\bibinfo  {journal} {Nature}\ }\textbf {\bibinfo {volume} {593}},\
  \bibinfo {pages} {528} (\bibinfo {year} {2021})}\BibitemShut {NoStop}%
\bibitem [{\citenamefont {Gold}\ \emph {et~al.}(2021)\citenamefont {Gold},
  \citenamefont {Knothe}, \citenamefont {Kurzmann}, \citenamefont
  {Garcia-Ruiz}, \citenamefont {Watanabe}, \citenamefont {Taniguchi},
  \citenamefont {Fal{'}ko}, \citenamefont {Ensslin},\ and\ \citenamefont
  {Ihn}}]{Gold2021Jul}%
  \BibitemOpen
  \bibfield  {author} {\bibinfo {author} {\bibfnamefont {C.}~\bibnamefont
  {Gold}}, \bibinfo {author} {\bibfnamefont {A.}~\bibnamefont {Knothe}},
  \bibinfo {author} {\bibfnamefont {A.}~\bibnamefont {Kurzmann}}, \bibinfo
  {author} {\bibfnamefont {A.}~\bibnamefont {Garcia-Ruiz}}, \bibinfo {author}
  {\bibfnamefont {K.}~\bibnamefont {Watanabe}}, \bibinfo {author}
  {\bibfnamefont {T.}~\bibnamefont {Taniguchi}}, \bibinfo {author}
  {\bibfnamefont {V.}~\bibnamefont {Fal{'}ko}}, \bibinfo {author}
  {\bibfnamefont {K.}~\bibnamefont {Ensslin}},\ and\ \bibinfo {author}
  {\bibfnamefont {T.}~\bibnamefont {Ihn}},\ }\bibfield  {title} {\bibinfo
  {title} {{Coherent Jetting from a Gate-Defined Channel in Bilayer
  Graphene}},\ }\href {https://doi.org/10.1103/PhysRevLett.127.046801}
  {\bibfield  {journal} {\bibinfo  {journal} {Phys. Rev. Lett.}\ }\textbf
  {\bibinfo {volume} {127}},\ \bibinfo {pages} {046801} (\bibinfo {year}
  {2021})}\BibitemShut {NoStop}%
\bibitem [{\citenamefont {Lindsay}\ and\ \citenamefont
  {Broido}(2010)}]{Lindsay2010}%
  \BibitemOpen
  \bibfield  {author} {\bibinfo {author} {\bibfnamefont {L.}~\bibnamefont
  {Lindsay}}\ and\ \bibinfo {author} {\bibfnamefont {D.~A.}\ \bibnamefont
  {Broido}},\ }\bibfield  {title} {\bibinfo {title} {{Optimized Tersoff and
  Brenner empirical potential parameters for lattice dynamics and phonon
  thermal transport in carbon nanotubes and graphene}},\ }\href
  {https://doi.org/10.1103/PhysRevB.81.205441} {\bibfield  {journal} {\bibinfo
  {journal} {Phys. Rev. B}\ }\textbf {\bibinfo {volume} {81}},\ \bibinfo
  {pages} {205441} (\bibinfo {year} {2010})}\BibitemShut {NoStop}%
\bibitem [{\citenamefont {Leven}\ \emph {et~al.}(2016)\citenamefont {Leven},
  \citenamefont {Maaravi}, \citenamefont {Azuri}, \citenamefont {Kronik},\ and\
  \citenamefont {Hod}}]{Leven2016}%
  \BibitemOpen
  \bibfield  {author} {\bibinfo {author} {\bibfnamefont {I.}~\bibnamefont
  {Leven}}, \bibinfo {author} {\bibfnamefont {T.}~\bibnamefont {Maaravi}},
  \bibinfo {author} {\bibfnamefont {I.}~\bibnamefont {Azuri}}, \bibinfo
  {author} {\bibfnamefont {L.}~\bibnamefont {Kronik}},\ and\ \bibinfo {author}
  {\bibfnamefont {O.}~\bibnamefont {Hod}},\ }\bibfield  {title} {\bibinfo
  {title} {{Interlayer Potential for Graphene/h-BN Heterostructures}},\ }\href
  {https://doi.org/10.1021/acs.jctc.6b00147} {\bibfield  {journal} {\bibinfo
  {journal} {J. Chem. Theory Comput.}\ }\textbf {\bibinfo {volume} {12}},\
  \bibinfo {pages} {2896} (\bibinfo {year} {2016})}\BibitemShut {NoStop}%
\bibitem [{\citenamefont {Kolmogorov}\ and\ \citenamefont
  {Crespi}(2005)}]{Kolmogorov2005}%
  \BibitemOpen
  \bibfield  {author} {\bibinfo {author} {\bibfnamefont {A.~N.}\ \bibnamefont
  {Kolmogorov}}\ and\ \bibinfo {author} {\bibfnamefont {V.~H.}\ \bibnamefont
  {Crespi}},\ }\bibfield  {title} {\bibinfo {title} {{Registry-dependent
  interlayer potential for graphitic systems}},\ }\href
  {https://doi.org/10.1103/PhysRevB.71.235415} {\bibfield  {journal} {\bibinfo
  {journal} {Phys. Rev. B}\ }\textbf {\bibinfo {volume} {71}},\ \bibinfo
  {pages} {235415} (\bibinfo {year} {2005})}\BibitemShut {NoStop}%
\bibitem [{\citenamefont {Rickhaus}\ \emph {et~al.}(2019)\citenamefont
  {Rickhaus}, \citenamefont {Zheng}, \citenamefont {Lado}, \citenamefont {Lee},
  \citenamefont {Kurzmann}, \citenamefont {Eich}, \citenamefont {Pisoni},
  \citenamefont {Tong}, \citenamefont {Garreis}, \citenamefont {Gold},
  \citenamefont {Masseroni}, \citenamefont {Taniguchi}, \citenamefont
  {Wantanabe}, \citenamefont {Ihn},\ and\ \citenamefont
  {Ensslin}}]{Rickhaus2019Dec}%
  \BibitemOpen
  \bibfield  {author} {\bibinfo {author} {\bibfnamefont {P.}~\bibnamefont
  {Rickhaus}}, \bibinfo {author} {\bibfnamefont {G.}~\bibnamefont {Zheng}},
  \bibinfo {author} {\bibfnamefont {J.~L.}\ \bibnamefont {Lado}}, \bibinfo
  {author} {\bibfnamefont {Y.}~\bibnamefont {Lee}}, \bibinfo {author}
  {\bibfnamefont {A.}~\bibnamefont {Kurzmann}}, \bibinfo {author}
  {\bibfnamefont {M.}~\bibnamefont {Eich}}, \bibinfo {author} {\bibfnamefont
  {R.}~\bibnamefont {Pisoni}}, \bibinfo {author} {\bibfnamefont
  {C.}~\bibnamefont {Tong}}, \bibinfo {author} {\bibfnamefont {R.}~\bibnamefont
  {Garreis}}, \bibinfo {author} {\bibfnamefont {C.}~\bibnamefont {Gold}},
  \bibinfo {author} {\bibfnamefont {M.}~\bibnamefont {Masseroni}}, \bibinfo
  {author} {\bibfnamefont {T.}~\bibnamefont {Taniguchi}}, \bibinfo {author}
  {\bibfnamefont {K.}~\bibnamefont {Wantanabe}}, \bibinfo {author}
  {\bibfnamefont {T.}~\bibnamefont {Ihn}},\ and\ \bibinfo {author}
  {\bibfnamefont {K.}~\bibnamefont {Ensslin}},\ }\bibfield  {title} {\bibinfo
  {title} {{Gap Opening in Twisted Double Bilayer Graphene by Crystal
  Fields}},\ }\href {https://doi.org/10.1021/acs.nanolett.9b03660} {\bibfield
  {journal} {\bibinfo  {journal} {Nano Lett.}\ }\textbf {\bibinfo {volume}
  {19}},\ \bibinfo {pages} {8821} (\bibinfo {year} {2019})}\BibitemShut
  {NoStop}%
\bibitem [{\citenamefont {Icking}\ \emph {et~al.}(2022)\citenamefont {Icking},
  \citenamefont {Banszerus}, \citenamefont
  {W{\ifmmode\ddot{o}\else\"{o}\fi}rtche}, \citenamefont {Volmer},
  \citenamefont {Schmidt}, \citenamefont {Steiner}, \citenamefont {Engels},
  \citenamefont {Hesselmann}, \citenamefont {Goldsche}, \citenamefont
  {Watanabe}, \citenamefont {Taniguchi}, \citenamefont {Volk}, \citenamefont
  {Beschoten},\ and\ \citenamefont {Stampfer}}]{Icking2022}%
  \BibitemOpen
  \bibfield  {author} {\bibinfo {author} {\bibfnamefont {E.}~\bibnamefont
  {Icking}}, \bibinfo {author} {\bibfnamefont {L.}~\bibnamefont {Banszerus}},
  \bibinfo {author} {\bibfnamefont {F.}~\bibnamefont
  {W{\ifmmode\ddot{o}\else\"{o}\fi}rtche}}, \bibinfo {author} {\bibfnamefont
  {F.}~\bibnamefont {Volmer}}, \bibinfo {author} {\bibfnamefont
  {P.}~\bibnamefont {Schmidt}}, \bibinfo {author} {\bibfnamefont
  {C.}~\bibnamefont {Steiner}}, \bibinfo {author} {\bibfnamefont
  {S.}~\bibnamefont {Engels}}, \bibinfo {author} {\bibfnamefont
  {J.}~\bibnamefont {Hesselmann}}, \bibinfo {author} {\bibfnamefont
  {M.}~\bibnamefont {Goldsche}}, \bibinfo {author} {\bibfnamefont
  {K.}~\bibnamefont {Watanabe}}, \bibinfo {author} {\bibfnamefont
  {T.}~\bibnamefont {Taniguchi}}, \bibinfo {author} {\bibfnamefont
  {C.}~\bibnamefont {Volk}}, \bibinfo {author} {\bibfnamefont {B.}~\bibnamefont
  {Beschoten}},\ and\ \bibinfo {author} {\bibfnamefont {C.}~\bibnamefont
  {Stampfer}},\ }\bibfield  {title} {\bibinfo {title} {{Transport Spectroscopy
  of Ultraclean Tunable Band Gaps in Bilayer Graphene}},\ }\href
  {https://doi.org/10.1002/aelm.202200510} {\bibfield  {journal} {\bibinfo
  {journal} {Adv. Electron. Mater.}\ }\textbf {\bibinfo {volume} {8}},\
  \bibinfo {pages} {2200510} (\bibinfo {year} {2022})}\BibitemShut {NoStop}%
\bibitem [{\citenamefont {Shintaku}\ \emph {et~al.}(2023)\citenamefont
  {Shintaku}, \citenamefont {Kareekunnan}, \citenamefont {Akabori},
  \citenamefont {Watanabe}, \citenamefont {Taniguchi},\ and\ \citenamefont
  {Mizuta}}]{Shintaku2023}%
  \BibitemOpen
  \bibfield  {author} {\bibinfo {author} {\bibfnamefont {T.}~\bibnamefont
  {Shintaku}}, \bibinfo {author} {\bibfnamefont {A.}~\bibnamefont
  {Kareekunnan}}, \bibinfo {author} {\bibfnamefont {M.}~\bibnamefont
  {Akabori}}, \bibinfo {author} {\bibfnamefont {K.}~\bibnamefont {Watanabe}},
  \bibinfo {author} {\bibfnamefont {T.}~\bibnamefont {Taniguchi}},\ and\
  \bibinfo {author} {\bibfnamefont {H.}~\bibnamefont {Mizuta}},\ }\bibfield
  {title} {\bibinfo {title} {{Berry curvature induced valley Hall effect in
  non-encapsulated hBN/Bilayer graphene heterostructure aligned with near-zero
  twist angle}},\ }\bibfield  {journal} {\bibinfo  {journal} {arXiv}\ }\href
  {https://doi.org/10.48550/arXiv.2301.02358} {10.48550/arXiv.2301.02358}
  (\bibinfo {year} {2023}),\ \Eprint {https://arxiv.org/abs/2301.02358}
  {2301.02358} \BibitemShut {NoStop}%
\bibitem [{\citenamefont {Yin}\ \emph {et~al.}(2022)\citenamefont {Yin},
  \citenamefont {Tan}, \citenamefont {Barcons-Ruiz}, \citenamefont {Torre},
  \citenamefont {Watanabe}, \citenamefont {Taniguchi}, \citenamefont {Song},
  \citenamefont {Hone},\ and\ \citenamefont {Koppens}}]{Yin2022}%
  \BibitemOpen
  \bibfield  {author} {\bibinfo {author} {\bibfnamefont {J.}~\bibnamefont
  {Yin}}, \bibinfo {author} {\bibfnamefont {C.}~\bibnamefont {Tan}}, \bibinfo
  {author} {\bibfnamefont {D.}~\bibnamefont {Barcons-Ruiz}}, \bibinfo {author}
  {\bibfnamefont {I.}~\bibnamefont {Torre}}, \bibinfo {author} {\bibfnamefont
  {K.}~\bibnamefont {Watanabe}}, \bibinfo {author} {\bibfnamefont
  {T.}~\bibnamefont {Taniguchi}}, \bibinfo {author} {\bibfnamefont {J.~C.~W.}\
  \bibnamefont {Song}}, \bibinfo {author} {\bibfnamefont {J.}~\bibnamefont
  {Hone}},\ and\ \bibinfo {author} {\bibfnamefont {F.~H.~L.}\ \bibnamefont
  {Koppens}},\ }\bibfield  {title} {\bibinfo {title} {{Tunable and giant
  valley-selective Hall effect in gapped bilayer graphene}},\ }\href
  {https://doi.org/10.1126/science.abl4266} {\bibfield  {journal} {\bibinfo
  {journal} {Science}\ }\textbf {\bibinfo {volume} {375}},\ \bibinfo {pages}
  {1398} (\bibinfo {year} {2022})}\BibitemShut {NoStop}%
\bibitem [{\citenamefont {Guinea}\ \emph {et~al.}(2010)\citenamefont {Guinea},
  \citenamefont {Katsnelson},\ and\ \citenamefont {Geim}}]{Guinea2010Jan}%
  \BibitemOpen
  \bibfield  {author} {\bibinfo {author} {\bibfnamefont {F.}~\bibnamefont
  {Guinea}}, \bibinfo {author} {\bibfnamefont {M.~I.}\ \bibnamefont
  {Katsnelson}},\ and\ \bibinfo {author} {\bibfnamefont {A.~K.}\ \bibnamefont
  {Geim}},\ }\bibfield  {title} {\bibinfo {title} {{Energy gaps and a
  zero-field quantum Hall effect in graphene by strain engineering}},\ }\href
  {https://doi.org/10.1038/nphys1420} {\bibfield  {journal} {\bibinfo
  {journal} {Nat. Phys.}\ }\textbf {\bibinfo {volume} {6}},\ \bibinfo {pages}
  {30} (\bibinfo {year} {2010})}\BibitemShut {NoStop}%
\bibitem [{\citenamefont {Wang}\ \emph {et~al.}(2021)\citenamefont {Wang},
  \citenamefont {Baumgartner}, \citenamefont {Makk}, \citenamefont {Zihlmann},
  \citenamefont {Varghese}, \citenamefont {Indolese}, \citenamefont {Watanabe},
  \citenamefont {Taniguchi},\ and\ \citenamefont
  {Sch{\ifmmode\ddot{o}\else\"{o}\fi}nenberger}}]{Wang2021}%
  \BibitemOpen
  \bibfield  {author} {\bibinfo {author} {\bibfnamefont {L.}~\bibnamefont
  {Wang}}, \bibinfo {author} {\bibfnamefont {A.}~\bibnamefont {Baumgartner}},
  \bibinfo {author} {\bibfnamefont {P.}~\bibnamefont {Makk}}, \bibinfo {author}
  {\bibfnamefont {S.}~\bibnamefont {Zihlmann}}, \bibinfo {author}
  {\bibfnamefont {B.~S.}\ \bibnamefont {Varghese}}, \bibinfo {author}
  {\bibfnamefont {D.~I.}\ \bibnamefont {Indolese}}, \bibinfo {author}
  {\bibfnamefont {K.}~\bibnamefont {Watanabe}}, \bibinfo {author}
  {\bibfnamefont {T.}~\bibnamefont {Taniguchi}},\ and\ \bibinfo {author}
  {\bibfnamefont {C.}~\bibnamefont
  {Sch{\ifmmode\ddot{o}\else\"{o}\fi}nenberger}},\ }\bibfield  {title}
  {\bibinfo {title} {{Global strain-induced scalar potential in graphene
  devices}},\ }\href {https://doi.org/10.1038/s42005-021-00651-y} {\bibfield
  {journal} {\bibinfo  {journal} {Commun. Phys.}\ }\textbf {\bibinfo {volume}
  {4}},\ \bibinfo {pages} {1} (\bibinfo {year} {2021})}\BibitemShut {NoStop}%
\bibitem [{\citenamefont {Choi}\ \emph {et~al.}(2010)\citenamefont {Choi},
  \citenamefont {Jhi},\ and\ \citenamefont {Son}}]{Choi2010}%
  \BibitemOpen
  \bibfield  {author} {\bibinfo {author} {\bibfnamefont {S.-M.}\ \bibnamefont
  {Choi}}, \bibinfo {author} {\bibfnamefont {S.-H.}\ \bibnamefont {Jhi}},\ and\
  \bibinfo {author} {\bibfnamefont {Y.-W.}\ \bibnamefont {Son}},\ }\bibfield
  {title} {\bibinfo {title} {{Effects of strain on electronic properties of
  graphene}},\ }\href {https://doi.org/10.1103/PhysRevB.81.081407} {\bibfield
  {journal} {\bibinfo  {journal} {Phys. Rev. B}\ }\textbf {\bibinfo {volume}
  {81}},\ \bibinfo {pages} {081407} (\bibinfo {year} {2010})}\BibitemShut
  {NoStop}%
\bibitem [{\citenamefont {Aktor}\ \emph {et~al.}(2021)\citenamefont {Aktor},
  \citenamefont {Garcia}, \citenamefont {Roche}, \citenamefont {Jauho},\ and\
  \citenamefont {Power}}]{Aktor2021}%
  \BibitemOpen
  \bibfield  {author} {\bibinfo {author} {\bibfnamefont {T.}~\bibnamefont
  {Aktor}}, \bibinfo {author} {\bibfnamefont {J.~H.}\ \bibnamefont {Garcia}},
  \bibinfo {author} {\bibfnamefont {S.}~\bibnamefont {Roche}}, \bibinfo
  {author} {\bibfnamefont {A.-P.}\ \bibnamefont {Jauho}},\ and\ \bibinfo
  {author} {\bibfnamefont {S.~R.}\ \bibnamefont {Power}},\ }\bibfield  {title}
  {\bibinfo {title} {{Valley Hall effect and nonlocal resistance in locally
  gapped graphene}},\ }\href {https://doi.org/10.1103/PhysRevB.103.115406}
  {\bibfield  {journal} {\bibinfo  {journal} {Phys. Rev. B}\ }\textbf {\bibinfo
  {volume} {103}},\ \bibinfo {pages} {115406} (\bibinfo {year}
  {2021})}\BibitemShut {NoStop}%
\bibitem [{\citenamefont {Renard}\ \emph {et~al.}(2014)\citenamefont {Renard},
  \citenamefont {Studer},\ and\ \citenamefont {Folk}}]{Renard2014}%
  \BibitemOpen
  \bibfield  {author} {\bibinfo {author} {\bibfnamefont {J.}~\bibnamefont
  {Renard}}, \bibinfo {author} {\bibfnamefont {M.}~\bibnamefont {Studer}},\
  and\ \bibinfo {author} {\bibfnamefont {J.~A.}\ \bibnamefont {Folk}},\
  }\bibfield  {title} {\bibinfo {title} {{Origins of Nonlocality Near the
  Neutrality Point in Graphene}},\ }\href
  {https://doi.org/10.1103/PhysRevLett.112.116601} {\bibfield  {journal}
  {\bibinfo  {journal} {Phys. Rev. Lett.}\ }\textbf {\bibinfo {volume} {112}},\
  \bibinfo {pages} {116601} (\bibinfo {year} {2014})}\BibitemShut {NoStop}%
\bibitem [{\citenamefont {Lee}\ \emph {et~al.}(2016)\citenamefont {Lee},
  \citenamefont {Wallbank}, \citenamefont {Gallagher}, \citenamefont
  {Watanabe}, \citenamefont {Taniguchi}, \citenamefont {Fal{'}ko},\ and\
  \citenamefont {Goldhaber-Gordon}}]{Lee2016}%
  \BibitemOpen
  \bibfield  {author} {\bibinfo {author} {\bibfnamefont {M.}~\bibnamefont
  {Lee}}, \bibinfo {author} {\bibfnamefont {J.~R.}\ \bibnamefont {Wallbank}},
  \bibinfo {author} {\bibfnamefont {P.}~\bibnamefont {Gallagher}}, \bibinfo
  {author} {\bibfnamefont {K.}~\bibnamefont {Watanabe}}, \bibinfo {author}
  {\bibfnamefont {T.}~\bibnamefont {Taniguchi}}, \bibinfo {author}
  {\bibfnamefont {V.~I.}\ \bibnamefont {Fal{'}ko}},\ and\ \bibinfo {author}
  {\bibfnamefont {D.}~\bibnamefont {Goldhaber-Gordon}},\ }\bibfield  {title}
  {\bibinfo {title} {{Ballistic miniband conduction in a graphene
  superlattice}},\ }\href
  {https://www-science-org.inp.bib.cnrs.fr/doi/10.1126/science.aaf1095}
  {\bibfield  {journal} {\bibinfo  {journal} {Science}\ } (\bibinfo {year}
  {2016})}\BibitemShut {NoStop}%
\bibitem [{\citenamefont {Berdyugin}\ \emph {et~al.}(2020)\citenamefont
  {Berdyugin}, \citenamefont {Tsim}, \citenamefont {Kumaravadivel},
  \citenamefont {Xu}, \citenamefont {Ceferino}, \citenamefont {Knothe},
  \citenamefont {Kumar}, \citenamefont {Taniguchi}, \citenamefont {Watanabe},
  \citenamefont {Geim}, \citenamefont {Grigorieva},\ and\ \citenamefont
  {Fal{'}ko}}]{Berdyugin2020}%
  \BibitemOpen
  \bibfield  {author} {\bibinfo {author} {\bibfnamefont {A.~I.}\ \bibnamefont
  {Berdyugin}}, \bibinfo {author} {\bibfnamefont {B.}~\bibnamefont {Tsim}},
  \bibinfo {author} {\bibfnamefont {P.}~\bibnamefont {Kumaravadivel}}, \bibinfo
  {author} {\bibfnamefont {S.~G.}\ \bibnamefont {Xu}}, \bibinfo {author}
  {\bibfnamefont {A.}~\bibnamefont {Ceferino}}, \bibinfo {author}
  {\bibfnamefont {A.}~\bibnamefont {Knothe}}, \bibinfo {author} {\bibfnamefont
  {R.~K.}\ \bibnamefont {Kumar}}, \bibinfo {author} {\bibfnamefont
  {T.}~\bibnamefont {Taniguchi}}, \bibinfo {author} {\bibfnamefont
  {K.}~\bibnamefont {Watanabe}}, \bibinfo {author} {\bibfnamefont {A.~K.}\
  \bibnamefont {Geim}}, \bibinfo {author} {\bibfnamefont {I.~V.}\ \bibnamefont
  {Grigorieva}},\ and\ \bibinfo {author} {\bibfnamefont {V.~I.}\ \bibnamefont
  {Fal{'}ko}},\ }\bibfield  {title} {\bibinfo {title} {{Minibands in twisted
  bilayer graphene probed by magnetic focusing}},\ }\href
  {https://doi.org/10.1126/sciadv.aay7838} {\bibfield  {journal} {\bibinfo
  {journal} {Sci. Adv.}\ }\textbf {\bibinfo {volume} {6}},\ \bibinfo {pages}
  {eaay7838} (\bibinfo {year} {2020})}\BibitemShut {NoStop}%
\bibitem [{\citenamefont {Taychatanapat}\ \emph {et~al.}(2013)\citenamefont
  {Taychatanapat}, \citenamefont {Watanabe}, \citenamefont {Taniguchi},\ and\
  \citenamefont {Jarillo-Herrero}}]{Taychatanapat2013}%
  \BibitemOpen
  \bibfield  {author} {\bibinfo {author} {\bibfnamefont {T.}~\bibnamefont
  {Taychatanapat}}, \bibinfo {author} {\bibfnamefont {K.}~\bibnamefont
  {Watanabe}}, \bibinfo {author} {\bibfnamefont {T.}~\bibnamefont
  {Taniguchi}},\ and\ \bibinfo {author} {\bibfnamefont {P.}~\bibnamefont
  {Jarillo-Herrero}},\ }\bibfield  {title} {\bibinfo {title} {{Electrically
  tunable transverse magnetic focusing in graphene}},\ }\href
  {https://doi.org/10.1038/nphys2549} {\bibfield  {journal} {\bibinfo
  {journal} {Nat. Phys.}\ }\textbf {\bibinfo {volume} {9}},\ \bibinfo {pages}
  {225} (\bibinfo {year} {2013})}\BibitemShut {NoStop}%
\bibitem [{\citenamefont {Trambly~de
  Laissardi{\ifmmode\grave{e}\else\`{e}\fi}re}\ \emph
  {et~al.}(2010)\citenamefont {Trambly~de
  Laissardi{\ifmmode\grave{e}\else\`{e}\fi}re}, \citenamefont {Mayou},\ and\
  \citenamefont {Magaud}}]{TramblydeLaissardiere2010}%
  \BibitemOpen
  \bibfield  {author} {\bibinfo {author} {\bibfnamefont {G.}~\bibnamefont
  {Trambly~de Laissardi{\ifmmode\grave{e}\else\`{e}\fi}re}}, \bibinfo {author}
  {\bibfnamefont {D.}~\bibnamefont {Mayou}},\ and\ \bibinfo {author}
  {\bibfnamefont {L.}~\bibnamefont {Magaud}},\ }\bibfield  {title} {\bibinfo
  {title} {{Localization of Dirac Electrons in Rotated Graphene Bilayers}},\
  }\href {https://doi.org/10.1021/nl902948m} {\bibfield  {journal} {\bibinfo
  {journal} {Nano Lett.}\ }\textbf {\bibinfo {volume} {10}},\ \bibinfo {pages}
  {804} (\bibinfo {year} {2010})}\BibitemShut {NoStop}%
\bibitem [{\citenamefont {Moon}\ and\ \citenamefont
  {Koshino}(2014)}]{Moon2014}%
  \BibitemOpen
  \bibfield  {author} {\bibinfo {author} {\bibfnamefont {P.}~\bibnamefont
  {Moon}}\ and\ \bibinfo {author} {\bibfnamefont {M.}~\bibnamefont {Koshino}},\
  }\bibfield  {title} {\bibinfo {title} {{Electronic properties of
  graphene/hexagonal-boron-nitride
  moir{\ifmmode\backslash\else\textbackslash\fi}'e superlattice}},\ }\href
  {https://doi.org/10.1103/PhysRevB.90.155406} {\bibfield  {journal} {\bibinfo
  {journal} {Phys. Rev. B}\ }\textbf {\bibinfo {volume} {90}},\ \bibinfo
  {pages} {155406} (\bibinfo {year} {2014})}\BibitemShut {NoStop}%
\bibitem [{\citenamefont {Zhou}\ and\ \citenamefont
  {Charlier}(2021)}]{Jiaqi2021}%
  \BibitemOpen
  \bibfield  {author} {\bibinfo {author} {\bibfnamefont {J.}~\bibnamefont
  {Zhou}}\ and\ \bibinfo {author} {\bibfnamefont {J.-C.}\ \bibnamefont
  {Charlier}},\ }\bibfield  {title} {\bibinfo {title} {Controllable spin
  current in van der waals ferromagnet
  ${\mathrm{fe}}_{3}{\mathrm{gete}}_{2}$},\ }\href
  {https://doi.org/10.1103/PhysRevResearch.3.L042033} {\bibfield  {journal}
  {\bibinfo  {journal} {Phys. Rev. Res.}\ }\textbf {\bibinfo {volume} {3}},\
  \bibinfo {pages} {L042033} (\bibinfo {year} {2021})}\BibitemShut {NoStop}%
\end{thebibliography}%

\section*{Acknowledgements}

 The authors acknowledge discussions with Ulf Gennser, Marco Polini, Herve Aubin, J.I.A. Li and Justin Song. R.R.-P. acknowledge the ERC starting grant TWISTRONICS. This work was done within the C2N micro nanotechnologies platforms and partly supported by the RENATECH network, the General Council of Essonne and the DIM-SIRTEC. V.-H.N. and J.-C.C. acknowledge financial support from the F\'ed\'eration Wallonie-Bruxelles through the ARC Grant (N$^{\circ}$ 21/26-116), from the European Unions Horizon 2020 Research Project and Innovation Program -  Graphene Flagship Core3 (N$^{\circ}$ 881603), from the Flag-Era JTC projects ``TATTOOS'' (N$^{\circ}$ R.8010.19) and ``MINERVA'' (N$^{\circ}$ R.8006.21), from the Pathfinder project ``FLATS'' (N$^{\circ}$ 101099139), from the F\'ed\'eration Wallonie-Bruxelles through the ARC Grant (N$^{\circ}$ 21/26-116) and the EOS project ``CONNECT'' (N$^{\circ}$ 40007563), and from the Belgium F.R.S.-FNRS through the research project (N$^{\circ}$ T.029.22F). Computational resources have been provided by the CISM supercomputing facilities of UCLouvain and the CE  CI consortium funded by F.R.S.-FNRS of Belgium (N$^{\circ}$ 2.5020.11). V.-H.N. thanks Dr. Xuan-Hoang TRINH for his helps in implementation of numerical codes to compute the lattice atomic structure relaxation.\\
 
 \section*{Author Contributions }
 
R.R.-P. and D.M. designed the experiment. E.A., M.D.L., Y.H. and L.F. fabricated the devices for electron transport measurements. G.M. and M.DL. fabricated the samples for structural characterization and performed the AFM measurements. E.A., M.D.L., Y.H., L.F.and R.R-P performed the electron transport experiments and analyzed the data.  T.T. and K.W. grew the crystals of hexagonal boron nitride. V.-H.N. and J.-C.C. performed the numerical simulations and participated to the data analysis. All authors participated to writing the paper. E.A. and V.-H.N. contributed equally to this work.
%\noin

\clearpage

%%%%%%%%%%%%%%%%%%%%%%%%%%%%%%%%%%%%%%%%%%%%%%%%%%%%%%%%%%%%%%%%%%%%%%%%%%%%%%%%%%%%%%%%
%\newpage
%\onecolumn
\newcommand{\bq}{{\bf q}}
\newcommand{\bp}{{\bf p}}
\newcommand{\br}{{\bf r}}
\newcommand{\bR}{{\bf R}}
\newcommand{\rhobar}{\bar{\rho}}
\newcommand{\nubar}{\bar{\nu}}

\renewcommand{\thefigure}{S\arabic{figure}}
\renewcommand{\thesubsection}{S\arabic{subsection}}
\renewcommand{\theequation}{S\arabic{equation}}
\setcounter{figure}{0} 
\setcounter{equation}{0}

\section*{Supplementary Information}

\section*{Samples used in this manuscript}

\begin{figure}[h!]
\centering
\includegraphics[scale=0.16]{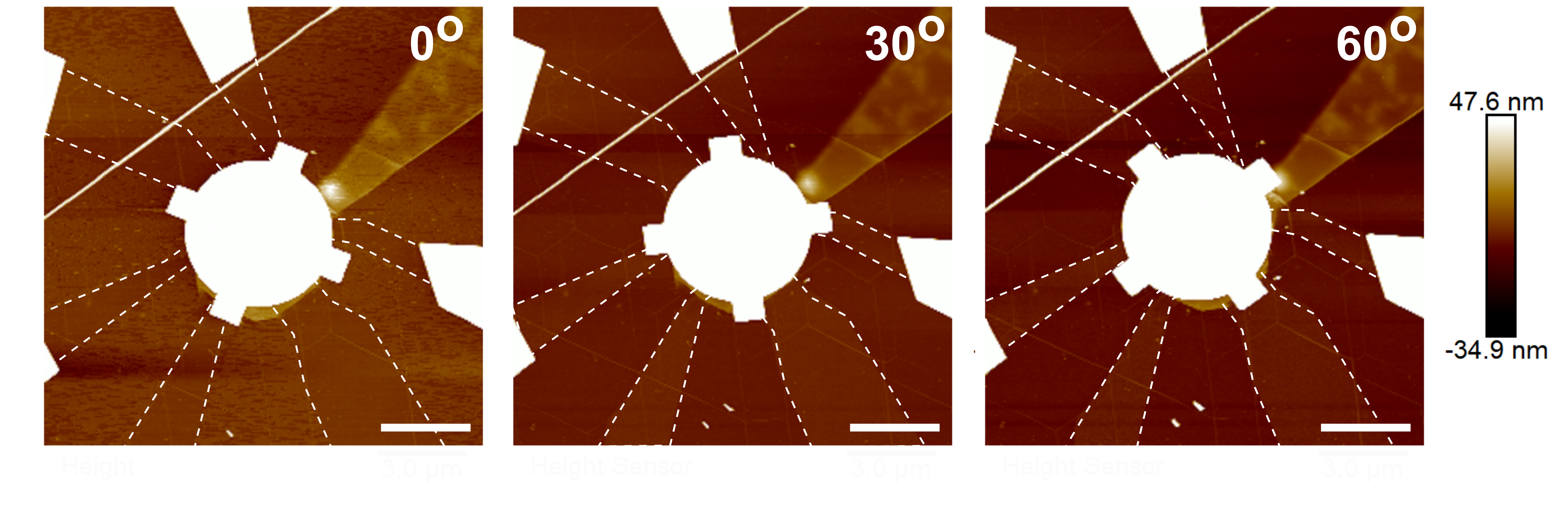}  
 \caption{{\bf Sample I.} Atomic force microscopy in tapping mode at $0^{\circ}$, $30^{\circ}$ and $60^{\circ}$ of alignment. Dashed lines highlight the Hall bar shape of the graphene layer. Scale bar 3 $\mu$m. }
\label{AFM-SampleI}
\end{figure}

\noindent {\it Sample I (called H038):} described in details in the the main text, it has a central graphite gate. In Fig. \ref{AFM-SampleI}  we present three different crystallographic alignments of the BN in this sample: 0$^{\circ}$, 30$^{\circ}$ and 60$^{\circ}$. The characteristic dimensions are: $W=$ 1.7 $\mu$m, $L=$ 2.3 $\mu$m, and the thickness of the bottom BN  is 24 nm. Capacitive coupling is $C_{\mathrm{g}}/e=6.4\times10^{15}$ V$^{-1}$ m$^{-2}$ (Hall measurements). 

\begin{figure}[h!]
\centering
\includegraphics[scale=0.28]{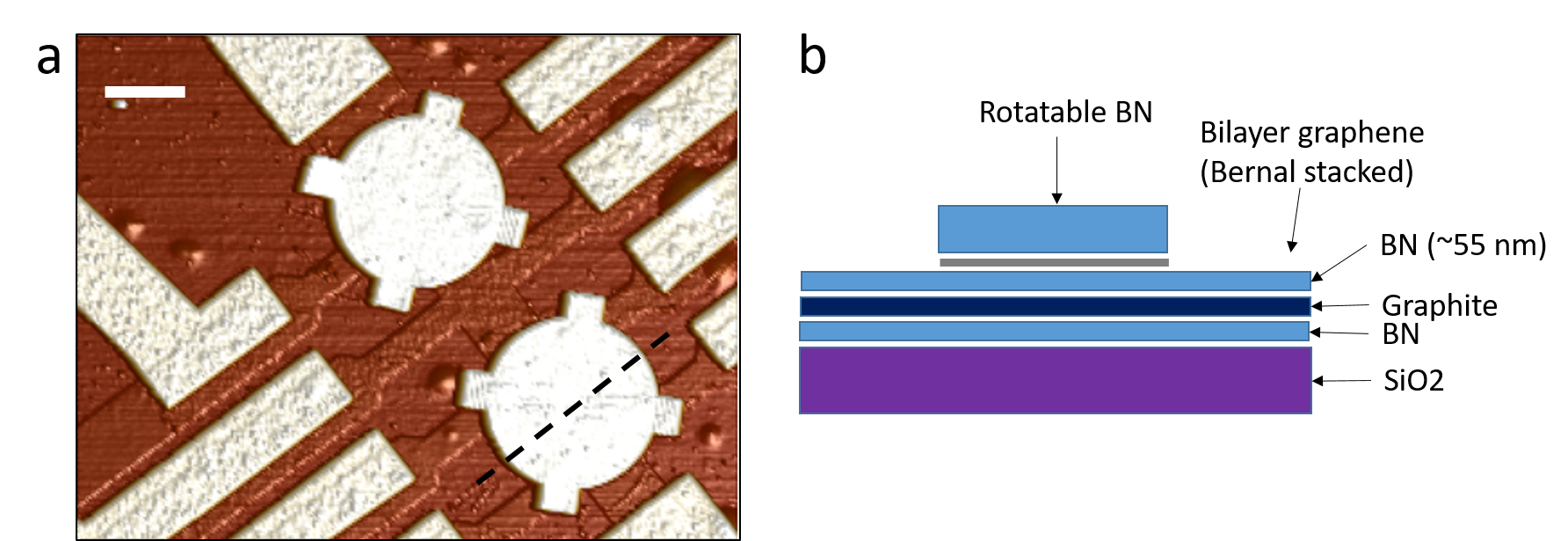}  
\caption{{\bf Sample II.} \textbf{a} Tapping mode AFM image of the sample.  Scale bar 2.5 $\mu$m. \textbf{b}, Schematic cross-section of the heterostructure at the position of the black dashed line.}
\label{AFM-SampleII}
\end{figure}

\noindent {\it Sample II (H012):} it was built using the same techniques but instead of a local graphite gate it has a global graphite gate, Fig. \ref{AFM-SampleII}. This sample is composed of two parts with two independent handles. For all the measurements presented here the  BN handle in the left  was misaligned. The characteristic dimensions are: $W=$ 1.8 $\mu$m, $L=$ 3 $\mu$m,  bottom BN thickness of 55 nm. Capacitive coupling is $C_{\mathrm{g}}/e=2.7\times10^{15}$ V$^{-1}$ m$^{-2}$ (Hall measurement).

\noindent {\it Sample III (MDL007a):} it has the same structure as sample I and it has been used to characterize the alignment inside the AFM. The characteristic dimensions are: $W=$ 1.2 $\mu$m, $L=$ 1.5 $\mu$m, and bottom BN thickness of 50 nm. Capacitive coupling is $C_{\mathrm{g}}/e=3.76\times10^{15}$ V$^{-1}$ m$^{-2}$ (plane capacitor calculation).

\noindent {\it Sample IV (SBG06):} with the same structure as sample I but different geometry. The characteristic dimensions are: $W=$ 2.3 $\mu$m, $L=$ 1.7 $\mu$m, and bottom BN thickness of 66 nm. Capacitive coupling $C_{\mathrm{g}}/e=3.02\times10^{15}$ V$^{-1}$ m$^{-2}$ (Hall measurement).

\section*{Crystallographic alignment characterization}

\begin{figure}[h!]
\centering
\includegraphics[scale=0.19]{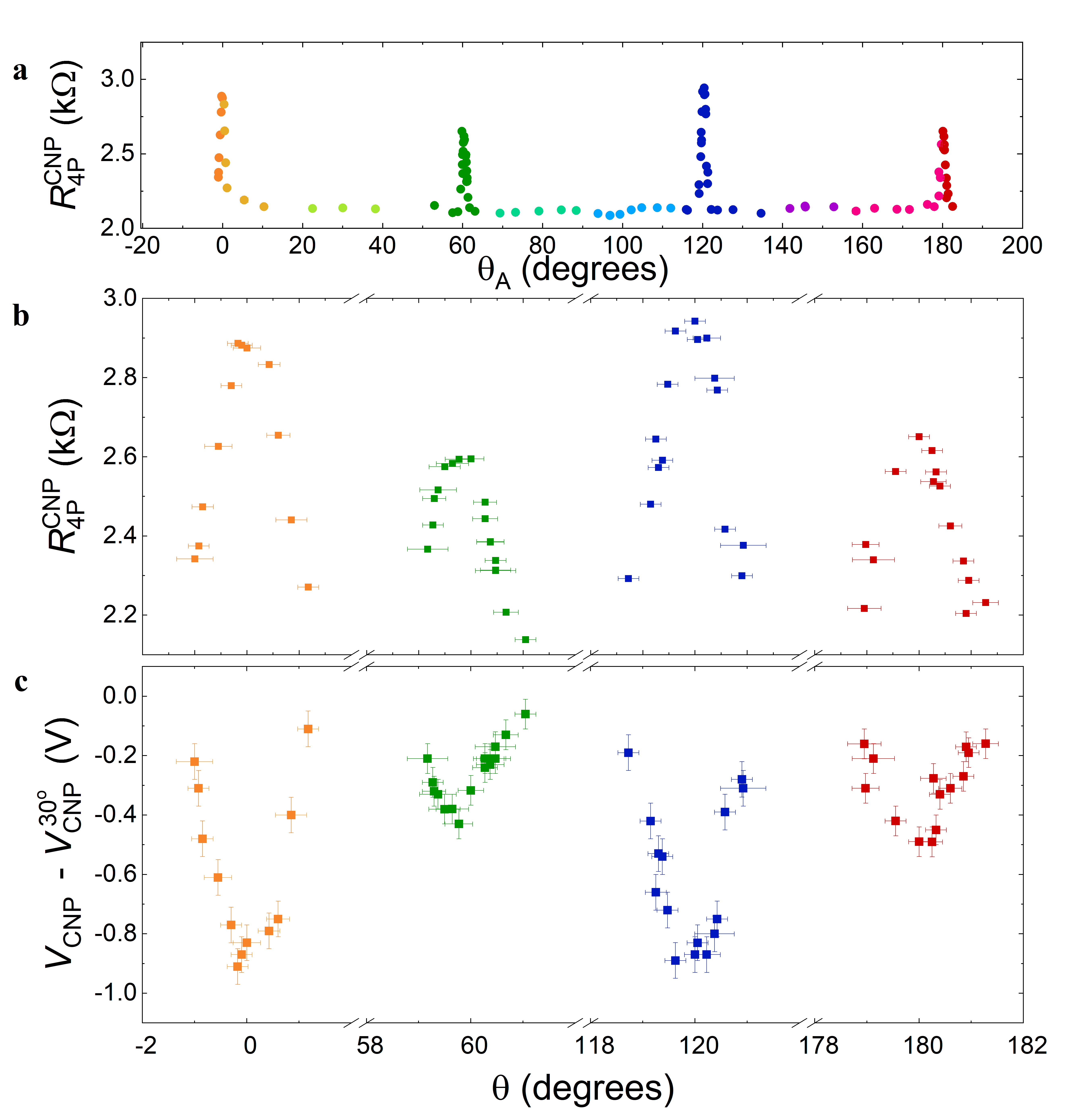}  
\caption{{\bf Crystallographic alignment at room temperature sample III.} \textbf{a}, Resistance of the CNP at as a function of the angular alignment,  measured with the AFM, from -2$^{\circ}$ to 182$^{\circ}$. \textbf{b}, Resistance value of the CNP as a function of the angular alignment around the aligned positions. \textbf{c}, Shift in voltage of the CNP, with respect to the misaligned position ($30^{\circ}$), as a function of the angular alignment around around the aligned positions.}
\label{Alignment-MDL07}
\end{figure}

As explained in the main text, we use an AFM tip to push the capping BN layer, by applying a force to one of the arms we are able to rotate it, as it can be seen in Fig. \ref{AFM-SampleI}. As we approach the position where graphene and BN are crystallographically aligned, the resistance peak around the CNP becomes larger. This enlargement appears every sixty degrees, Fig. \ref{Alignment-MDL07}. However,  we can see that the enlargement is not the same every sixty degrees, instead it has a hundred and twenty degrees periodicity, that can be seen even at room temperature. In Fig. \ref{Alignment-MDL07}a, b and c, we can clearly see the similarities between the curves: at 0$^{\circ}$ and 120$^{\circ}$ we observe similar height ($R^{CNP}_{4P}$) and shift on the voltage of the CNP ($V_{CNP}-V^{30^{\circ}}_{CNP}$), while they are different from the ones at 60$^{\circ}$ and 180$^{\circ}$. 

\begin{figure}[h!]
\centering
\includegraphics[scale=0.2]{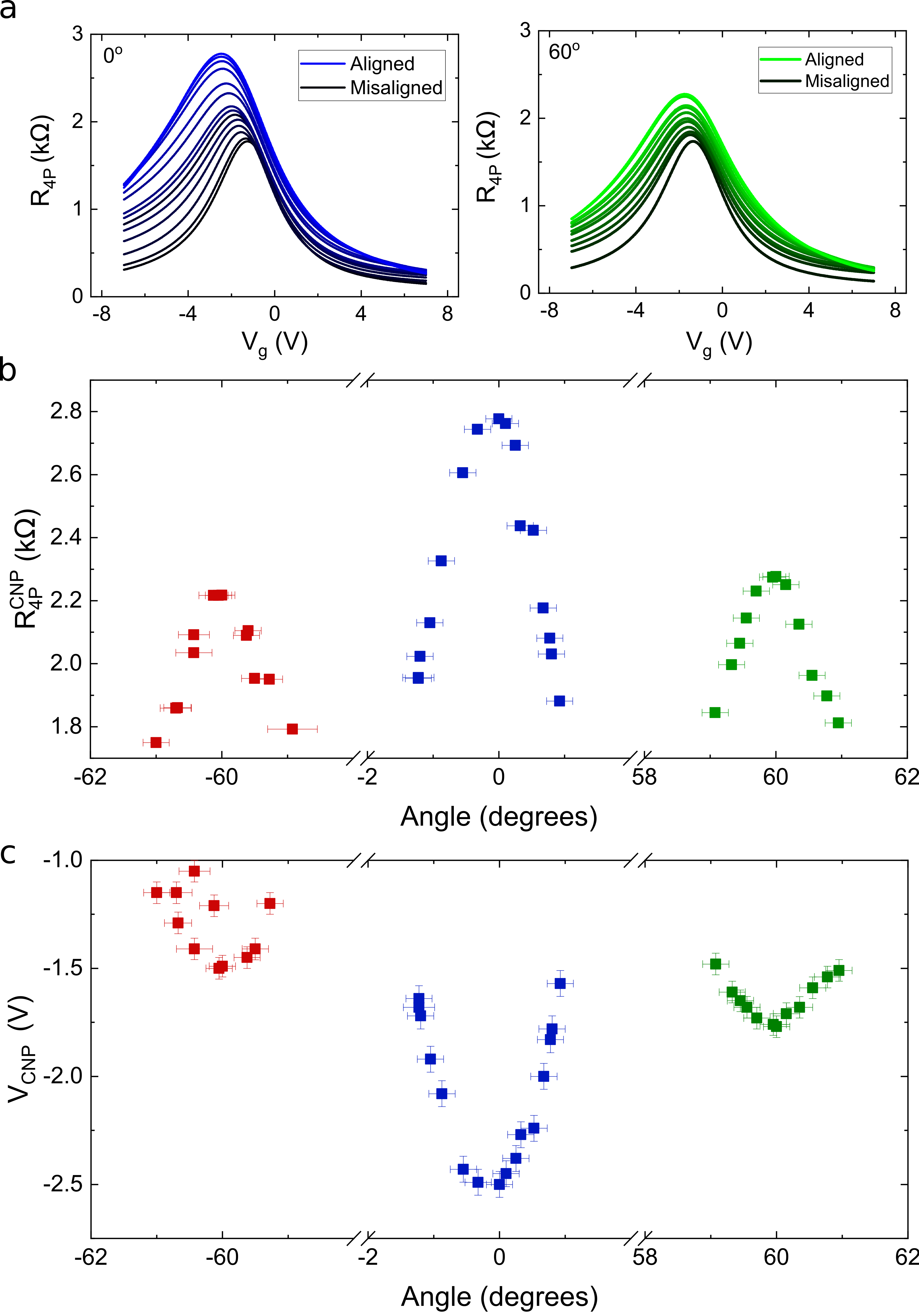}  
\caption{{\bf Crystallographic alignment at room temperature sample II.} \textbf{a}, Resistance as a function of the gate voltage for different crystallographic alignments around 0$^{\circ}$ and 60$^{\circ}$. \textbf{b}, Resistance value of the CNP as a function of the angular alignment measured with the AFM around -60$^{\circ}$, 0$^{\circ}$ and 60$^{\circ}$. \textbf{c}, Position of the CNP as a function of the angular alignment around -60$^{\circ}$, 0$^{\circ}$ and 60$^{\circ}$.}
\label{Alignment-H012}
\end{figure}

The periodicity of this behavior is observed in all the measured samples, see Fig. \ref{Alignment-MDL07}, \ref{Alignment-H012}  and Fig 2c of the main text. In all measurements the height of the resistance peak and the position of the CNP in voltage are periodic every hundred and twenty degrees. 

It is important to highlight that the differences between 0$^{\circ}$ and 60$^{\circ}$ alignment becomes evident in our experiments only because we are able to measure the same sample with different crystallographic alignments. In other words, if we had two different samples with these  characteristics, the difference would be attributed to sample-to-sample variation and not to a real effect of the angular alignment.

In an intuitive picture we can expect that if both aligned positions are obtained in the same sample and share about the same properties at room temperature the difference between them is coming from a more subtle difference, a different effective mass (see discussion in the main text). 

The change in position of the CNP, different for 0$^{\circ}$ and 60$^{\circ}$, reflects different levels of strain in the system. It has been demonstrated experimentally \cite{Wang2021}, and explained theoretically\cite{Choi2010}, that  strain will modify the work function of graphene. This will be reflected in a change in the position of the CNP as a function of the strain. In our experiments, the existence of the commensurate state generates strain inside the moir\'e cell. This  has the same  effect in our samples, a shift on the position of the CNP in gate voltage.

\section*{Sample characterization}

\begin{figure}
\centering
\includegraphics[scale=0.35]{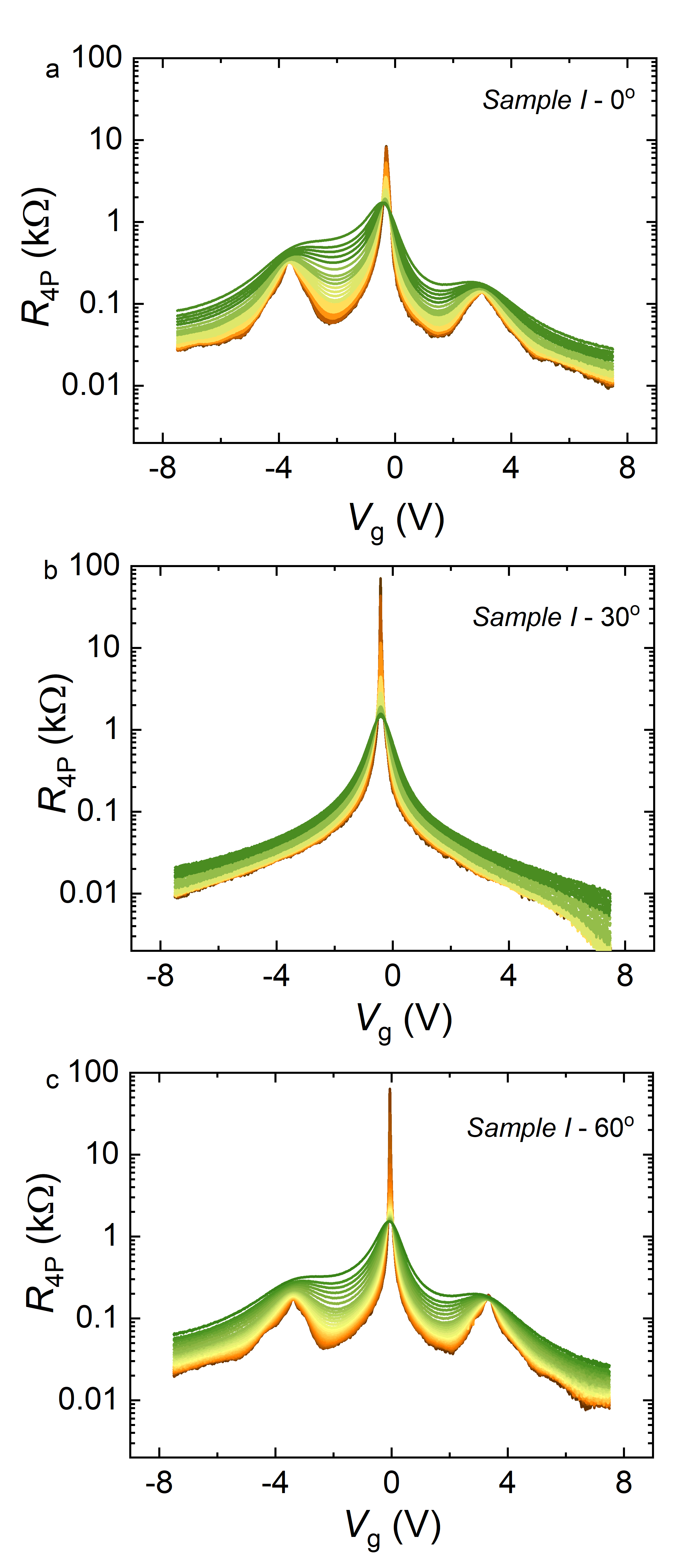}  
\caption{{\bf Temperature dependence for the resistance of sample I.} Temperature dependence of the resistance as a function of the gate voltage, in semi-Log scale for \textbf{a} 0$^{\circ}$,  \textbf{b} 30$^{\circ}$ and  \textbf{c} 60$^{\circ}$ of alignment. Measurements between 1.4 K (brown) and 220 K (green).}
\label{FigLocalTemp-H038}
\end{figure}

\begin{figure}
\centering
\includegraphics[scale=0.35]{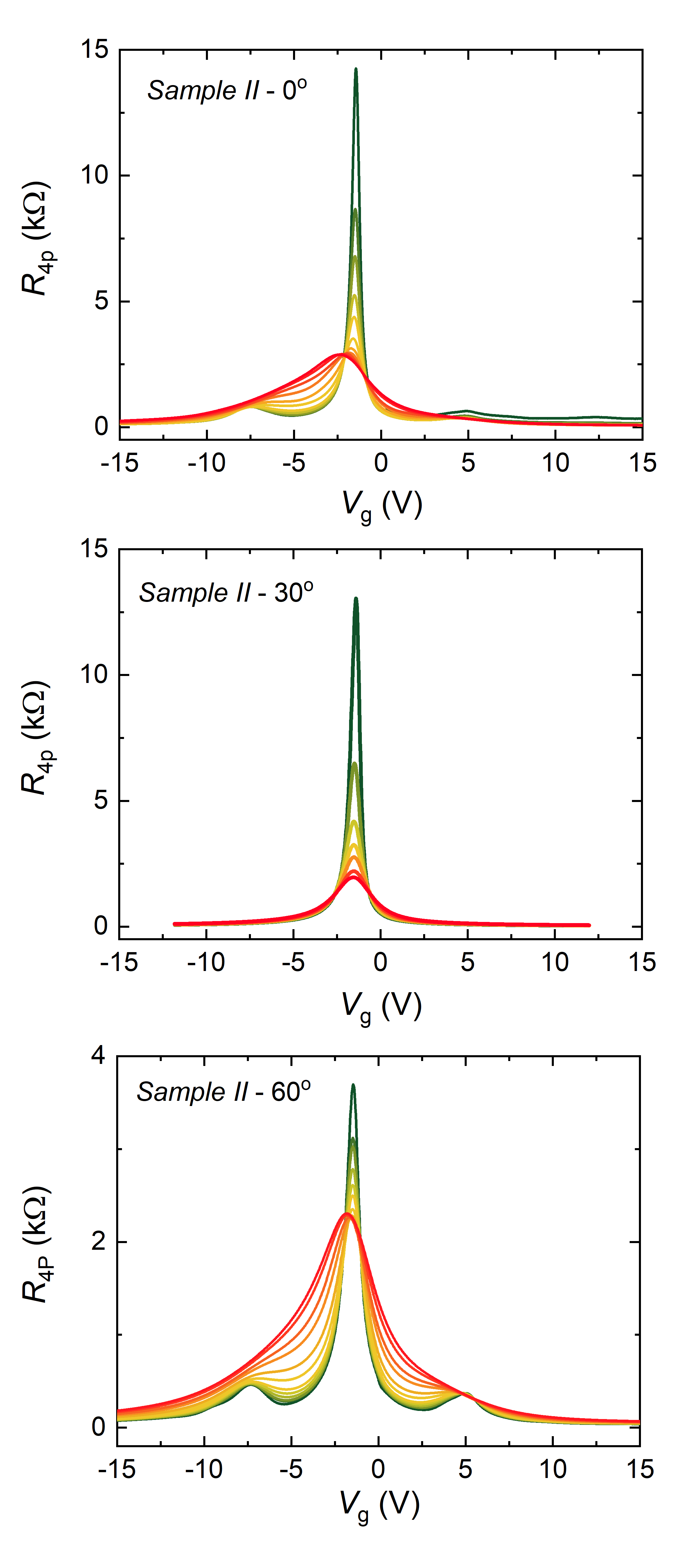}  
\caption{{\bf Temperature dependence sample II.} Temperature dependence of the resistance as a function of the gate voltage for \textbf{a} 0$^{\circ}$,  \textbf{b} 30$^{\circ}$ and  \textbf{c} 60$^{\circ}$ of alignment. Measurements between 20 K (green) and 200 K (red).}
\label{FigLocalTemp-H012}
\end{figure}

The temperature dependence of the resistance as a function of the gate voltage, Figs. \ref{FigLocalTemp-H038} and \ref{FigLocalTemp-H012}, shows clearly the presence of satellite peaks at low temperature. Its evolution in temperature shows that  the broadening of the $R(V_{\mathrm{g}})$ curves at room temperature is an indication of the appearance of the satellite peaks, hidden by thermal broadening.

\begin{figure}
\centering
\includegraphics[scale=0.35]{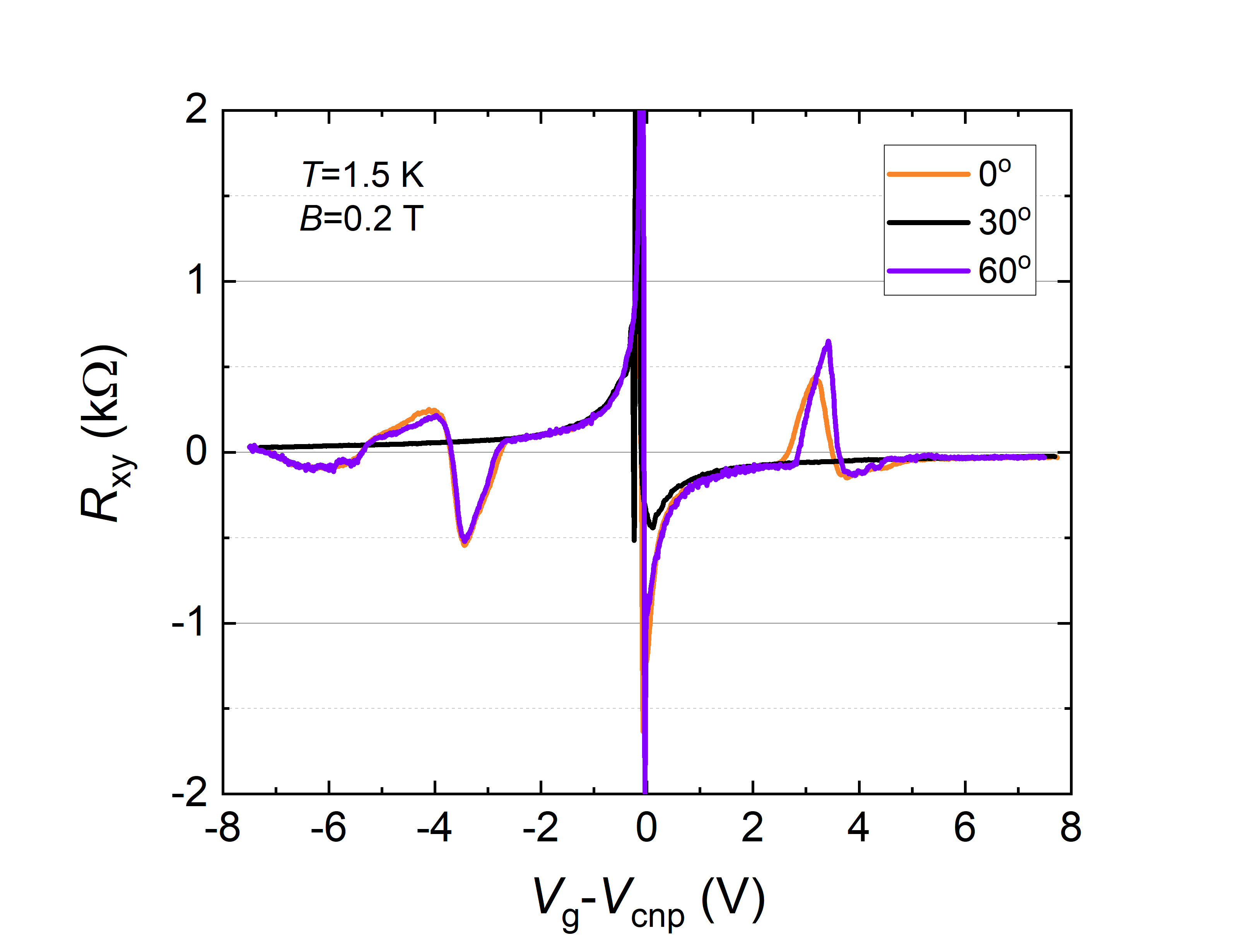}  
\caption{{\bf Hall resistance measurement, sample I}. Measurements at low temperature and 0.2 T of the Hall resistance for the three crystallographic alignments $0^{\circ}$, $30^{\circ}$ and $60^{\circ}$.}
\label{Hall resistance}
\end{figure}

When performing Hall resistance measurements in the presence of a low magnetic field (0.2 T), both the satellite peaks and the CNP are accompanied by sign inversion of the Hall resistance, Fig. \ref{Hall resistance}, both evident at 0$^{\circ}$ and 60$^{\circ}$ alignment. At $30^{\circ}$  case, full misalignment, there are no satellite peaks and therefore the Hall resistance stays close to zero in such regions.

%%%%%%%%%%%%%%%%%%%%%%%%%%%%%%%%%%%%%%

\subsection*{Mean free path at different angular alignments}

We calculate the mean free path versus temperature from the measurements of resistance as a function of gate voltage at different temperatures, for example Fig. \ref{FigLocalTemp-H038} and \ref{FigLocalTemp-H012}, by using the expression:

\begin{equation}
l_{\mathrm{mfp}}=\frac{\sigma h}{2e^2\sqrt{\pi n}}.
\end{equation}

\begin{figure}
\centering
\includegraphics[scale=0.3]{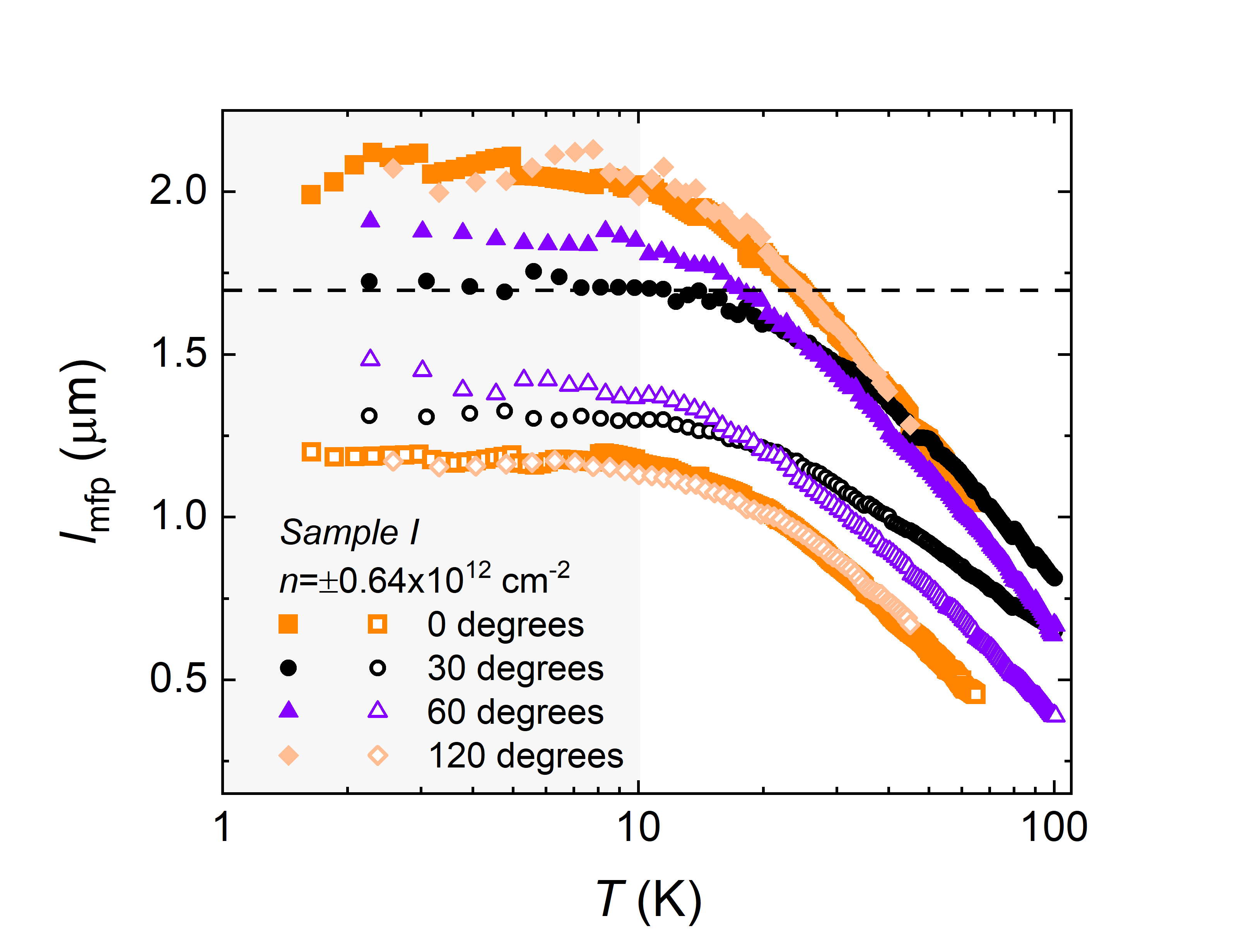}  
\caption{{\bf Mean free path, for sample I,} calculated from the resistance as a function of $V_{\mathrm{g}}$ at different temperatures for a density of $+0.64\times10^{12}$ cm$^{-2}$ filled symbols and $-0.64\times10^{12}$ cm$^{-2}$ (empty symbols) for different angular alignments. Dashed horizontal line represents the width of our sample. The shaded area corresponds to the values of temperature where the mean free path is not varying with temperature.}
\label{MFP_sample_I}
\end{figure}

In Fig. \ref{MFP_sample_I} we can see that sample I is in a ballistic regime for $T<10$ K, shadow area. We also remarked that the temperature dependence of the mean free path for the misaligned ($30^{\circ}$) case turns out to be different for the aligned case. The understanding of this regime is out of the scope of our manuscript, since it will need to extend the measurements to higher temperatures. 

\begin{figure}
\centering
\includegraphics[scale=0.3]{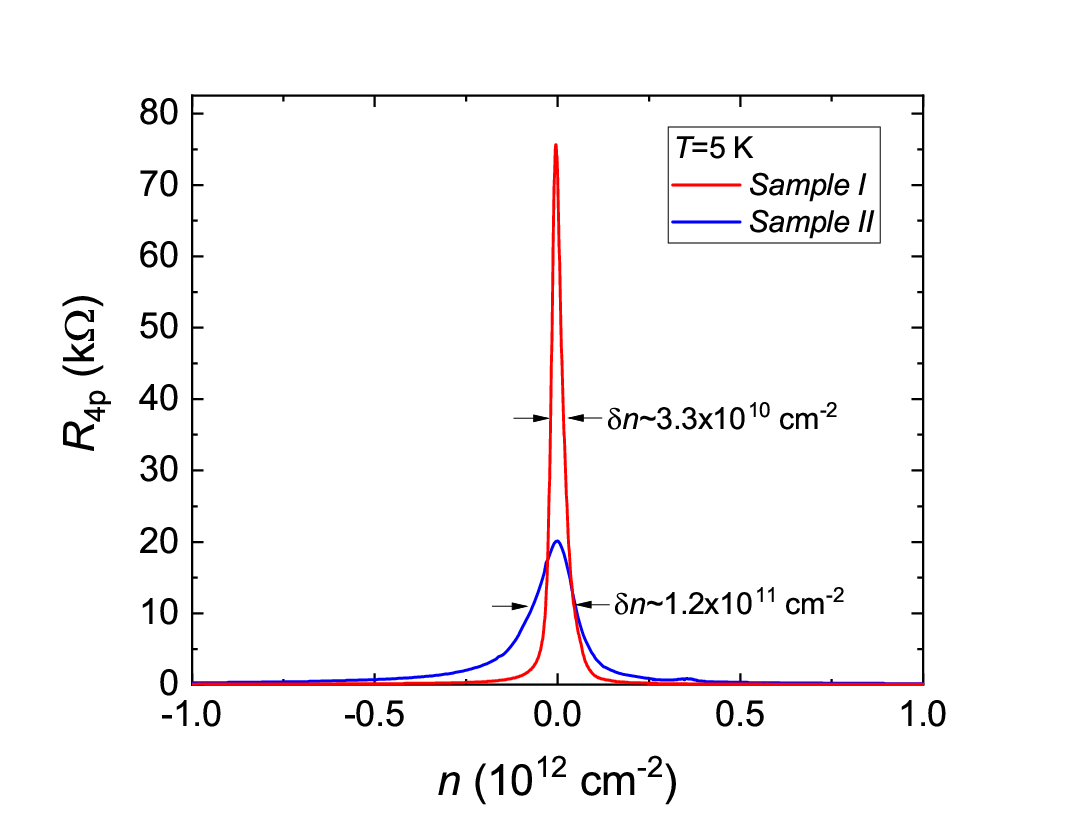}  
\caption{{\bf Residual carrier density for sample I and II}. Four probe resistance measurements as a function of the carrier density for samples I and II at $T=5$ K in the misaligned position 30$^{\circ}$. The full width at half maximum of the resistance peak at zero magnetic field provides a measurement of sample quality.}
\label{Disorder}
\end{figure}

From  measurements of resistance as a function of the carrier density we can have an idea of the sample quality by extracting the residual carrier density, $\delta n$, from the full width at half maximum of the CNP peak. In Fig. \ref{Disorder} we can see that the residual carrier density for sample II is about one order of magnitude larger than for sample I. This is only a comparative measurement, unfortunately we cannot extract a quantitative value of disorder.

Also notice that this qualitative comparison refers mostly to bulk disorder and not to edge disorder.

%%%%%%%%%%%%%%%%%%%%%%%%%%%%%%%%%%%%%%%%%%%%

%\newpage

\section*{Non-local measurements}

\subsection*{Measurement configurations}

\begin{figure}
\centering
\includegraphics[scale=0.3]{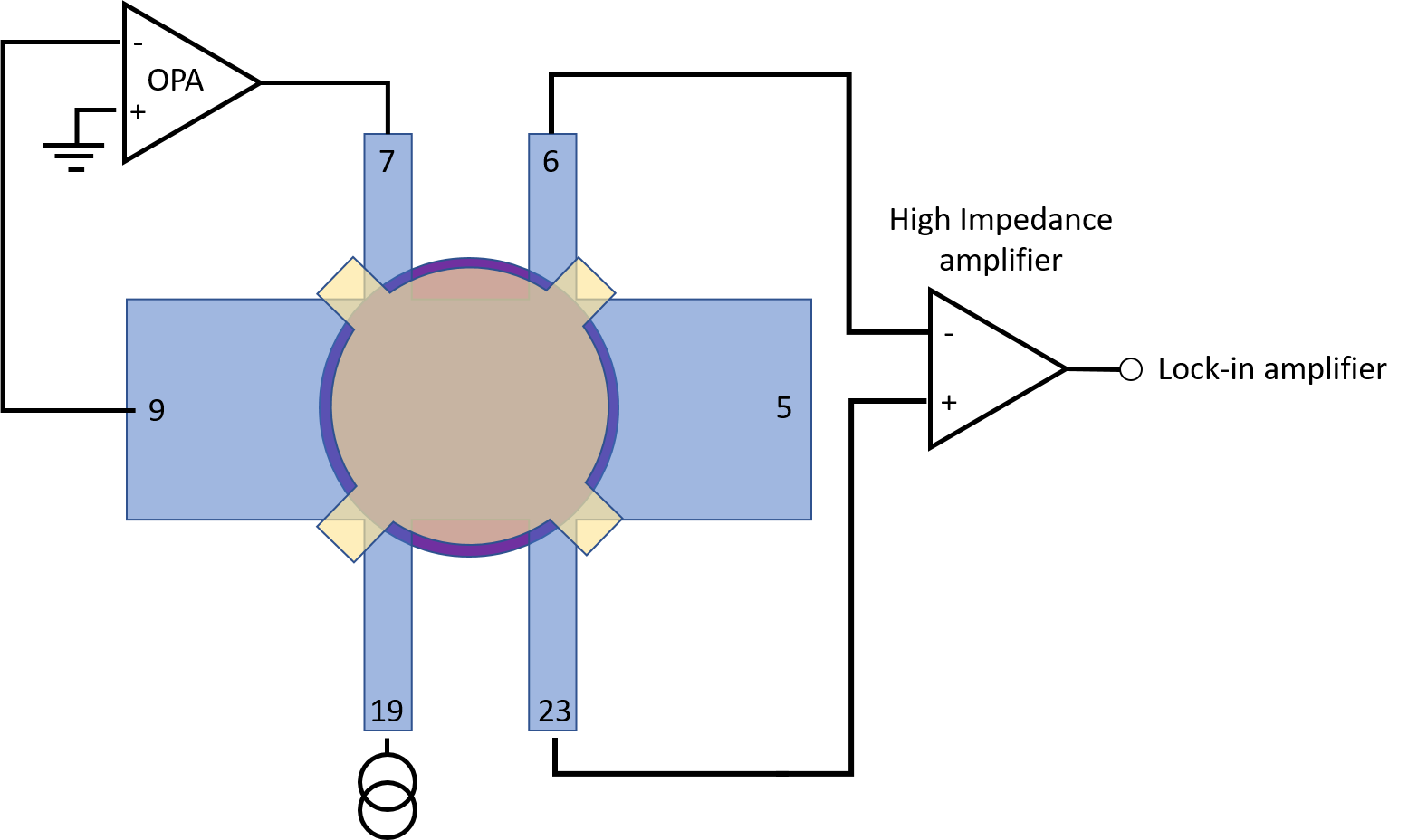}  
\caption{{\bf Non-local measurements}. Non-local configuration measurement using an operational amplifier and high input impedance voltage amplifiers.}
\label{Non-localConf}
\end{figure}

To avoid any common grounds and spurious signals we have followed the measurement scheme developed in \cite{Shimazaki2015}. This setup consists of an operational amplifier to keep the voltage of our sample balanced and high impedance amplifiers (CELIANS EPC-1B) to avoid any current leak, see Figure \ref{Non-localConf}.

Our non-local measurements are performed at low current (10 nA) to avoid Joule heating at the injector that could cause heat to flow into and past the detector region. The resulting temperature gradient along the detector region would give rise to a non-local voltage across the detector contacts via the Nernst effect, quantified by the transverse thermopower coefficient. This temperature gradient is proportional to the heating power, quadratic in current, and therefore contributes to the non-local voltage only at the second harmonic of the excitation frequency. As a result, Joule heating would not affect the first harmonic data presented in this report \cite{Renard2014}

\subsection*{Current dependence of non local measurements}

\begin{figure}
\centering
\includegraphics[scale=0.25]{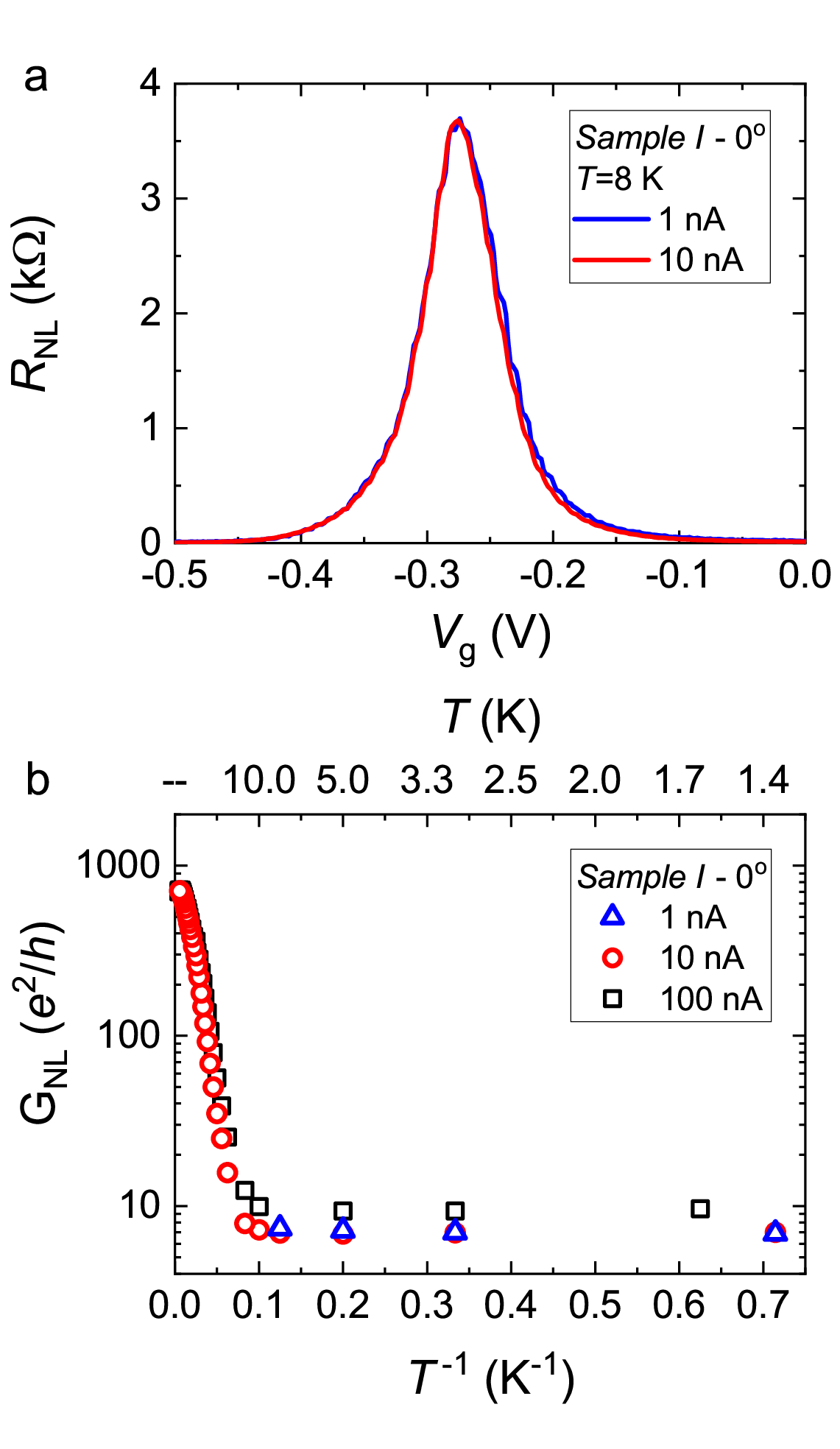}  
\caption{{\bf Non-local measurements for different applied current, sample I}. \textbf{a}, non-local resistance as a function of the gate voltage for 1 nA and 10 nA. \textbf{b}, Non-local conductivity as a function 1/$T$ for 1 nA, 10 nA and 100 nA.}
\label{Non-local-current}
\end{figure}

To ensure that our non-local measurements are not affected by the amplitude of the current we inject, we have measured the non-local signal at different injection currents, Fig. \ref{Non-local-current}a. The resemblance between the curves tells us that the magnitude of the current is not affecting our measurements. We have also performed the same measurements for different temperatures and extracted the non-local conductance (inverse of the non-local resistance at the CNP) as a function of the temperature for 1 nA, 10 nA and 100 nA, Fig. \ref{Non-local-current}b. We can see that at low temperatures for the highest measured current (100 nA) there is a slight effect reflected by a small increase of the non-local conductivity. 

%%%%%%%%%%%%%%%%%%%%%%%%%%%%%%%%%%%%%%%%%%%%

\newpage

\section*{Determining the moir\'e wavelength by magneto transport measurements}

\begin{figure}
\centering
\includegraphics[scale=0.6]{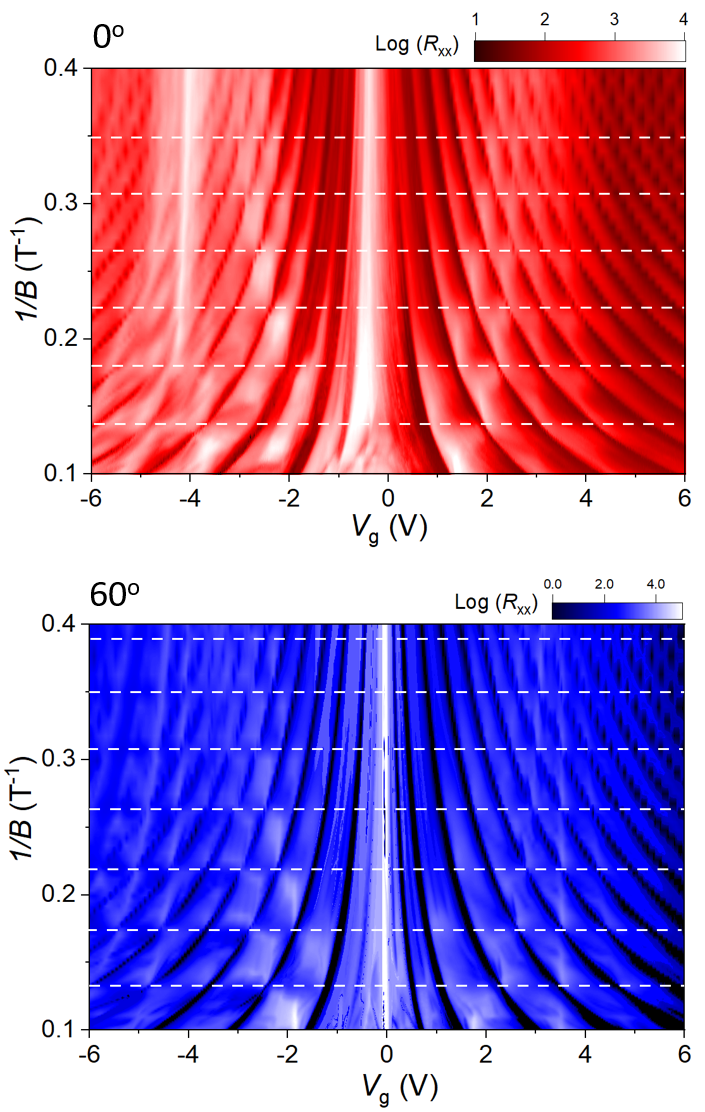}  
\caption{{\bf Magneto transport measurements, sample I}. Longitudinal resistance, in log scale, as a function of the applied gate voltage and the inverse magnetic field for \textbf{a}  $0^{\circ}$ and \textbf{b} $60^{\circ}$ of alignment at 1.4 K. Horizontal dashed lines are guides for the eyes.}
\label{Magneto-transport}
\end{figure}

By using magneto transport measurements we can determine the magnetic field at which one flux quantum threads the supperlattice unit cell. To obtain this we plot the longitudinal resistance as a function of the gate voltage (or carrier density) and 1$/B$, Fig. \ref{Magneto-transport}. In this plot we can see the appearance of equally spaced horizontal lines that appear when the Landau fans coming from the charge neutrality point and the satellite peaks intercept each other. It is important to remark that this is a purely geometrical effect and gives therefore a very accurate estimation of the moir\'e superlattice size \cite{Hunt2013}.

We obtain a space between these horizontal lines corresponding to $B_{0}=24.16$ T and $B_{0}=22.69$ T. This corresponds to $\lambda$=14.1$\pm$0.4 nm and $\lambda$=14.5$\pm$0.3 nm for $0^{\circ}$ and $60^{\circ}$, respectively.

%%%%%%%%%%%%%%%%%%%%%%%%%%%%%%%%%%%%%%%%%
\newpage

\section*{Temperature dependence and thermally activated regimes for the local and non-local resistance}

\begin{figure}
\centering
\includegraphics[scale=0.35]{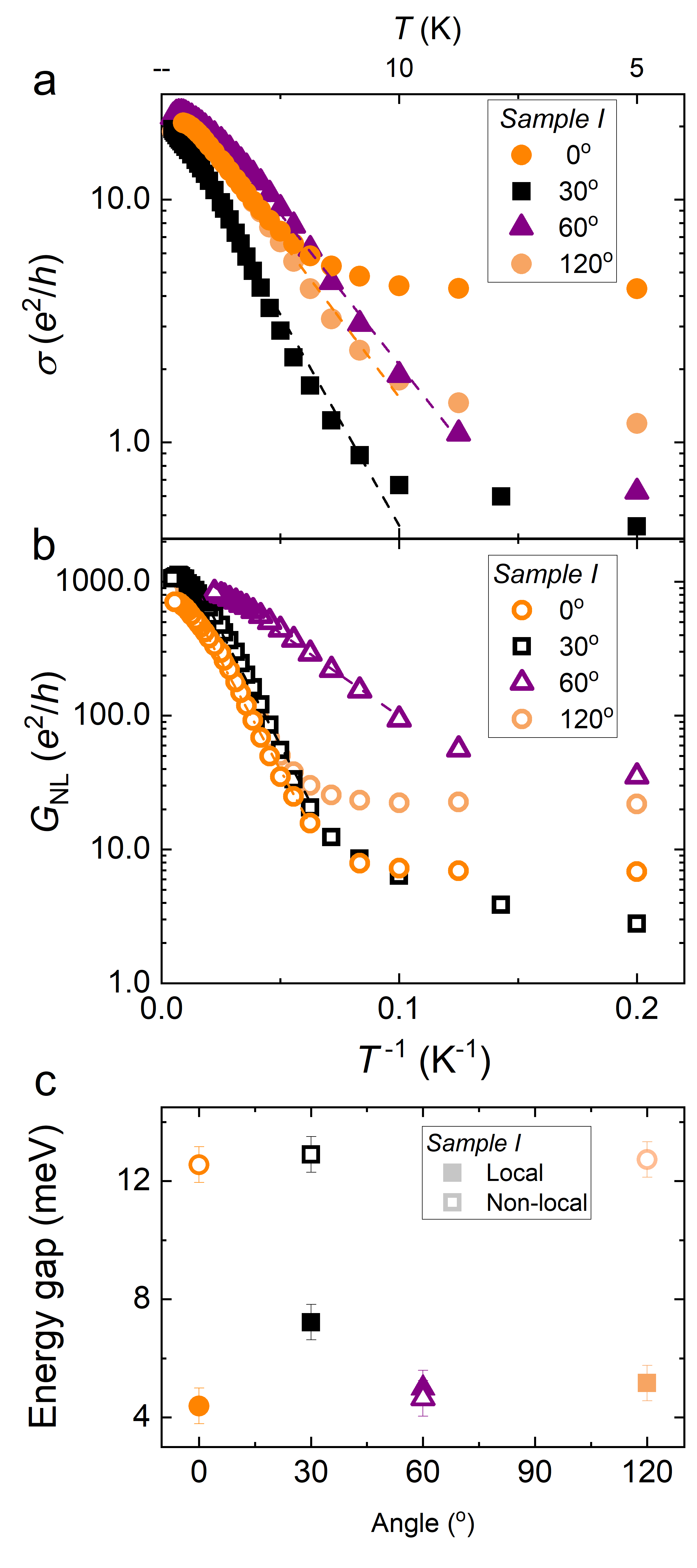} \caption{{\bf Temperature dependence of the local and non-local measurements at the CNP for $0^{\circ}$, $30^{\circ}$ and $60^{\circ}$, sample I.} \textbf{a} and \textbf{b}, Arrhenius plot for the energy gap at the CNP for the local and non-local measurements, respectively. \textbf{c} Energy gaps for the sample described in the main text.}
\label{activationZero}
\end{figure}

In our local measurements we observe different transport regimes at the CNP: first a thermally activated regime, fit with an Arrhenius law, and a hopping regime for temperatures lower than $10$ K. In Fig. \ref{activationZero}, we plot an example for the activation regime of the CNP for the measurements at 0$^{\circ}$, 30$^{\circ}$, 60$^{\circ}$ and 120$^{\circ}$ in local (a) and non-local (b) configuration, sample I. The values for the local and non-local energy gaps at the CNP for different alignments are summarized in the main text and in Fig. \ref{activationZero}c.

If we compare the energy gap obtained from the local and non-local measurements, for the $0^{\circ}$ and $120^{\circ}$ measurements,  we can easily see that, as expected, there is approximately a factor of three of difference between these energy gaps, in agreement with \cite{Yamamoto2015,Endo2019}, and supporting of the cubic relation between the local and non-local signals. It is also important to notice that the extracted values for the energy gaps, local and non-local, are at least a factor of two larger than previously reported \cite{Endo2019}, reaffirming the high quality of our samples.

\begin{figure}
\centering
\includegraphics[scale=0.35]{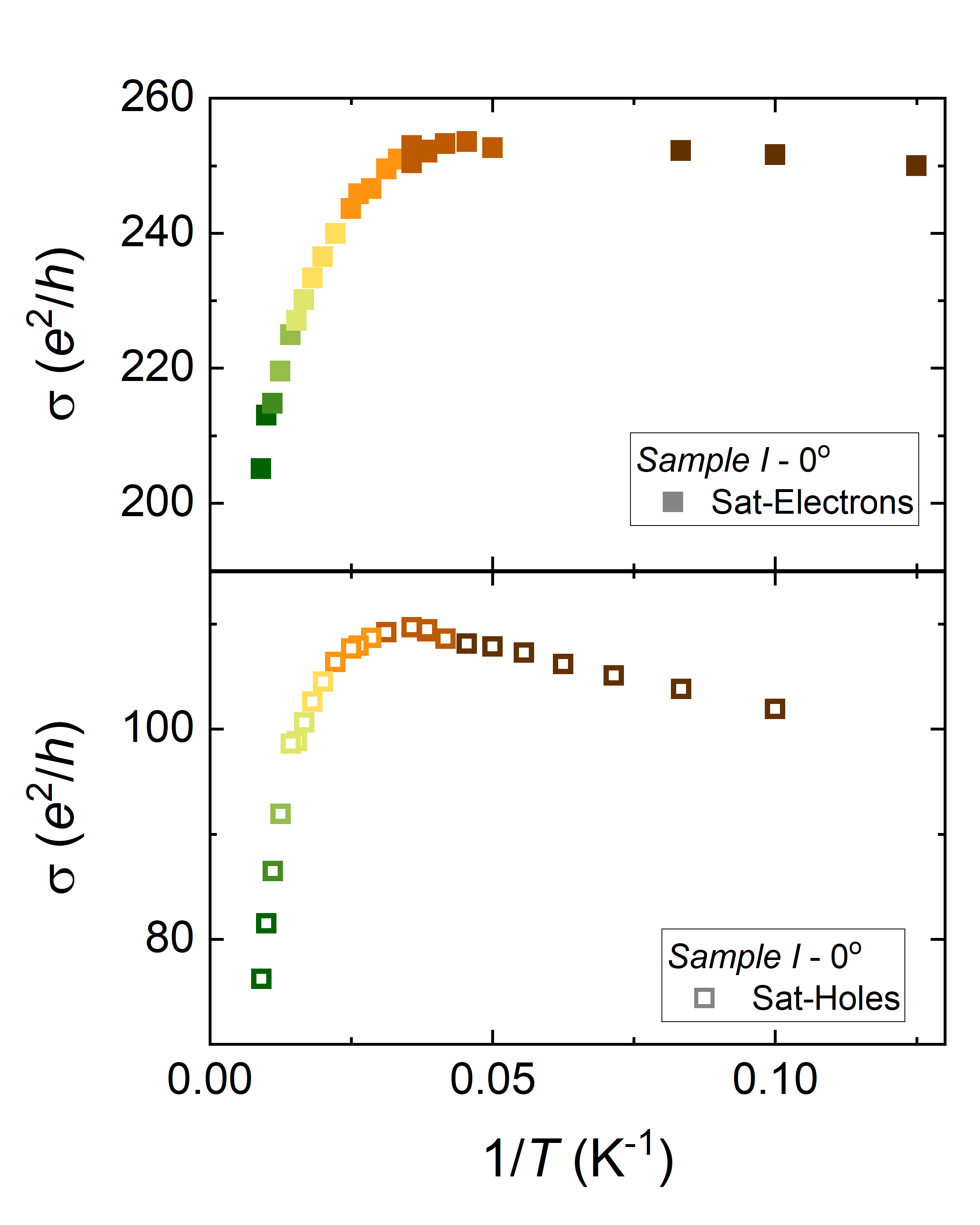} 
\includegraphics[scale=0.35]{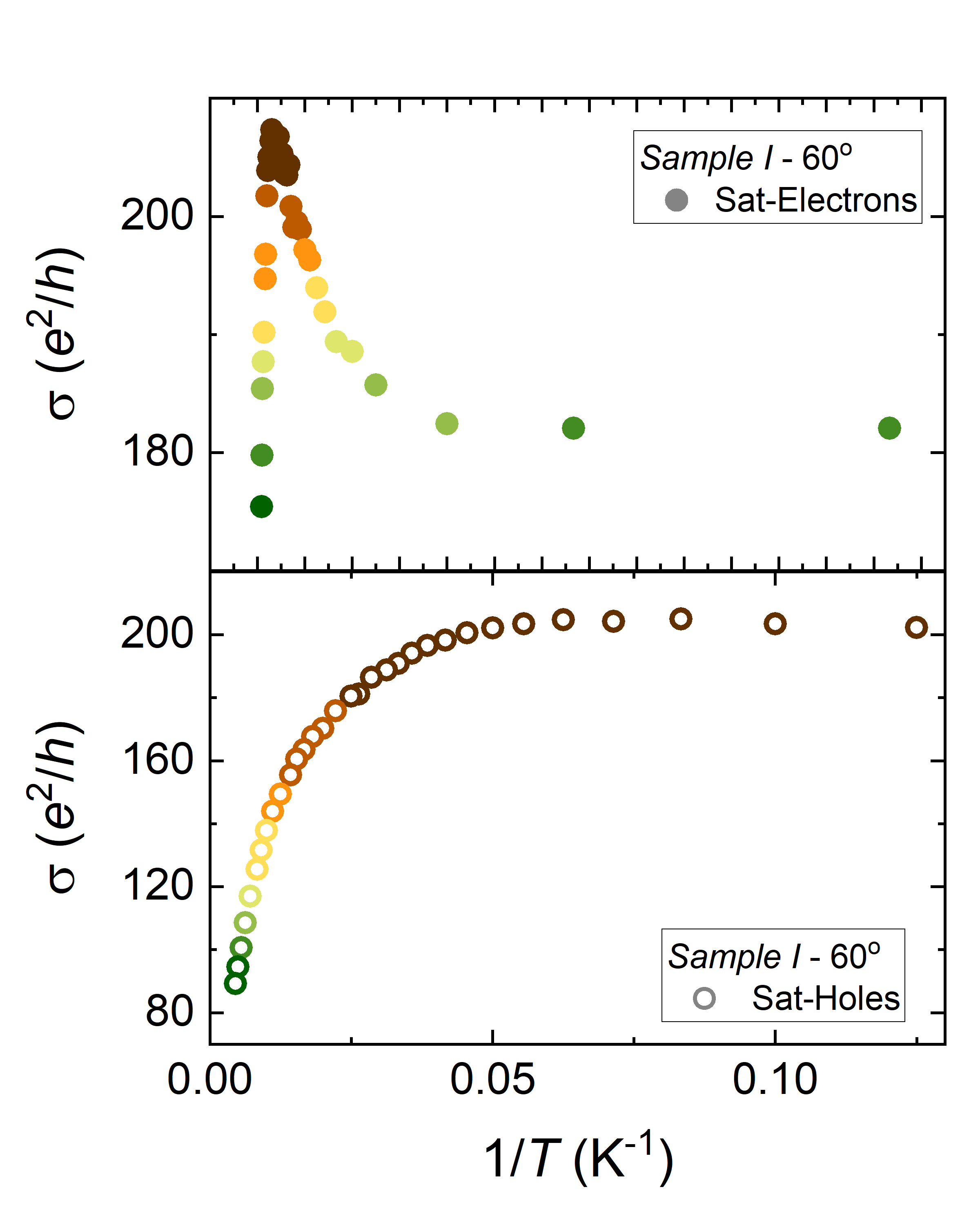} 
\caption{{\bf Temperature dependence local measurements at the satellite peaks, sample I}. Measurements at the satellite peak, for 0$^{\circ}$ (\textbf{top}) and 60$^{\circ}$ (\textbf{bottom}), for negative (electrons) and positive (holes) values of the gate voltage applied to the graphite gate, while the Si gate is kept at high doping.}
\label{activationSatSampleI}
\end{figure}

\begin{figure}
\centering
\includegraphics[scale=0.45]{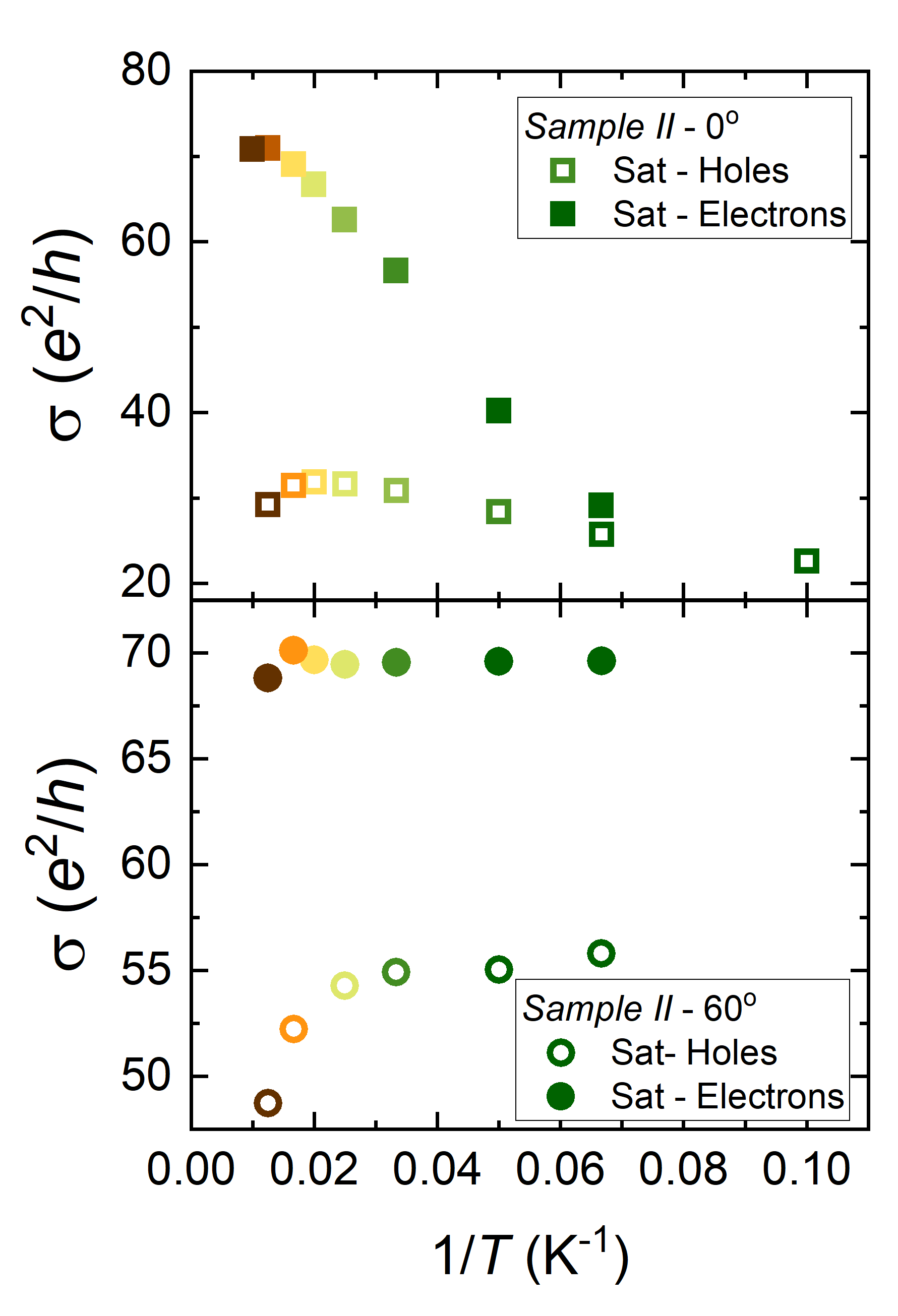} 
\caption{{\bf Temperature dependence local measurements at the satellite peaks, sample II}. Measurements at the satellite peak, for 0$^{\circ}$ (\textbf{top}) and 60$^{\circ}$ (\textbf{bottom}), for negative (electrons) and positive (holes) values of the voltage applied to the graphite gate.}
\label{activationSatSampleII}
\end{figure}

In contrast, we do not observe any activation regime at the satellite peaks , Fig. \ref{activationSatSampleI}, for sample I nor sample II Fig. \ref{activationSatSampleII}. However, we do observe a change in behavior in the electron side but for opposite alignments (60$^{\circ}$ in sample I and 0$^{\circ}$ for sample II). This behavior remains to be investigated and could be associated to the different intrinsic displacement fields in the samples.

%%%%%%%%%%%%%%%%%%%%%%%%%%%%%%%%%%%%%%%%%%%%%%%%

\subsection*{Valley Hall around the charge neutrality point}

\begin{figure}
\centering
\includegraphics[scale=0.3]{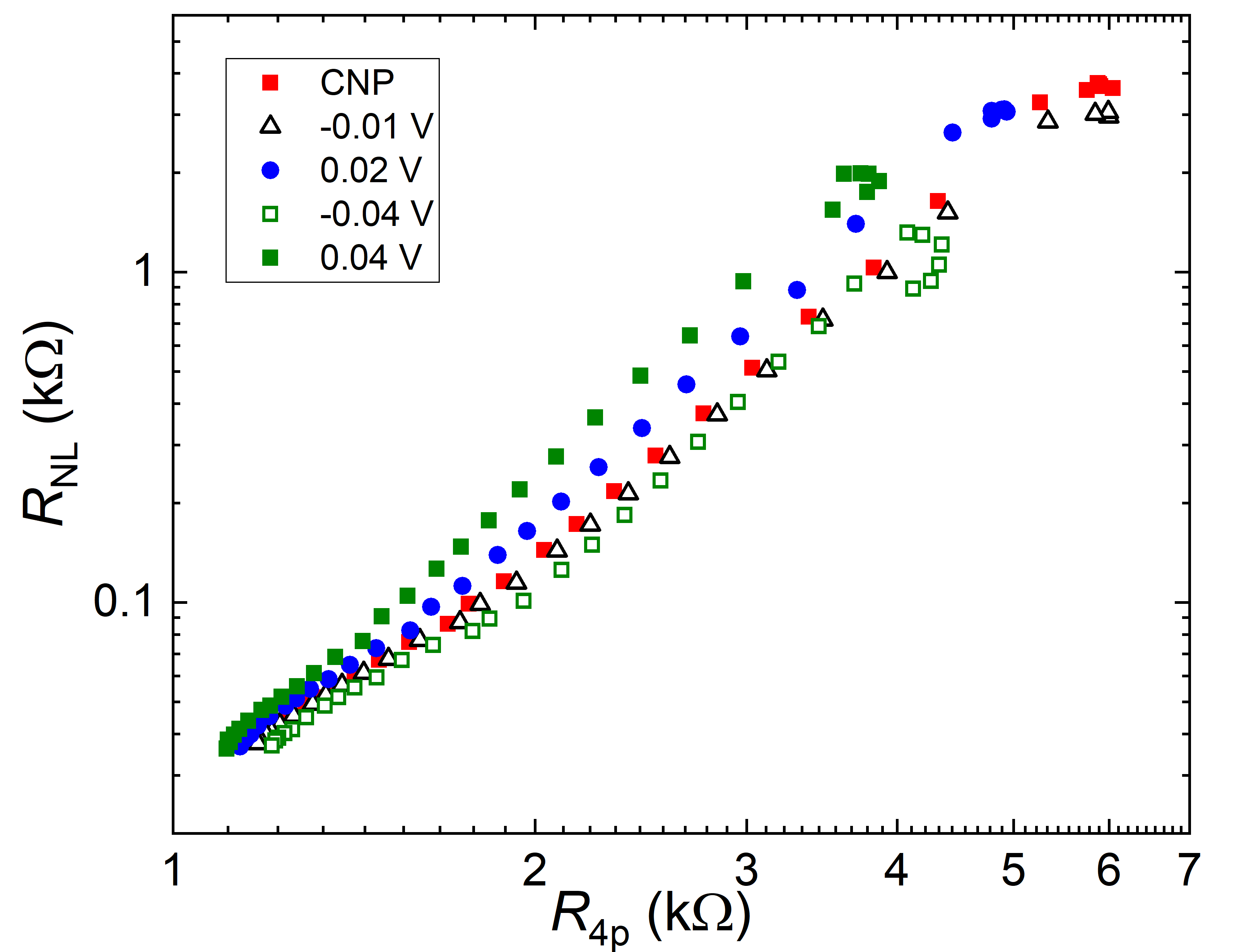}  
\caption{{\bf Cubic relation around the CNP for $0^{\circ}$, sample I}. The cubic relation can be observed for values of gate voltage close to the CNP.}
\label{activationSatdiffVg}
\end{figure}

The cubic relation of the non-local resistance as a function of the local resistance can be observed inside the conduction and valence band also. In Fig. \ref{activationSatdiffVg} we show the same cubic relation for values of gate voltage around the CNP.

\subsection*{Measurements at 120 degrees}

\begin{figure}[h!]
\centering
\includegraphics[scale=0.35]{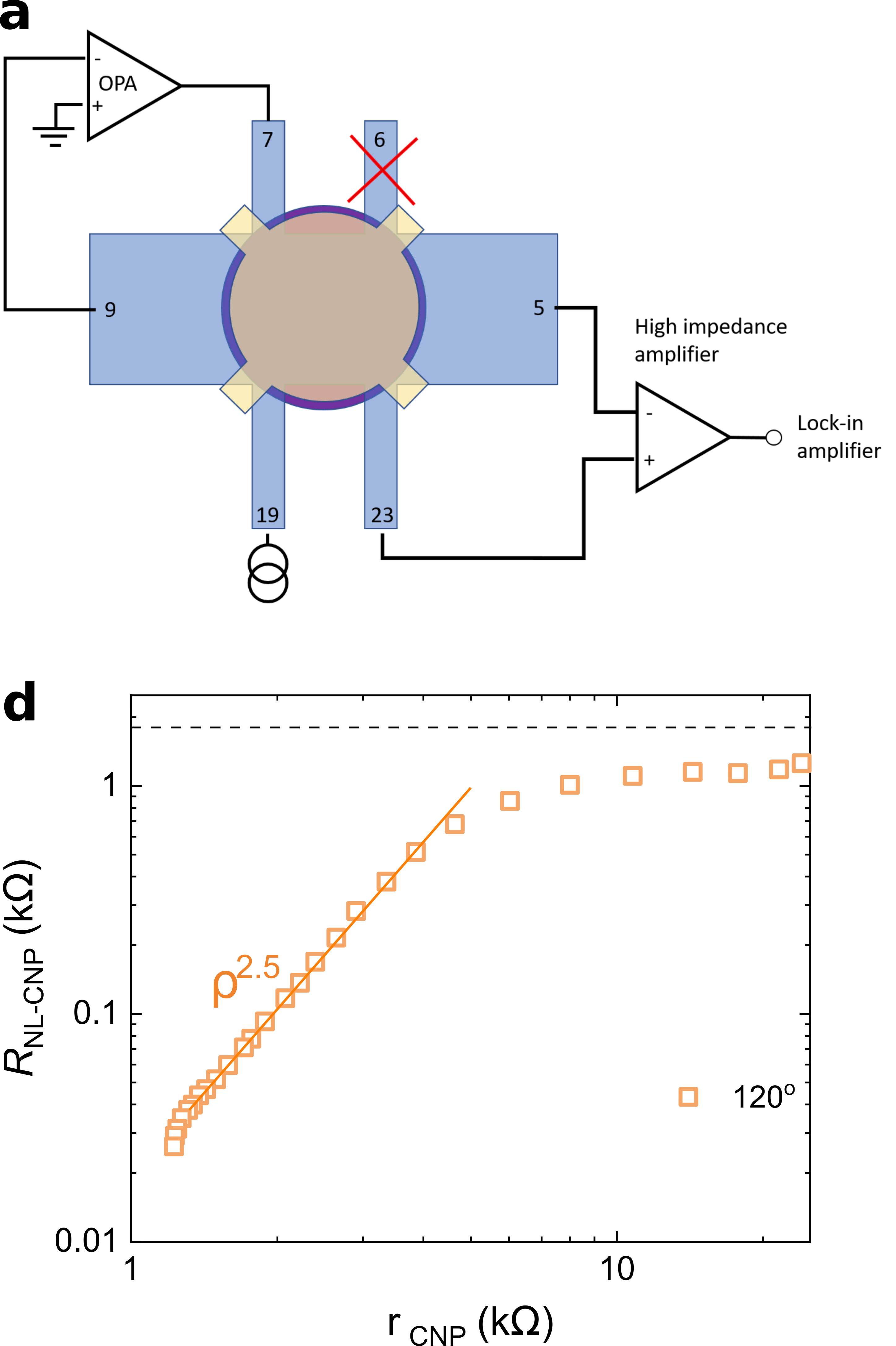}  
\caption{{\bf Non-local resistance versus local resistivity at 120$^{\circ}$ of alignment in sample I}. \textbf{a}, Measurements configuration after one contact got broken. \textbf{b}, Non-local resistance versus local resistivity for $120^{\circ}$ of alignment, sample I. Measurements between 4 K and 110 K.}
\label{120deg}
\end{figure}

We have performed local versus non-local measurements for  $120^{\circ}$ alignment, sample I. However, at the moment of the measurements contact \#6 was broken, Fig.\ref{120deg}a. This clearly impacted the saturation regime of our measurements, since it changes the geometrical configuration of our sample. However, a nearly cubic relation, $\rho^{2.5}$ is still visible, Fig. \ref{120deg}b. Considering that the width of our sample is divided by  a factor of two given the change in configuration, and using equation (2) of the main text, the expected value for the non-local resistance in the fully developed valley Hall regime is $\approx1.8$ k$\Omega$ (dashed line in Fig. \ref{120deg}b), instead we obtain  $\approx1.2$ k$\Omega$, which we consider in good agreement.

It is also important to remark that, in the same fashion as for the measurements at 0$^{\circ}$, for temperatures higher than 70 K the $\rho^{2.5}$ dependence is lost.

%%%%%%%%%%%%%%%%%%%%%%%%%%%%%%%%

\subsection*{Angular dependence in other samples}

For Sample II, the global graphite gate  generates a very large contact resistance when passing through the CNP, since there is a part of our graphene flake that is exposed. Therefore this sample is not as clean as the part covered by the BN handles. This sample architecture was improved by the ones presented in the main text (sample I and  III). 

\begin{figure}[h!]
\centering
\includegraphics[scale=0.3]{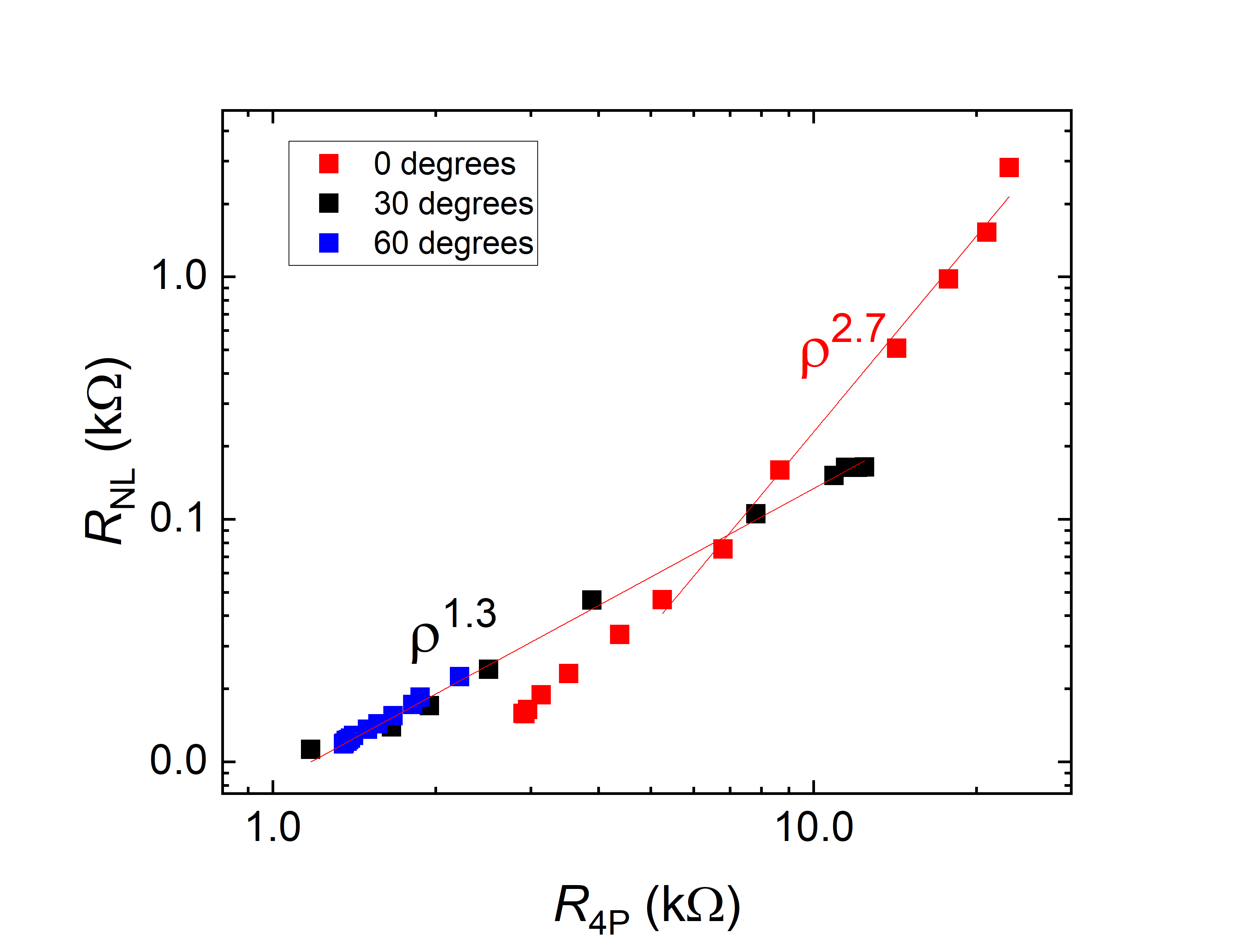}  
\caption{{\bf Non-local versus local resistance for sample II}. Non-local resistance versus four probes resistance at 0$^{\circ}$, 30$^{\circ}$ and 60$^{\circ}$. Temperature range between 8 K and 150 K.}
\label{device2}
\end{figure}

\begin{figure}[h!]
\centering
\includegraphics[scale=0.3]{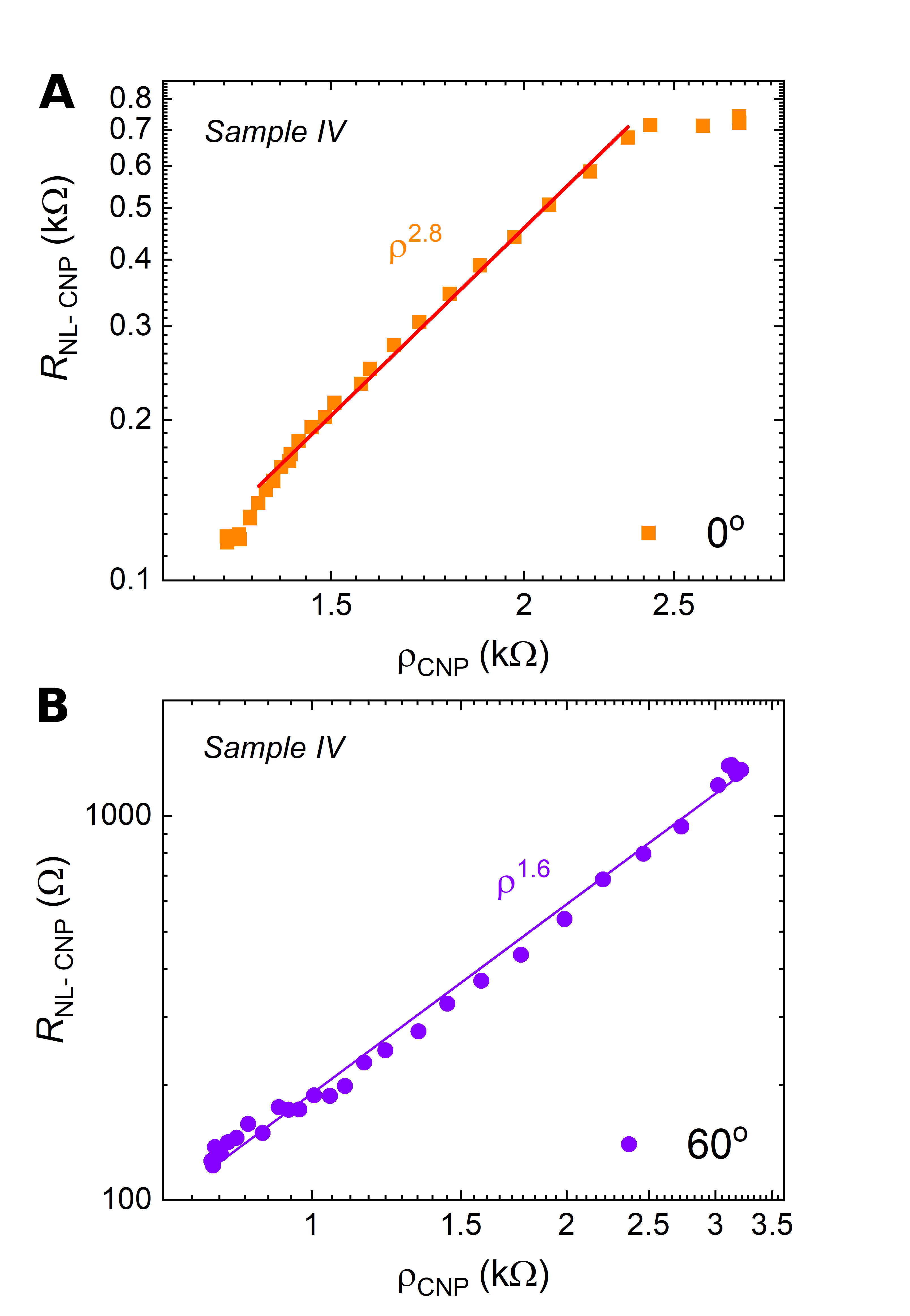}  
\caption{{\bf Non-local versus local resistance for sample IV}. Non-local resistance versus four probes resistance at 0$^{\circ}$ (A), and 60$^{\circ}$ (B). Temperature range between 1.4 K and 90 K.}
\label{device3}
\end{figure}

In sample II  (and IV) we can still observe a nearly cubic relation of the non-local and local resistance, $R_{\mathrm{NL}}\propto \rho^{2.7}$ ($R_{\mathrm{NL}}\propto \rho^{2.8})$. This relation is strongly modified for $30^{\circ}$ and $60^{\circ}$ of alignment, Fig. \ref{device2} (Fig. \ref{device3}). It is important to notice that technical problems prevent us from using the Si gate of Sample IV and therefore the tuning of the contacts, gated by the Si gate, is not performed in this measurements.

It is important to point out that  measurements on sample II were performed in a dry $4$ K cryostat with a resistive coil with max $\pm300$ mT. As we should expect the resistive coil to has no residual field, the non-local signal cannot be attributed to remnant field in the magnet.

%%%%%%%%%%%%%%%%%%%%%%%%%%%%%%%%%%%%

\subsection*{Non-local signal in magnetic field}

Non-local measurements as a function of gate voltage and magnetic field are similar to a transverse magnetic focusing (TMF) measurement but in a symmetric configuration. This means that the observed focusing lines are a reflection of the matching of cyclotron orbits with the distance between electrodes. This measurements are highly sensitive to modifications on the band structure and can reflect for example the presence of van Hove singularities at saddle points in the band structure \cite{Lee2016}. Although the detailed analysis of these 2D plots is out of the scope of this manuscript, we would like to point out the main differences in these 2D plots, which reflect differences in the electronic band structure of the different alignments \cite{Lee2016,Berdyugin2020}.  

\begin{figure*}
\centering
\includegraphics[scale=0.22]{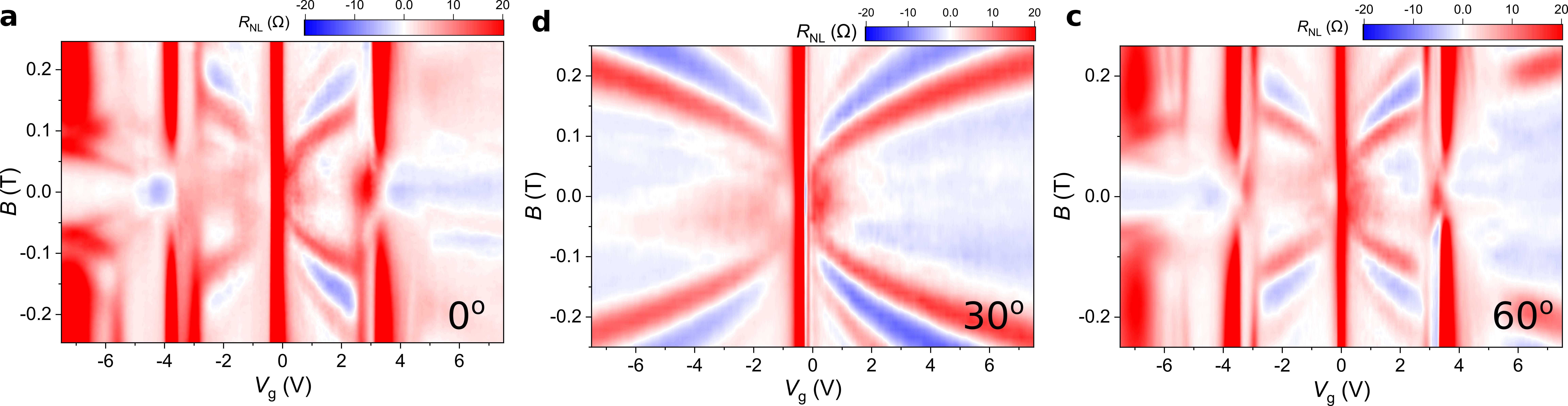}  
\caption{{\bf Non-local resistance as a function of the magnetic field and density, sample I}. \textbf{a}, 0$^{\circ}$; \textbf{b}, 30$^{\circ}$ and  \textbf{c}, 60$^{\circ}$. All measurements are taken at 10 K.}
\label{TMF-sup}
\end{figure*}

The strong differences in the non-local response resembles the one of  systems with very different electronic band structures, as we proposed in the main text.  In Fig. \ref{TMF-sup} we can observe the TMF 2D plots for three crystallographic alignments: 0$^{\circ}$, 30$^{\circ}$ and 60$^{\circ}$.  We highlight the main features that support the hypothesis of strongly different electronic band structures. The 2D  map observed at 30$^{\circ}$, Fig.\ref{TMF-sup}b, is the expected one for the unaltered band structure of bilayer graphene\cite{Taychatanapat2013}. In this we observe uninterrupted magneto focusing lines through all the carrier density range. In the case of 0$^{\circ}$ and 60$^{\circ}$ we observe an increase of the resistance at the gate voltage which corresponds to the satellite peak. This has been attributed to the presence of saddle points in the band structure \cite{Lee2016}. Between the two satellite peaks the TMF signal do not show any particular behaviour. However, for carrier densities beyond the satellite peaks we observe strong differences between the two alignments, the most remarkable one being the fact that the magnetic focusing peaks of the central band seem to propagate beyond the satellite peak in the 60$^{\circ}$ of alignment, Fig. \ref{TMF-sup}c. The strong difference between these TMF plots is a clear indication of the existence of different electronic band structures.

%%%%%%%%%%%%%%%%%%%%%%%%%%%%%%%%%%%%

\subsection*{Stacking configurations for BN/bilayer graphene}

\begin{figure}[h!]
\centering
\includegraphics[scale=0.15]{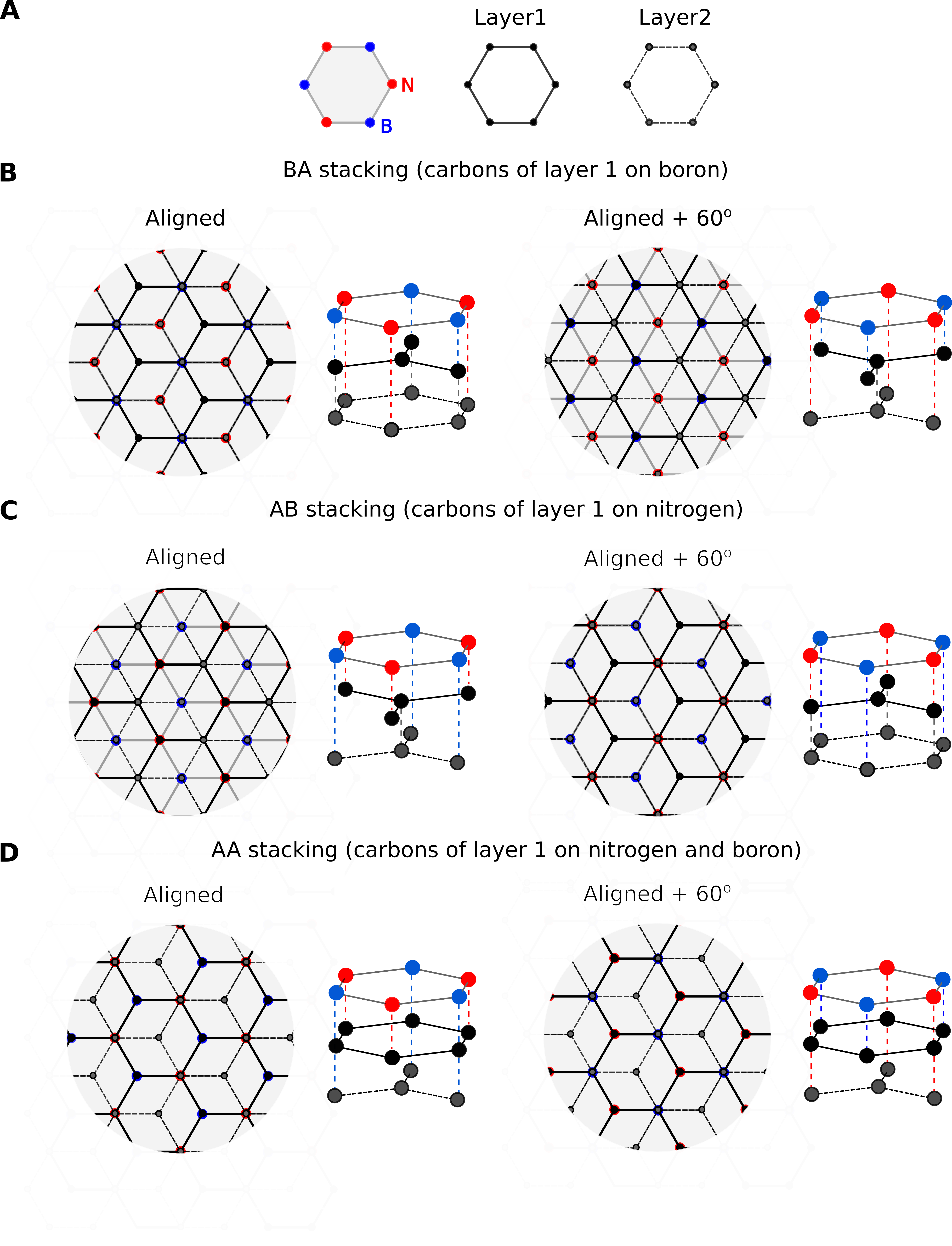}  
\caption{{\bf Stacking configurations for BN/bilayer graphene}. {\bf A}, decription of the differnet layers. {\bf B, C and D}, Atomic configurations inside the moir\'e cell, in the commensurate state, for different stacking configurations of the first layer with respect to the BN: BA - carbon on boron; AB - carbon on nitrogen and AA - carbon in boron and nitrogen. Each is represented for both aligned positions 0$^{\circ}$ and 60$^{\circ}$.}
\label{AtomicConf}
\end{figure}

In Fig. \ref{AtomicConf} we show all the possible atomic configurations, from the more energetically favorable \cite{Jung2015} BA to the least one AA. For all of those we can see that a sixty degrees rotation of the BN layer gives rise to a different atomic configuration. 

%%%%%%%%%%%%%%%%%%%%%%%%%%%%%%%%%%%%

\subsection*{Sample for structural characterization}

\begin{figure}[h!]
\centering
\includegraphics[scale=0.5]{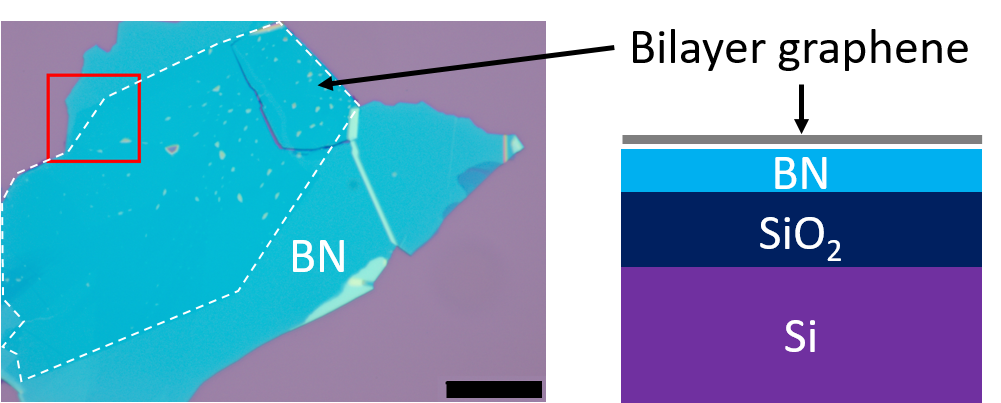}  
\caption{{\bf Sample for structural characterization}. Optical micrograph of the sample (left). Dashed line represents the graphene area. Scale bar 20 $\mu$m. Cross section of the sample (right).}
\label{Gaia001}
\end{figure}

We have prepared a sample to measure with the AFM the out of plane atomic displacement, wrinkles around the moir\'e cell are formed as a consequence of the atomic in plane displacement inside the moir\'e cell. This sample consist of BN and bilayer graphene on surface. The two crystals have been pre-aligned at the moment of the preparation by aligning their crystallographic edges. A picture of this sample and its cross section are shown in Fig. \ref{Gaia001}.  

The bilayer character of the sample has been tested by Raman spectroscopy. The AFM measurement in this sample, shown in the main text shows the clear presence of a moir\'e superlattice where the in-plane atomic displacement gives rise to relaxation regions around the moire cell.

%%%%%%%%%%%%%%%%%%%%%%%%%%%%%%%%%%%%%%
\subsection*{Electronic models}

To compute the electronic structure of the considered graphene/hBN systems, we employed the $p_z$ tight-binding Hamiltonian, similar to those presented in \cite{TramblydeLaissardiere2010,Moon2014}. In particular, the Hamiltonian is written as
\begin{equation}
    H_{tb} = \sum_{n} V_n a^\dag_n a_n + \sum_{n,m} t_{nm} a^\dag_n a_m \nonumber
\end{equation}
where the on-site energies $V_n$ = 0, 3.34 eV, and -1.4 eV for carbon, boron, and nitride atoms, respectively. The hopping energies $t_{nm}$ are determined using the standard Slater-Koster formula
\begin{eqnarray}
	t_{nm} (r_{nm}) &=& V_{pp\pi} \sin^2 \phi_{nm} + V_{pp\sigma} \cos^2 \phi_{nm} ,  \nonumber \\
	V_{pp\pi} &=& V^0_{pp\pi} \exp \left( (a_0 - r_{nm})/r_0 \right), \nonumber \\
	V_{pp\sigma} &=& V^0_{pp\sigma} \exp \left( (d_0 - r_{nm})/r_0 \right)  \nonumber	
\end{eqnarray}
where the direction cosine of ${\vec r}_{nm}$ along Oz axis is $\cos \phi_{nm} = z_{nm}/r_{nm}$, $r_0 = 0.184a$, $a_0 = a/\sqrt 3$, and $d_0 = 3.35$\AA ~ while $a \simeq 2.49$\AA.

%%%%%%%%%%%%%%%%%%%%%%%%%%%%%%%%%%%%%%%%%%%%%%%%%%%%
\subsection*{Atomic structure relaxation and electronic band structures}

\begin{figure*}[h!]
\centering
\includegraphics[width=0.9\textwidth]{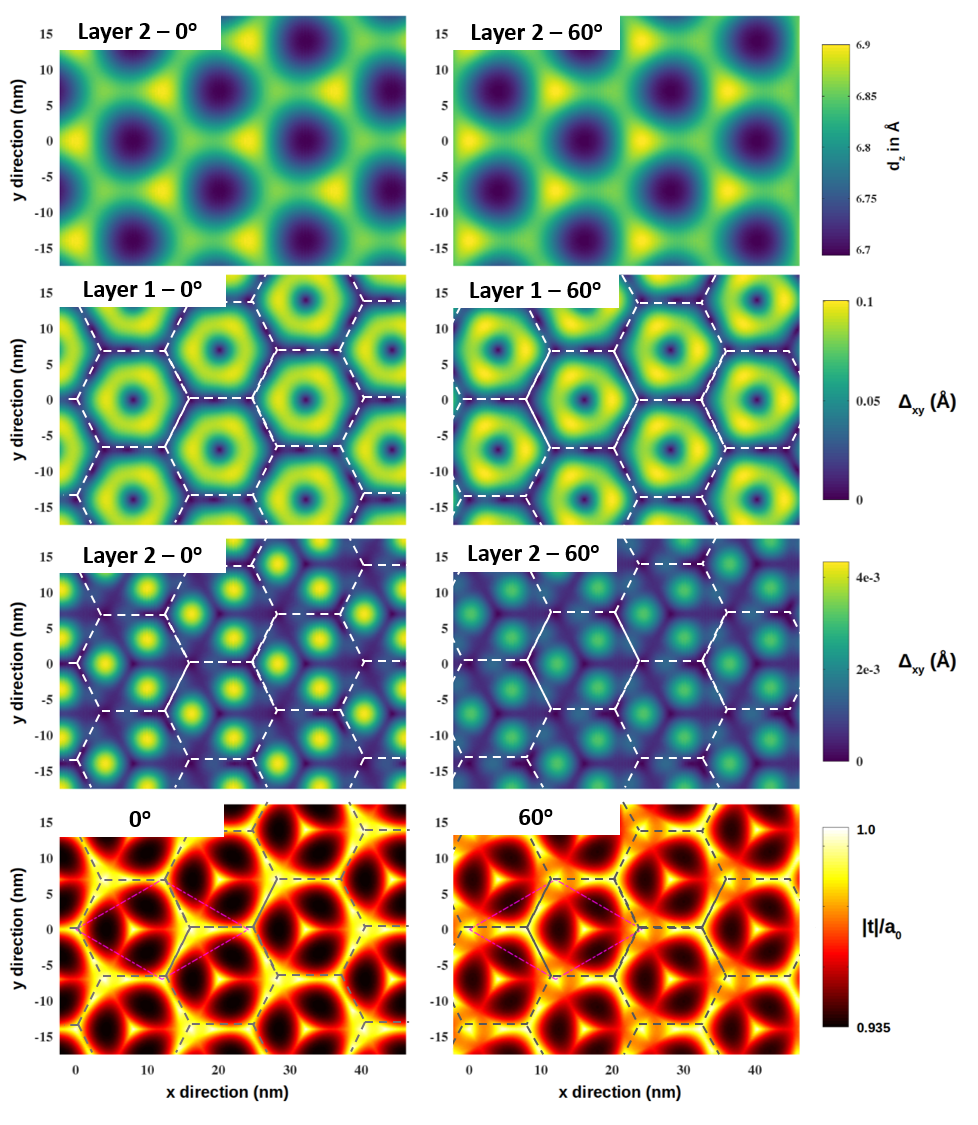}  
\caption{{\bf Atomic structure relaxation}. Bilayer graphene/hBN are aligned at $0^{\circ}$ (left) and $60^{\circ}$ (right). The variation of interlayer distance $d_z$ between the second graphene layer  and hBN one is presented in the two top images. The four images below illustrate the variations of the in-plane displacements $D_{xy}$ in both graphene layers. At last, the two bottom images display the modification of stacking configuration between two graphene layers, by illustrating the variation of stacking vector \textbf{t}. \textbf{t} is determined as the translation vector applied locally to a C-ring of one graphene layer to recover the AA stacking configuration at the considered position, i.e., $|$\textbf{t}$|$ = $a_0$ and 0 for the local AB and AA stacking configurations, respectively.}
\label{SimulationSI}
\end{figure*}

\begin{figure*}[h!]
\centering
\includegraphics[width=0.7\textwidth]{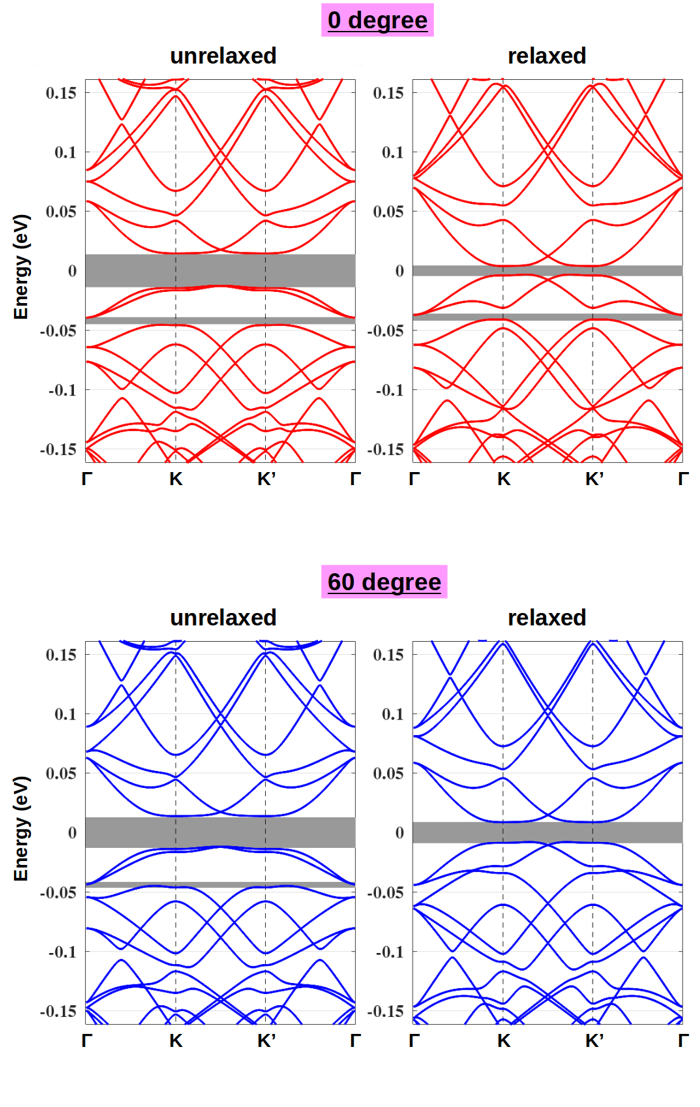}  
\caption{{\bf Electronic band structures}. Bilayer graphene/hBN are aligned at $0^{\circ}$ (top) and $60^{\circ}$ (bottom). Ideal pristine atomic structure of the graphene layers are preserved on the left, while atomic structure relaxations are considered on the right. Numerical simulation without the presence of displacement field.}
\label{SimulationSI2}
\end{figure*}

More details related to atomic structure relaxation are presented in Fig. \ref{SimulationSI}, including additionally the variation of interlayer distance between graphene/hBN layers and the modification of stacking structure, that is due to atomic reconstruction effects. In Fig. \ref{SimulationSI2}, the electronic band structures computed for both unrelaxed and relaxed lattices are displayed. While it is almost negligible in the unrelaxed case, the presented results clearly demonstrate that the effects of atomic structure relaxation essentially govern the significant difference in the electronic structures of $0^\circ$- and $60^\circ$- alignments. 

\newpage

\subsection*{Other theories applied to the valley Hall effect}

The valley Hall effect can also be interpreted in terms of two leading theories:

{\it Berry curvature driven valley currents:} If we consider  the existence of a Berry curvature, $\Omega(\mathbf{k})$ \cite{Gorbachev2014,Komatsu2018,Shimazaki2015,Sui2015,Li2020}, which acts as a pseudo-magnetic field, this will give rise to an ``anomalous'' velocity, perpendicular to the external longitudinal electric field $\mathbf{E}$. The electron velocity, $\mathbf{v}$,  then becomes:

\begin{equation}
    \mathbf{v}(\mathbf{k})= \frac{\partial \epsilon(\mathbf{k})}{\hbar\partial\mathbf{k}} -\frac{q}{\hbar}\mathbf{E}\times\Omega(\mathbf{k}), 
    \label{anomalousVelocity}
\end{equation}

\noindent where $k$ is the wavevector, $\epsilon(\mathbf{k})$ is the band energy, $q$ is the carrier charge and $\hbar$ is the reduced Planck constant. From equation \ref{anomalousVelocity} we can see that the larger the Berry curvature, the stronger will be the anomalous velocity. The Berry curvature in bilayer graphene is given by \cite{Yin2022}:

\begin{equation}
    \Omega(\mathbf{p})=\frac{2\hbar^2\nu^4\gamma\Delta |\mathbf{p}|^2}{(\nu^4\mathbf{p}^4+\gamma^2\Delta^2)^{3/2}},
\end{equation}

\noindent where $\Delta$ is the value of the energy gap, $\gamma$ is the interlayer coupling, $\nu$ is the Fermi velocity in monolayer and $\mathbf{p}$ is momentum. This means that the Berry curvature will spike to a maximum at very small values of the energy gap and then reduce rapidly as the energy gap increases. We could therefore explain the contrast between $0^{\circ}$ and $60^{\circ}$ by a difference in the energy gaps at the CNP as predicted by our numerical simulations without displacement field, Fig \ref{SimulationSI2}. However, this is not consistent with our experimental observations where the energy gap is, between error bars, the same for all the alignments. In the following section we present calculations of the Berry curvature from the electronic band structures presented in the main text.\\

{\it Spatially varying regions of broken sublattice symmetry}: Recent theoretical calculations propose that the valley Hall effect observed in monolayer graphene aligned with BN \cite{Gorbachev2014,Komatsu2018,Li2020} originates from the spatial variation of the broken sublattice symmetry\cite{Aktor2021}. If this effect is at the origin of the valley Hall effect in monolayer graphene the picture becomes more complicated when dealing with bilayer graphene. Following the results of our numerical simulations we can  say that the spatial variations of broken sublattice symmetry will be different between the two layers, and it will always exist for the first layer. It is then not evident why the valley effect is observed for only one of the two layer alignments, and clearly, further numerical investigations would be needed to clarify the situation.

\subsection*{Berry curvature calculation}

\begin{figure*}[h!]
\centering
\includegraphics[width=0.7\textwidth]{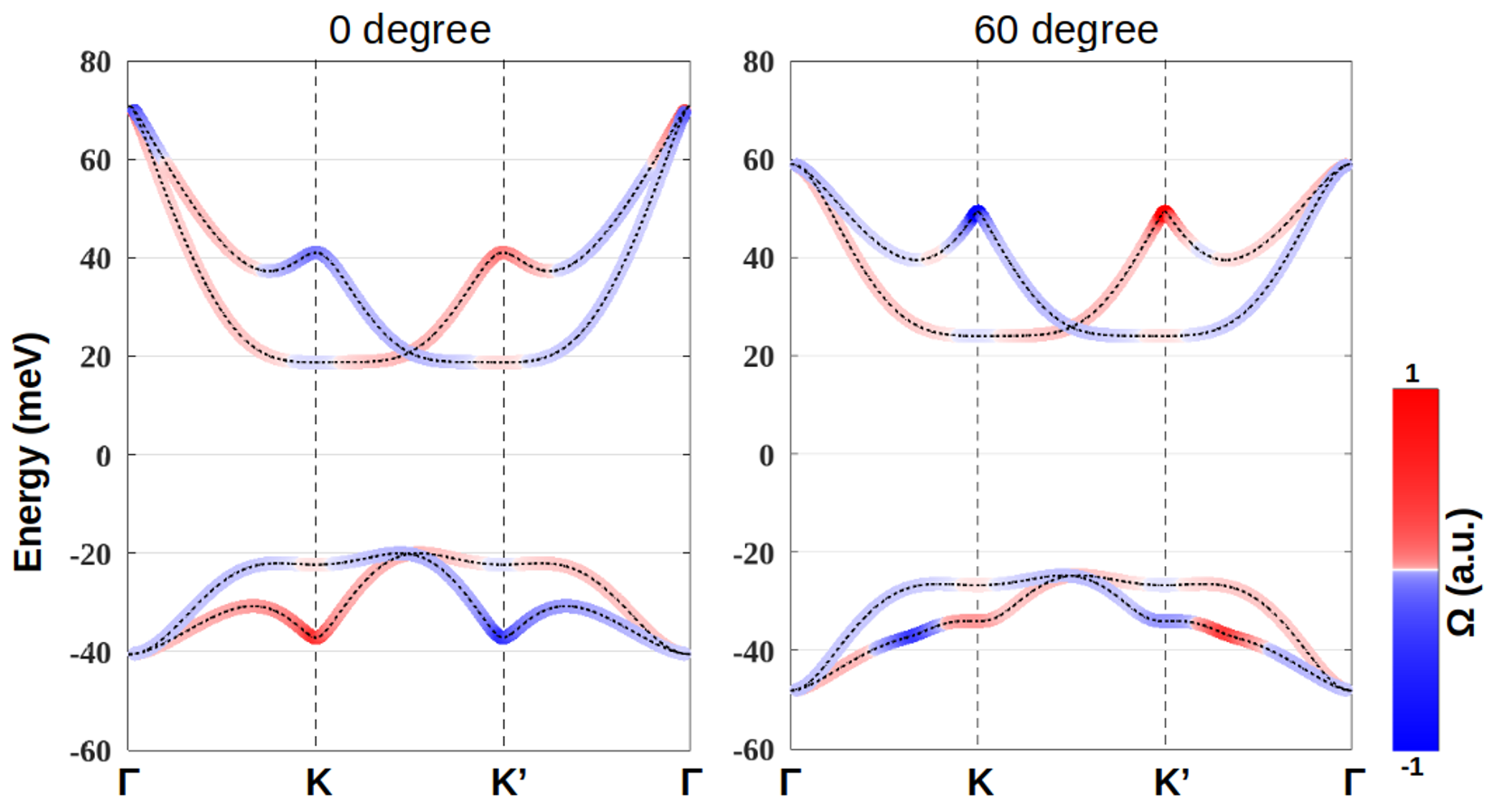}  
\caption{{\bf Berry curvature} computed near the zero energy point of fully aligned structures and when a displacement field of -0.1 V/nm is applied (see Fig.4 in the main text). }
\label{SimulationBerryCurvature}
\end{figure*}

The Berry curvature is numerically computed using the following equation \cite{Jiaqi2021}
\begin{equation}
\Omega_n (\mathbf{k}) = \hbar^2 \sum_{m \neq n} \frac{-2  \mathrm{Im} [\left \langle n\mathbf{k} \left | \hat{v}_x \right | m\mathbf{k} \right \rangle \left \langle m\mathbf{k} \left | \hat{v}_y \right | n\mathbf{k} \right \rangle]}{(\varepsilon_{n\mathbf{k}}-\varepsilon_{m\mathbf{k}})^2}
\end{equation}
where $\left | n\mathbf{k} \right \rangle$ and $\varepsilon_{n\mathbf{k}}$ are the eigenwavefunctions and eigenvalues, respectively, computed using the tight binding Hamiltonian presented above. The numerical resutls for each alignment are presented in Fig. \ref{SimulationBerryCurvature} for both alignments.

\subsection*{Effective mass}

\begin{figure*}[h!]
\centering
\includegraphics[width=0.65\textwidth]{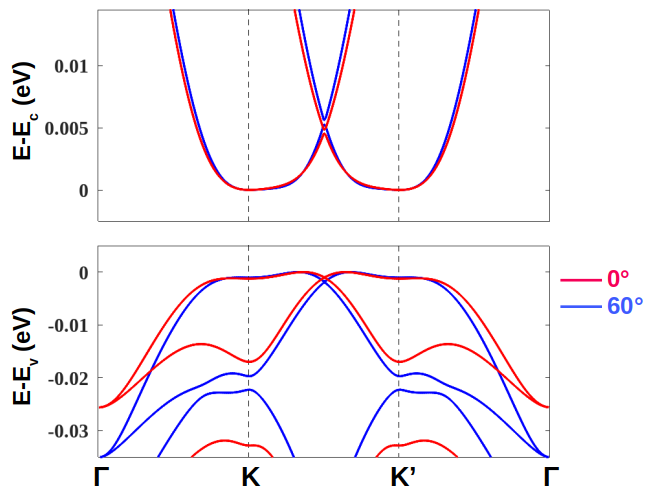}  
\caption{{\bf Effective mass comparison:} the bandstructures of both aligned cases in Fig.4 are zoomed in around the bandgap.}
\label{Effective mass}
\end{figure*}

Given their complexity as seen in Fig.4, the low energy bands of two $0^\circ$ and $60^\circ$ aligned cases can not be described using a simple effective model. Hence, these bands are zoomed in and presented in Fig. \ref{Effective mass} and by this way, their effective masses can be estimated and roughly compared. In particular, by considering the band curvature, we can conclude that the charge carriers in the $0^\circ$ case are heavier than those in the $60^\circ$ one.

%\bibliography{references}

\end{document}